%% file: paper.tex
\documentclass[a4paper,11pt]{article}

\usepackage{jheppub}
\usepackage{graphicx}
\usepackage{amsmath}
\usepackage{slashed}
\usepackage{braket}
\usepackage[export]{adjustbox}
\usepackage{enumitem}
\usepackage{placeins}
\usepackage[normalem]{ulem}
\usepackage{multirow}
\usepackage{hyperref}

\definecolor{mycolor}{rgb}{0.984375, 0.398438, 0.}
\hypersetup{
    pdftitle={Collinear Approximations for LHC Cross Sections: Factorization and Resummation},
    colorlinks=true,
    linkcolor=mycolor,
    citecolor=mycolor,
    filecolor=mycolor,      
    urlcolor=mycolor
}

\newcommand{\refcite}[1]{ref.~\cite{#1}}
\newcommand{\refscite}[1]{refs.~\cite{#1}}

\newcommand{\eq}[1]{eq.~\eqref{eq:#1}}
\newcommand{\eqs}[2]{eqs.~\eqref{eq:#1} and \eqref{eq:#2}}

\renewcommand{\sec}[1]{section~\ref{sec:#1}}
\newcommand{\Sec}[1]{Section~\ref{sec:#1}}
\newcommand{\secs}[2]{sections~\ref{sec:#1} and \ref{sec:#2}}

\newcommand{\app}[1]{appendix~\ref{app:#1}}
\newcommand{\fig}[1]{figure~\ref{fig:#1}}

\newcommand{\strictlim}{\lim\limits_{\text{strict }n-\text{coll.}}}

\newcommand{\muprime}{\mu^{\prime}}
\newcommand{\df}{\mathrm{d}}

\newcommand{\bn}{{\bar n}}

\newcommand{\cI}{\mathcal{I}}

\newcommand{\cO}{\mathcal{O}}

\newcommand{\wa}{{w_1}}
\newcommand{\wb}{{w_2}}

\newcommand{\as}{\alpha_s}
\newcommand{\GammaC}{\Gamma_{\rm cusp}}
\newcommand{\GammaCS}{-\Gamma^r_{\rm cusp}}
\newcommand{\gammaS}{\gamma^r_S}
\newcommand{\GammaCB}{\Gamma^i_{\rm cusp}}
\newcommand{\gammaB}{\gamma^i_B}
\newcommand{\nn}{\nonumber}

\def\beq{\begin{equation}}
\def\eeq{\end{equation}}
\def\bea{\begin{eqnarray}}
\def\eea{\end{eqnarray}}

\newcommand{\rapidityfactorization}{factorization for the collinear approximation}

\catcode`\@=11
\font\manfnt=manfnt
\def\Watchout{\@ifnextchar [{\W@tchout}{\W@tchout[1]}}
\def\W@tchout[#1]{{\manfnt\@tempcnta#1\relax%
  \@whilenum\@tempcnta>\z@\do{%
    \char"7F\hskip 0.3em\advance\@tempcnta\m@ne}}}
\let\foo\W@tchout
\def\dubious{\@ifnextchar[{\@dubious}{\@dubious[1]}}

\def\@dubious[#1]{%
  \setbox\@tempboxa\hbox{\@W@tchout#1}
  \@tempdima\wd\@tempboxa
  \list{}{\leftmargin\@tempdima}\item[\hbox to 0pt{\hss\@W@tchout#1}]}
\def\@W@tchout#1{\W@tchout[#1]}
\catcode`\@=12

\preprint{CERN-TH-2025-041,SLAC-PUB-250203}

\title{Collinear Approximations for LHC Cross Sections: Factorization and Resummation}

\author[a]{Bernhard Mistlberger,}
\emailAdd{bernhard.mistlberger@gmail.com}
\author[b]{Gherardo Vita}
\emailAdd{gherardo.vita@cern.ch}

\affiliation[a]{SLAC National Accelerator Laboratory, Stanford University, Stanford, CA 94039, USA}
\affiliation[b]{CERN Theory Division, CH-1211, Geneva 23, Switzerland}

\abstract{
We explore a factorization theorem for color singlet production cross sections at the LHC in the limit of additional radiation becoming collinear to the direction of either of the colliding protons. 
The resulting formula approximates the cross section as a function of the Born variables of the color singlet final state, specifically its mass and rapidity.
We analyze the quality of this approximation and study its limitations at the example of gluon-fusion Higgs boson production and Drell-Yan production through next-to-next-to-leading order in QCD perturbation theory.
Furthermore, we resum logarithms enhanced in the collinear limit to next-to-next-to-next-to-leading logarithmic accuracy for our two example processes.
We conclude that this framework of collinear approximation is a natural successor to the threshold approximation and resummation for inclusive and differential observables in color-neutral processes.
}

\begin{document}
\maketitle
\input{Chapters/Introduction.tex}
\input{Chapters/Setup.tex}
\input{Chapters/FixedOrderXS.tex}

\input{Chapters/PhenoStudy.tex}
\input{Chapters/Resummation.tex}
\input{Chapters/Conclusions.tex}

\section{Acknowledgements}
We would like to thank Stefano Forte, Martin Beneke, Alexander Huss, and Lance Dixon for fruitful discussions.
BM is supported by the United States Department of Energy, Contract DE-AC02-76SF00515.
\appendix
\input{Chapters/Transforms.tex}
\input{Chapters/OpDefandMellin.tex}

\addcontentsline{toc}{section}{References}
\bibliographystyle{jhep}
\bibliography{refs}

\end{document}

%% file: Chapters/Introduction.tex
\section{Introduction}
As collider experiments grow increasingly precise, accurate theoretical predictions become a critical factor in testing the Standard Model. 
A key role in the theory precision program for colliders, such as the Large Hadron Collider (LHC), is played by higher order calculations in perturbative QCD.
In the last few years,  significant progress has been made in this direction allowing to take the level of accuracy for certain inclusive LHC cross sections to next-to-next-to-next-to-leading order (N$^3$LO) ~\cite{Anastasiou:2015ema,Dreyer:2016oyx,Mistlberger:2018etf,Dulat:2018bfe,Dreyer:2018qbw,Duhr:2019kwi,Duhr:2020seh,Duhr:2020kzd,Baglio:2022wzu}. 
Aiming for even more realistic predictions, even differential computations for for LHC processes at N$^3$LO in QCD have become available (see for example refs.~\cite{Chen:2022xnd,Chen:2021vtu,Cieri:2018oms,Buccioni:2025bkl,Chen:2021isd,Chen:2022lpw,Chen:2022cgv,Gehrmann-DeRidder:2023urf}).
However, extending such results to more complex processes at this level of accuracy, particularly for differential distributions, remains an extremely challenging task.
Several obstacles need to be overcome to systematize the techniques used to achieve N$^3$LO accuracy for these processes across a broader range of cases. For a recent review of this topic, we refer the interested reader to \refcite{Caola:2022ayt}.

Given the challenges of obtaining differential distributions at high perturbative order it is important to develop methods that can quantitively capture sizeble contributions to these higher order corrections via physically motivated approximations.
Moreover, given the lack of a robust methodology for estimating the size of missing higher order corrections, these approximations can give important quantitative insights in this direction, allowing to possibly highlight accidental behaviours in the perturbative series at previous orders.

In this work, we study a framework for obtaining approximations to LHC cross sections by capturing the contribution to rapidity distributions coming from soft and collinear radiation. 
We will refer to this framework as \rapidityfactorization.
Ref.~\cite{Lustermans:2019cau}  first introduced this expansion (referred to as "generalized threshold expansion" in this reference) and presented a factorization theorem for LHC cross sections.
This \rapidityfactorization~constitutes a step forward in applying the collinear expansion of color singlet cross sections, a methodology previously developed in \cite{Ebert:2020lxs}, to approximate a host of LHC cross sections. 
This methodology has already been applied to the calculation of N$^3$LO beam functions for $N$-Jettiness \cite{Ebert:2020unb} and $q_T$ \cite{Ebert:2020yqt}, as well as final state quantities like the N3LO TMD Fragmentation Functions and Energy-Energy Correlation jet functions~\cite{Ebert:2020qef,Ebert:2020sfi}.
We demostrate how this approximation naturally encodes and extends the well known threshold approximation for fully inclusive cross sections~\cite{Ahrens:2010rs,Ahrens:2009cxz,Ahrens:2008qu,Catani:2003zt,Catani:1990rp,Catani:1989ne,Sterman:1986aj} and rapidity distributions~\cite{Ravindran:2006bu,Becher:2007ty,Ravindran:2007sv}. 
We extend this framework by including channels sub-leading in the accuracy of our approximation and relating its building blocks to physical partonic LHC cross sections, therefore providing a clear path for their calculation to high perturbative order.
Furthermore, we perform additional phenomenological analysis of \rapidityfactorization~showcasing its applicability to approximate LHC cross sections.

The paper is organized as follows: in \Sec{setup} we setup the fixed order expansion for color-singlet rapidity distributions in the collinear limit. 
We define the rapidity beam function and soft function as the objects encoding the contribution of soft, collinear radiation to this limit. 
We show that the rapidity beam function can be related to the collinear limit of partonic cross sections in terms of a one dimensional measurement constraint. 
Compared to the original definition \cite{Lustermans:2019cau} as shifted integral over fully differential beam functions~\cite{Jain:2011iu,Gaunt:2014xxa,Gaunt:2020xlc} this definition paves a way for a more straightforward calculation of this object at N$^3$LO.
In \Sec{fo}, we discuss the resulting collinear approximation at fixed order. 
We analyze the structure of the partonic cross section differential in rapidity and its collinear approximation. 
We demonstrate how it relates to the inclusive cross section and how the leading collinear behavior fully incorporates the leading and next-to-leading term of the threshold expansion. 
We show qualitatively why the range of validity of the \rapidityfactorization is dramatically extended thanks to the behavior of the parton luminosity compared to a naive estimate based solely on kinematical considerations. 
We then quantitatively asses the validity of our approximation up to NNLO by comparing it to the full analytic result both for Higgs and Drell-Yan.
In \Sec{resummation}, we resum logarithmically enhanced contributions to the partonic cross section via Renormalization Group Equations (RGEs) of the rapidity beam and soft functions and discuss their phenomenological impact.
We conclude in \Sec{conclusions}.

%% file: Chapters/SetUp.tex
\section{Factorization for the Collinear Approximation}
\label{sec:setup}
We consider the production of a color neutral final state in the scattering process of two initial state hadrons.
Alongside the desired final state $h$ we allow any additional radiation. 
While our discussion is general, we in particular have a Drell-Yan lepton pair or a Higgs boson in mind as the color neutral final state.
\beq
P(P_1)+P(P_2)\to h(p_h)+X(k).
\eeq
We are interested in the description of the production probability for $h$ as a function of the rapidity Y and define the variables
\beq
\label{eq:hadronicvars}
S=(P_1+P_2)^2,\hspace{1cm}\tau=\frac{Q^2}{S},\hspace{1cm}Y=\frac{1}{2}\log\left(\frac{2P_2p_h}{2P_1p_h}\right),\hspace{1cm}Q^2=p_h^2.
\eeq
Here, we assume that the proton with momentum $P_1$ travels in the positive $z$-direction and consequently that the rapidity of the final state $h$ scattered in the direction of this proton is positive.
The scattering cross section differential in the rapidity and invariant mass of $h$ is then given by
\begin{align}
\label{eq:hadY}
	\frac{\df \sigma_{P\,P\rightarrow h+X}}{\df Y \df Q^2}(\xi_1,\xi_2)&=\tau \sum_{i,j} \int_{\xi_1}^1 \frac{\df z_1}{z_1} \int_{\xi_2}^1 \frac{\df z_2}{z_2}  f_i\left(\frac{\xi_1}{z_1},\mu^2\right) f_j\left(\frac{\xi_2}{z_2},\mu^2\right) \eta_{ij}(z_1,z_2,\mu^2) \nn\\
	&=\tau \sum_{i,j}  f_i(\xi_1,\mu^2)\otimes_{\xi_1} \eta_{ij}(\xi_1,\xi_2,\mu^2) \otimes_{\xi_2} f_j(\xi_2,\mu^2),
\end{align}
where we made use of the Mellin convolution defined in \app{trafos} and we introduced the short-hand notation
\beq
\xi_1=\sqrt{\tau} e^{-Y},\hspace{1cm}
\xi_2=\sqrt{\tau} e^{Y}.
\eeq
Above, $f_i(x)$ is a parton distribution function (PDF) of flavour $i$ evaluated at momentum fraction $x$ and the sum runs over all parton species and $\eta_{ij}(\xi_1,\xi_2,\mu^2) $ is the partonic coefficient function that can be computed in perturbative QFT. 
The above formula is valid up to hadronic power corrections.
We display the functional dependence of the above quantities on the perturbative scale $\mu^2$ to indicate that these are finite, renormalized quantities. 
We do not explicitly indicate the dependence of individual functions on $Q^2$.
Throughout this article we work in the $\overline{\text{MS}}$ scheme using dimensional regularization with the space time dimension being equal to $d=4-2\epsilon$.

One of the main results of this article is an approximation of the above cross section that is valid as either one of the variables $\xi_i$ tend towards one. 
We will show that in this limit the cross section can be described by the following factorization theorem.
\bea
\label{eq:raplimitdef}
Q^2 \frac{\df \sigma_{P\,P\rightarrow h+X}}{\df Q^2  \df Y}(\xi_1,\xi_2)&=&\tau \sum_{i,j}  H_{ij}(Q^2,\mu^2) \left[
 B^{Y}_{i}(\xi_1,\mu^2;\bar \xi_2) \otimes_{\xi_1;\xi_2} S^{r_{ij}}(\mu^2;\bar \xi_1 \bar \xi_2) \otimes_{\xi_2;\xi_1} B^{Y}_{i}(\xi_2,\mu^2;\bar \xi_1)
 \right]\nonumber\\
&+&\mathcal{O}\left(\bar \xi_1^0,\bar \xi_2^0\right).
\eea
Here, we made use of the universal rapidity beam function $B^Y_i$ (referred to as "generalized threshold beam function" in ref.~\cite{Lustermans:2019cau}), the rapidity soft function $S^{r_{ij}}$, and the short-hand notation $\bar \xi_i=1-\xi_i$.
The symbol $\mathcal{O}\left(\bar \xi_1^0,\bar \xi_2^0\right)$ indicates that the above formula is accurate up to power corrections that start either with $\bar \xi_1$ or $\bar \xi_2$ raised to zero$^{\text{th}}$ power.
We use the semi-colon notation in the beam functions and soft function in order to underscore the role of the argument to the right of it as the "observable" of the beam function.
This is done in order to highlight the resemblance to other beam functions dependent for example on the transverse momentum of $h$ or the 0-Jettiness beam function.
Above, we introduced notation indicating a double convolution that is a combination of a Laplace convolution and a Mellin convolution. 
This double convolution acts on two functions f and g in the following way.
\beq
\label{eq:dconvdef}
f(x_1,x_2)\otimes_{x_1;x_2} g(x_1,x_2)=\int_{x_1}^1 \frac{\df y_1}{y_1} \int_{x_2}^1 \df y_2 f(y_1,y_2) g\left(\frac{x_1}{y_1},1+x_2-y_2\right).
\eeq
In eq.~\eqref{eq:raplimitdef} it is implied that all Laplace convolutions are performed first.
The superscript $r_{ij}$ of the soft function refers to the fact that the soft function only depends on the external partons i and j via their color representation (fundamental representation for quarks and adjoint representation for gluons in QCD).
Furthermore, $H_{ij}(\mu^2)$ is the process-dependent hard function.
These newly introduced quantities depend on the perturbative scale $\mu^2$, which indicates that they are renormalized, finite quantities.
We derive the above formula in the detail below.

\subsection{Derivation of the Approximate Rapidity Distribution}
In the limit $\xi_1\to 1$ the cross section of eq.~\eqref{eq:hadY} can be written using the following factorization theorem.
\beq
\label{eq:raplimdef}
Q^2 \frac{\df \sigma_{P\,P\rightarrow h+X}}{\df Y\df Q^2}(\xi_1,\xi_2)=\tau \sum_{i,j}  H_{ij} f_{i}(\xi_1) \otimes_{\xi_1} B^Y_j(\xi_2;\bar \xi_1)+\mathcal{O}(\bar{ \xi}_1^0).
\eeq
We refer to $B^Y_j(\xi_2;\bar \xi_1)$ as the rapidity beam function.
The above factorization theorem was first stated explicitly in ref.~\cite{Lustermans:2019cau} and can be recovered from the study of the limiting behavior of cross sections as in refs.~\cite{Dulat:2018bfe,Ebert:2020lxs}.
Note, that in contrast to the previous section we are now using bare, unrenormalized quantities and consequently dropped the dependence of the functions on $\mu^2$.
As $\xi_1=\sqrt{\frac{Q^2}{S}} e^{-Y}$, the limit $\xi_1\to 1$ is approached as the virtuality of the state $h$ is increased or as it gets boosted towards negative rapidity. 
Radiation produced on top of $h$ consequently must be either be boosted in the positive $Y$ direction or be very low energetic. 

The kinematic limit $\xi_1 \to 1$ is consequently described by the collinear limit of the scattering cross section and we follow the approach of ref.~\cite{Ebert:2020lxs} to derive this limit for the Higgs boson and Drell-Yan production cross section. 
Using these techniques we can then extract the rapidity beam function matching kernel $I^Y_{ij}(\xi_2;\xi_1) $ and relate the rapidity beam function to the usual collinear PDF using the following identity.
\beq
 \label{eq:beam_master}
B^Y_{i}(\xi_2;\bar \xi_1)= \sum_{j} I^Y_{ij}(\xi_2; \bar \xi_1) \otimes_{\xi_2} f_j(\xi_2).
\eeq
The beam function matching kernel is related to the strict collinear limit of a production cross section via 
\begin{align} 
I^Y_{ij}(\xi_2;\xi_1) &
 =
  \int_0^1 \df x \int_0^\infty \df \wa\df \wb \, \delta\left[\xi_2-(1-w_2)\right]
   \nn\\&\qquad\times
   \strictlim \left\{\delta\left[\bar\xi_1-\frac{2-w_2-w_2 x}{2(1-w_2)}w_1 \right] \frac{\df\eta_{j \bar i}}{ \df Q^2  \df \wa \df \wb \df  x}\right\}
\,.\end{align}
We refer the reader to ref.~\cite{Ebert:2020lxs} for detailed definitions of the partonic coefficient function $\frac{\df\eta_{j \bar i}}{ \df Q^2  \df \wa \df \wb \df  x}$ and just repeat the definition of the variables for convenience.
\beq
w_1=-\frac{2p_1 k}{2p_1 p_2},\hspace{1cm}
w_2=-\frac{2p_2 k}{2p_1 p_2},\hspace{1cm}
x=\frac{k^2}{s w_1 w_2}.
\eeq
Above, $p_1$, $p_2$ and $k$ are the momenta of the scattering partons and of the radiation, such that $k=-p_1-p_2-p_h$.
The perturbative matching kernels were previously obtained in refs.~\cite{Gaunt:2020xlc,Lustermans:2019cau} through NNLO by projection of a triple differential object, the fully-differential beam functions \cite{Jain:2011iu}.
Here, we recompute the matching kernel based on the methods developed in ref.~\cite{Ebert:2020lxs} and as a single variable expansion of explicit partonic coefficient functions for the Higgs boson and Drell-Yan production cross section.

Naturally, we may consider the limit where $\xi_2\to 1$ instead of $\xi_1\to 1$.
The problem is largely symmetric under the exchange of the variables and we find an equivalent formula for the hadronic rapidity distribution in that limit. 
We can then combine the two limits and find
\bea
\label{eq:raphaddef1}
Q^2 \frac{\df \sigma_{P\,P\rightarrow h+X}}{\df Y \df Q^2}(\xi_1,\xi_2)&=&\tau \sum_{i,j}  H_{ij} \left[
 f_{i}(\xi_1) \otimes_{\xi_1} B_j(\xi_1;\bar \xi_2)
 +B_i(\xi_2;\bar \xi_1)  \otimes_{\xi_2} f_{j}(\xi_2) \right.\\
&-&\left.f_{i}(\xi_1) \otimes_{\xi_1} S^{-1,\,r_{ij}}(\bar \xi_1 \bar \xi_2)\otimes_{\xi_2} f_{j}(\xi_2)\nonumber
 \right]\\
&+&\mathcal{O}\left(\bar \xi_1^0,\bar \xi_2^0\right).\nonumber
\eea
Above, we refer to $S^{-1,\,r_{ij}}(\bar \xi_1 \bar \xi_2)$ as the inverse of the rapidity soft function we introduced in eq.~\eqref{eq:raplimitdef}, which is related to the rapidity beam function matching kernel by
\beq
S^{-1,\,r_{ij}}(\bar \xi_1 \bar \xi_2)=\lim\limits_{\xi_2\to 1} I_{ij}(\xi_2;\bar \xi_1).
\eeq
Note that this quantity is sometimes referred to as the \emph{double differential} threshold soft function and its relation to the eikonal limit of beam functions was studied in \refscite{Lustermans:2019cau,Billis:2019vxg}.

We would like to further simplify eq.~\eqref{eq:raphaddef1} and define to this end $S^{r_{ij}}(\bar \xi_1 \bar \xi_2)$ such that
\beq
\label{eq:softinverse}
S^{r_{ij},-1}(\bar \xi_1 \bar \xi_2) \circ_{\xi_1,\xi_2}S^{r_{ij}}(\bar \xi_1 \bar \xi_2) =1.
\eeq
Above, we made use of the definition of the double Laplace transformation, see \app{trafos}.
With this we may now rewrite eq.~\eqref{eq:raphaddef1} up to terms that are suppressed in the limits we consider as
\bea
Q^2 \frac{\df \sigma_{P\,P\rightarrow h+X}}{\df Y\df Q^2}(\xi_1,\xi_2)&=&\tau \sum_{i,j}  H_{ij} \left[
 B^{Y}_{i}(\xi_1;\bar \xi_2) \otimes_{\xi_1;\xi_2} S^{r_{ij}}(\bar \xi_1 \bar \xi_2) \otimes_{\xi_2;\xi_1} B^{Y}_{j}(\xi_2;\bar \xi_1)
 \right]\nonumber\\
&+&\mathcal{O}\left(\bar \xi_1^0,\bar \xi_2^0\right).\nonumber
\eea
The above formula is written in terms of bare, unrenormalized quantities and we discuss their renormalization and properties below.

\subsection{The Hard Function}\label{sec:HFsetup}
The hard function $H_{ij}$ refers to purely virtual contributions for the process under consideration and arises from squared scattering amplitudes integrated over the Born phase space. 
For Higgs boson and Drell-Yan production the hard function is currently known through four loops thanks to the computation of the virtual form factors~\cite{vonManteuffel:2020vjv,Lee:2021uqq,Lee:2022nhh,Gehrmann:2010ue,Gehrmann:2010tu,Gehrmann2005,Baikov:2009bg,Agarwal:2021zft}. In case of the Higgs boson produced via gluon fusion in the heavy top quark effective theory~\cite{Wilczek1977,Spiridonov:1988md,Shifman1978,Inami1983} also the knowledge of a Wilson coefficient is required~\cite{Schroder:2005hy,Kramer:1996iq,Chetyrkin:2005ia,Chetyrkin:1997un}.

The bare hard function is related to its renormalized counter part by
\beq
 H_{ij}(\mu^2)=Z_H(\mu^2){\bf Z_{\alpha_S}} H_{ij},\nonumber\\
\eeq
As above, we indicate renormalized quantities by explicitly writing their dependence on the unphysical renormalization scale $\mu^2$.
The operator ${\bf Z_{\alpha_S}}$ implements the renormalization of the strong coupling constant and can be constructed from the QCD $\beta$-function currently known through five loops~\cite{Tarasov:1980au,Larin:1993tp,vanRitbergen:1997va,Czakon:2004bu,Herzog:2017ohr,Baikov:2016tgj}.
The factor $Z_H$ is a multiplicative $\overline{\text{MS}}$ counter term that absorbs all poles of the bare hard function.

\subsection{The Soft Function}\label{sec:SFsetup}
The rapidity soft function contains all contributions towards the rapidity distribution at threshold arising due to real radiation~\cite{Ravindran:2006bu,Becher:2007ty,Ravindran:2007sv,Li:2016axz,Billis:2019vxg}.
This quantity can be computed from the inclusive threshold soft function without any additional information and is therefore currently known to third loop order~\cite{Anastasiou:2014vaa,Li:2014afw}.
Logarithmically enhanced terms can even be determined to fourth loop order (see for example refs.~\cite{Das:2020adl,Das:2019btv,Duhr:2022cob}).
The renormalized soft function is related to its bare pendant via 
\beq
S^{r,-1}(\mu^2 ;\bar \xi_1\bar \xi_2)=Z_H^{-1}(\mu^2)\Gamma^r_s(\xi_1,\mu^2)\circ_{\xi_1}\left[ {\bf Z_{\alpha_S}}  S^{r,-1}(\bar \xi_1 \bar \xi_2)\right]\circ_{\xi_2}\Gamma^r_s(\xi_2,\mu^2).
\eeq
Above, we introduced 
\beq
\Gamma^r_s(x)=\lim_{x\to1} \Gamma_{i_r i_r}(x),
\eeq
where $ \Gamma_{i_r i_r}$ is the standard mass factorization counter term constructed out of DGLAP~\cite{Gribov:1972ri,Dokshitzer:1977sg,Altarelli:1977zs} splitting functions~\cite{Vogt:2004mw,Moch:2004pa,Moch:2014sna,Ablinger:2014nga,Moch:2015usa,Moch:2017uml,Ablinger:2017tan,Vogt:2018miu,Moch:2018wjh,Das:2020adl,Das:2019btv,Blumlein:2021enk,Moch:2021qrk} and $i_{\text{adjoint}}=g$ and $i_{\text{fundamental}}=q$.
Note, that above we made used of the Laplace convolution (see \app{trafos}).
The renormalized soft function and its inverse are now easily related in analogy to eq.~\eqref{eq:softinverse}.

\subsection{The Matching Kernel and the Rapidity Beam Function}\label{sec:BFsetup}
The matching kernel for the rapidity beam function was first introduced in ref.~\cite{Lustermans:2019cau} and computed through NNLO~\cite{Gaunt:2020xlc,Lustermans:2019cau}.
The renormalized matching kernel is given by
\beq
I_{ij}(\xi_1,\mu^2;\bar \xi_2)=Z_H^{-1}(\mu^2) \sum_{k}\Gamma^{r_{ii}}_s(\xi_1,\mu^2)\circ_{\xi_1}\left[ {\bf Z_{\alpha_S}}  I_{ik}(\xi_1,\xi_2)\right]\otimes_{\xi_2}\Gamma_{kj}(\xi_2,\mu^2).
\eeq
Note, that above a Laplace convolution as well as a Mellin convolution was used. 
This particular assignment guarantees that no power suppressed terms beyond the collinear limit are introduced due to the convolution.
The renormalized rapidity beam function is then given by
\beq
 \label{eq:beam_master}
B_{i}(\xi_2,\mu^2;\bar \xi_1)= \sum_{j} I_{ij}(\xi_2,\mu^2; \bar \xi_1) \otimes_{\xi_2} f_j(\xi_2,\mu^2),
\eeq
where the standard $\overline{\text{MS}}$ PDF is given by
\beq
f_i(x,\mu^2)=\sum\limits_{j}\Gamma^{-1}_{ij}(x,\mu^2)\otimes_{x}f_j(x).
\eeq

%% file: Chapters/FixedOrderXS.tex
\section{Fixed Order Cross Sections}
\label{sec:fo}
We discuss the properties of the approximate partonic coefficient function derived above for the purposes of our factorization theorem, eq.~\eqref{eq:raplimitdef}.
Next, we relate the partonic coefficient function to its counter part for the inclusive production cross section of the same process. 
Finally, we explore how well the Drell-Yan and gluon-fusion Higgs boson production cross sections are described at a given order in perturbation theory by the approximation proposed above.

\subsection{Structure of the Fixed Order Partonic Coefficient Function}

In eq.~\eqref{eq:hadY} we introduced the hadronic rapidity distribution of a colorless final state and in eq.~\eqref{eq:raplimitdef} an approximation for the same cross section in the limit where either of the variables $\xi_i$ tend towards one.
We conclude that eq.~\eqref{eq:raplimitdef} contains an approximate partonic coefficient function, $\eta_{ij}^{Y_\text{approx.}}$, that approximates its full counter part in the discussed limit.
Explicitly, we find 
\beq
\label{eq:foapprox}
\frac{\df \sigma_{P\,P\rightarrow h+X}}{\df Y}(\xi_1,\xi_2)=\tau \sum_{i,j}   f_i(\xi_1,\mu)\otimes_{\xi_1} \eta^{Y_\text{approx.}}_{ij}(\xi_1,\xi_2,\mu) \otimes_{\xi_2} f_j(\xi_2,\mu)
+\mathcal{O}\left(\bar \xi_1^0,\bar \xi_2^0\right),
\eeq
where the approximate partonic coefficient function is given by
\beq
\eta_{ij}^{Y_\text{approx.}}(\xi_1,\xi_2,\mu^2)=\sum_{k,l} H_{kl}(\mu^2) I_{ik}(\xi_1,\mu^2;\bar \xi_2) \otimes_{\xi_1;\xi_2} S^{r_{kl}}(\mu^2;\bar \xi_1 \bar \xi_2) \otimes_{\xi_2;\xi_1} I_{lj}(\xi_2,\mu^2;\bar \xi_1).
\eeq
The above approximation can be expanded in the strong coupling constant and then compared order by order to the known fixed order results for the exact partonic coefficient function $\eta_{ij}$.
In particular, both coefficient functions contain distributions of the form
\beq
\left[\frac{\log^ n \left(1-\xi_i\right)}{1-\xi_i}\right]_+\hspace{1cm} \text{ and }\hspace{1cm}\delta(1-\xi_i) .
\eeq
Above, $n\geq 0$. These distributions act on in the usual way on the parton distribution functions. 
By construction, our approximation reproduces all distribution valued terms for the exact partonic coefficient function for the rapidity distribution.
In addition to the distributions, our approximated cross section depends on the variables $\xi_i$ via rational and algebraic functions and via generalized poly logarithms~\cite{Goncharov1998}.

Note, that due to the convolutions of two beam function matching kernels as well as the use of Mellin convolutions power suppressed terms beyond our approximation are introduced.
While these terms are beyond the desired accuracy, it is still interesting to study their consequences.
First, note that as the beam function matching kernel is not diagonal in flavor, other channels than the Born combination of initial state parton flavours are introduced.
For example, in the case of gluon fusion Higgs boson production only the gluon initial state contributes at Born level. 
Distributions in $(1-\xi_i)$ are present in the gluon-gluon and gluon-qark initial state and correctly reproduced by our formula.
As a consequence, eq.~\eqref{eq:raplimitdef} therefor correctly reproduces the entire next-to-leading power term in a threshold expansion of the partonic cross section at fixed order (The fact that the collinear expansion contains this was already pointed out in ref.~\cite{Lustermans:2019cau}). 

In addition, our formula produces non-vanishing partonic cross sections for initial state with two quarks if the Born cross section is initiated by two gluons and vice versa.
These terms are free of distributions in $(1-\xi_i)$ and beyond the accuracy of our approximation.
Here, we make an interesting observation: The first non-vanishing term in the simultaneous expansion of the partonic coefficient functions in both $(1-\xi_1)$ and $(1-\xi_2)$ is correctly reproduced by our approximation at NNLO.
In the threshold expansion of the partonic cross section these terms would appear at next-to-next-to-leading power.

Below, we want to study the phenomenological implications of our approximation. 
To study the quality of the approximation we want to consider not only the first term in the expansion our cross sections around either $\xi_1\to 1$ or $\xi_2 \to 1$ but also subleading terms.
In analogy to eq.~\eqref{eq:raphaddef1} we introduce an approximate partonic coefficient function that is correct up to the $n^{th}$ term in the desired expansion.
\beq
\label{eq:expansiondef}
\eta_{ij}^{Y_{exp}(n)}(\xi_1,\xi_2,\mu^2)=\eta_{ij}^{(n,\infty)}(\xi_1,\xi_2)+\eta_{ij}^{(\infty,n)}(\xi_1,\xi_2)-\eta_{ij}^{(n,n)}(\xi_1,\xi_2).
\eeq
Above, $\eta_{ij}^{(n,m)}$ corresponds to a generalized power series expansion\footnote{We speak of a generalized power series since the first term is distribution valued and the coefficients contain logarithms of the variables $1-\xi_i$,} of the partonic coefficient function around $\xi_1\to 1$ and in $\xi_2\to 1$ including the first $n$ and $m$ terms respectively. 
The approximation we introduced above in eq.~\eqref{eq:raplimitdef} is formally accurate only up to the first term in this approximation, $n=-1$.
\beq
\label{eq:etaaprox}
\eta_{ij}^{Y_\text{approx.}}(\xi_1,\xi_2,\mu^2)=\eta_{ij}^{Y_{exp}(-1)}(\xi_1,\xi_2,\mu^2)+\mathcal{O}\left(\bar\xi_1^0,\bar \xi_2^0\right).
\eeq

\subsection{Relation to the Inclusive Cross Section}\label{sec:relationtoInc}
The inclusive partonic cross section for gluon fusion Higgs boson production and Drell-Yan production was computed through N$^3$LO in refs.~\cite{Mistlberger:2018etf,Duhr:2021vwj,Duhr:2020seh,Anastasiou:2015ema}. 
Relating the partonic coefficient function for the rapidity distribution and the inclusive cross section serves as a strong cross check and the computational methods for the calculation of the rapidity beam functions used here benefit from this relation. 

In order to obtain the inclusive partonic cross section, we first introduce the following variable transformation.
\beq
\label{eq:incvar}
\bar \xi_1=\bar x \bar z,\hspace{1cm}\bar \xi_2=\bar z\frac{(1-\bar x)}{1-\bar x \bar z}.
\eeq
Explicitly, we find the relation
\beq
\label{eq:increl}
\eta_{ij}^{\text{inclusive}}(\bar z)=\int_0^1 \df \bar x \frac{\bar z}{1-\bar x\bar z} \eta_{ij}\left(1-\bar x \bar z,\frac{1-\bar z}{1-\bar x\bar z}\right).
\eeq
The variable $\bar z$ corresponds to the standard choice of variable used for inclusive partonic cross sections.
The variable $\bar z$ tends to zero as the production threshold for the hard colorless final state particle is approached.
Inspecting eq.~\eqref{eq:incvar}, we see that an expansion in $\bar z$ corresponds to a simultaneous expansion in $\bar \xi_1$ and $\bar \xi_2$.
The first term in the simultaneous expansion of both $\xi_{1,2}\to 1$ is comprised only of distributions in both our variables and is consequently fully contained in our collinear approximation.
The second in this double expansion is at least distribution valued in one of the variables $\xi_i$ and is consequently also fully contained in our approximation. 
As a result, our approximate partonic coefficient function $\eta_{ij}^{Y_\text{approx.}}$ contains the first and second term of an expansion of the inclusive partonic coefficient function in the variable $\bar z$ at fixed order (see also ref.~\cite{Lustermans:2019cau}).

%% file: Chapters/PhenoStudy.tex
\section{Phenomenological Study of the Collinear Approximation}\label{sec:pheno}

\subsection{Partonic vs Hadronic Threshold in Rapidity Distributions}
\begin{figure*}[!h]
\centering
\includegraphics[width=0.49\textwidth]{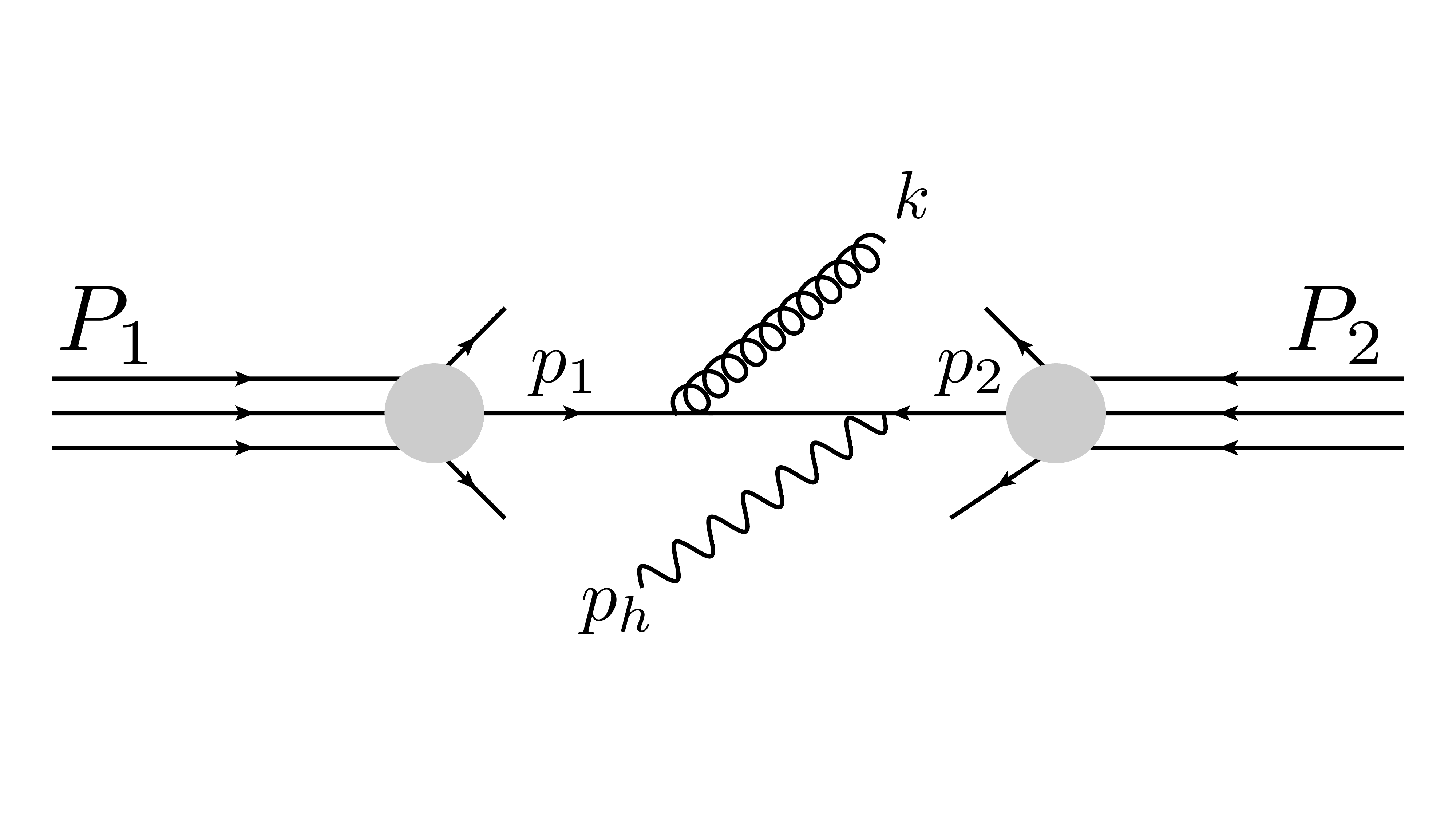}
\includegraphics[width=0.49\textwidth]{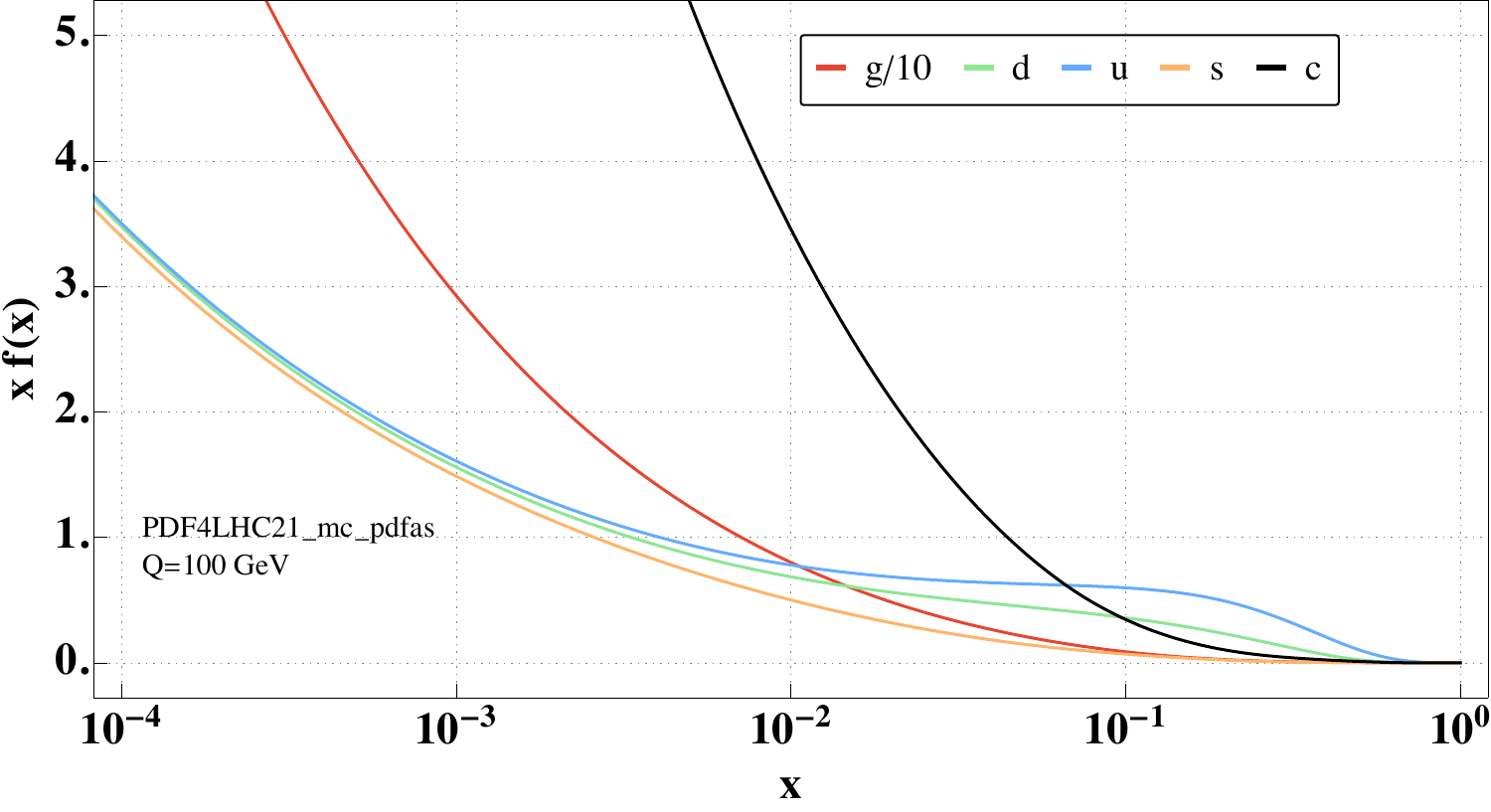}
\caption{\label{fig:PDFplot}
The left figure shows a schematic of a scattering process of two protons producing a virtual photon and forward radiation.
The figure on the right shows the PDF4LHC21 parton distribution functions~\cite{Cridge:2021qjj} as a function of the momentum fraction $x$ and for a resolution scale $Q=100$ GeV. The gluon PDF has been rescaled by a factor of $0.1$ to enhance legibility. 
}
\end{figure*}

The approximation derived in this article is related to the hadronic limit in which one $\xi_i \to 1$. 
For most lepton pairs or Higgs boson produced at the LHC this limit is not satisfied.
For centrally produced final states ($Y=0$) with an invariant mass of about $100$ GeV we find that
\beq
\xi_i=\sqrt{\frac{Q^2}{S}}e^{\pm Y}\sim 0.01,
\eeq
which is very far from one.
To approach the kinematic region where the approximation is clearly valid we may consider one of the following two scenarios.
One option is to increase the virtuality to very large values (for $\xi_1=0.9$ we have to choose a Q=12.24 TeV final state).
Another option is to consider hadronic final states boosted to very large rapidities (for $\xi_1 =0.9$ we require a rapidity of $- 4.8$ for $Q=100$ GeV). 
However, the validity of our approach is greatly enhanced by a particular feature of parton distribution functions and the fact that our approximation is performed at the partonic level. Let's take another look at eq.~\eqref{eq:raplimdef} describing the hadronic cross section in the limit $\xi_1\to 1$.
\beq
Q^2 \frac{\df \sigma_{P\,P\rightarrow h+X}}{\df Y \df Q^2}(\xi_1,\xi_2)=\tau \sum_{i,j}  H_{ij} \int_{\xi_1}^1 \frac{\df x_1}{x_1}f_{i}\left(\frac{\xi_1}{x_1}\right)  B^Y_j(\xi_2,\mu^2;\bar x_1)+\mathcal{O}(\bar \xi_1^0).
\eeq
At the partonic level, the Beam Function $B^Y_j(\xi_2,\mu^2;\bar x_1)$ contains the leading term in the limit $x_1\to 1$. 
The integration boundaries enforce $x_1\geq \xi_1$, so if $\xi_1\to 1$ then we automatically find $x_1\to 1$.
If the hard final state $h$ is produced without radiation boosted by the initial proton with momentum $P_1$ then we have $x_1=1$ and the argument of the PDF $f_i$ is at its minimum $\xi_1$.
Additionally emitted radiation (see for example the left-hand panel of fig.~\ref{fig:PDFplot}) will require $x_1$ to be different from one and consequently increase the argument of the PDF to $\frac{\xi_1}{x_1}$.
However, parton distribution functions at the LHC fall steeply as their argument is increased. 
The part of the partonic coefficient function contributing most to the hadronic cross section is consequently located close to $x_1\sim 1$.
This is exactly the region in which our expansion is accurate.

Figure~\ref{fig:PDFplot} shows the PDF4LHC21 parton distributions~\cite{Cridge:2021qjj} as a function of their argument $x$. 
We see in particular that the gluon distribution function is rapidly falling with increasing values of $\xi$ (labelled as $x$ in the figure). Valence quarks (u and d) have a more significant plateau until fairly large values of $\xi\sim 0.3$ as they carry a substantial fraction of the total proton momentum.
At low x valence and see quark distributions alike exhibit rapidly falling behavior.

\begin{figure*}[!h]
\centering
\includegraphics[width=0.49\textwidth]{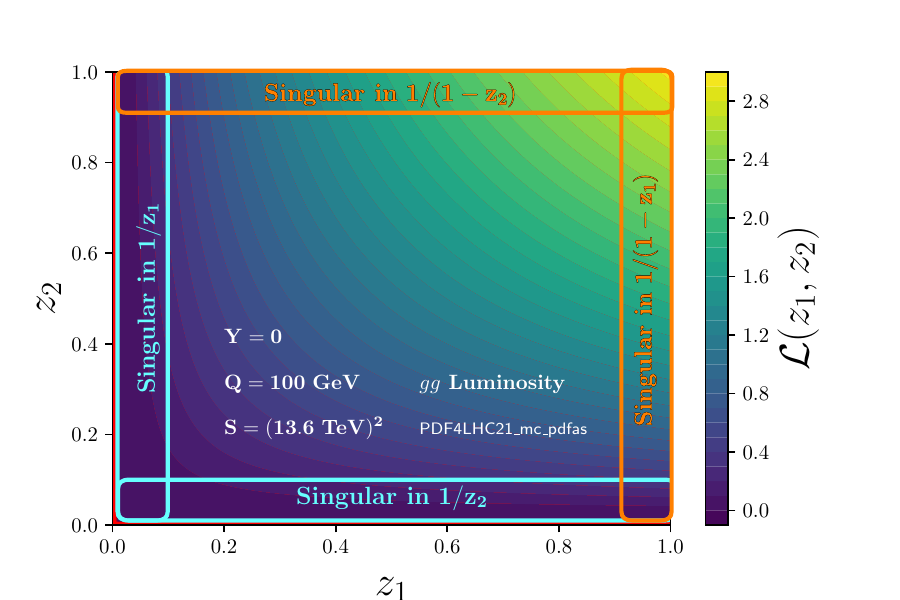}
\includegraphics[width=0.49\textwidth]{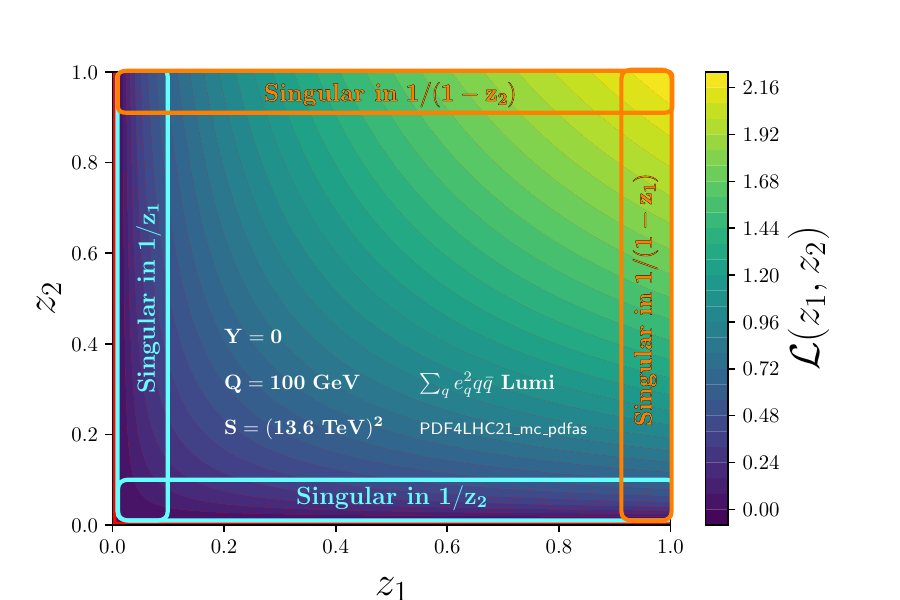}
\caption{\label{fig:Lumi}
The figures show the parton luminosity for the Higgs boson and Drell-Yan production cross section as a function of $z_1$ and $z_2$ for the LHC at $13.6$ TeV for the production of central (Y=0) final state with a virtuality of $Q=100$ GeV. 
The shade of the color represents the magnitude of the luminosity with lighter colors representing larger luminosity.
The orange and tyrkois boxes indicate where the partonic coefficient function is kinematically enhanced.
}
\end{figure*}
Figure~\ref{fig:Lumi} displays the parton luminosity for Higgs boson and Drell-Yan production at $Y=0$ and $Q=100$ GeV.
\beq
\mathcal{L}_{\text{Higgs}}(z_1,z_2)\sim \frac{1}{z_1}\frac{1}{z_2} f_g\left(\frac{\xi_1}{z_1}\right)f_g\left(\frac{\xi_2}{z_2}\right),\hspace{1cm}
\mathcal{L}_{\text{DY}}(z_1,z_2)\sim \frac{1}{z_1}\frac{1}{z_2}\sum\limits_{i\neq g}e_i^2 f_i\left(\frac{\xi_1}{z_1}\right)f_{-i}\left(\frac{\xi_2}{z_2}\right).
\eeq
Above, $e_i$ is the electric charge of a quark of flavor $i$.
Lighter colors indicate larger parton luminosity. It is easy to see that the largest luminosities can be found in the vicinity of $z_1 \sim 1$ or $z_2 \sim 1$, i.e. the region of validity of our approximation.
The partonic coefficient function are kinematically enhanced and contain singularities of the form
\beq
\frac{\log^n(1-z_i)}{1-z_i},\hspace{1cm}
\frac{\log^n(z_i)}{z_i}.
\eeq
Our approximation exactly reproduces terms of the first type in the above equation. 
Terms of the second type, which are not covered by our approximation, are suppressed by the parton luminosity. 
In combination, the large parton luminosity and kinematic enhancement of the partonic coefficient function in the region of our approximation motivates its application to precision LHC phenomenology. 
Note, that these findings are well known to the community and were for example pointed out in ref.~\cite{Bonvini:2012an} in a Mellin space analysis.

The crude phenomenology analysis above leads us to expect the following scenarios in which the collinear approximation of hadronic cross sections should perform well. 
Gluon initiated processes should be favored due to the rapidly falling gluon PDFs. In order to benefit similarly for quark initiated PDFs it is necessary to consider large invariant masses of the final state. 
In general, the approximation is going to improve as the invariant mass $Q$ is increased.
Equally, the approximation should perform well when very boosted kinematic configurations are considered.
In any case, since this approximation can capture both off-diagonal channels and terms singular in a single $z_i \to 1$ limit, it is expected to perform significantly better than the threshold approximation.
Below we will explicitly demonstrate this by studying the gluon fusion Higgs boson and Drell-Yan production cross sections. 
However, we would like to point out that this framework should be tested in the future on a larger class of processes and the quality of the approximation may vary (see for example ref.~\cite{Bern:2002pv} for related observations).

\subsection{Approximating the Higgs and Drell-Yan Rapidity Distribution at NNLO}
The Higgs boson and Drell-Yan rapidity distribution have been computed many times (see for example refs.~\cite{Anastasiou:2002qz,Anastasiou:2003ds,Anastasiou:2003yy}) and are currently known at N$^3$LO in QCD perturbation theory~\cite{Chen:2021vtu,Chen:2022xnd,Cieri:2018oms,Dulat:2018bfe}. 
Here, we compute the partonic coefficient functions and implement them into a numerical code to study the expansion proposed in this article. 
We evaluate our cross sections numerically using PDF4LHC21 parton distribution functions~\cite{Cridge:2021qjj}. 
We derive predictions for the LHC with a center-of-mass energy $S=13.6$ TeV. 
In the case of the Drell-Yan cross section, we consider QCD corrections to the production of a virtual photon and its subsequent decay to an electron-positron pair of virtuality $Q$. 
In the case of the gluon fusion Higgs boson cross section, we consider the on-shell production cross section of a Higgs boson with a mass $m_h=Q$.
Throughout this study, these two processes are meant to serve as an example of gluon- or quark induced processes.
Consequently, varying the Higgs boson mass should not be seen as an exercise in beyond the Standard Model physics but rather using these predictions as a proxy process for other color neutral final states produced in the fusion of two gluons.

\begin{figure*}[!h]
\centering
\includegraphics[width=0.49\textwidth]{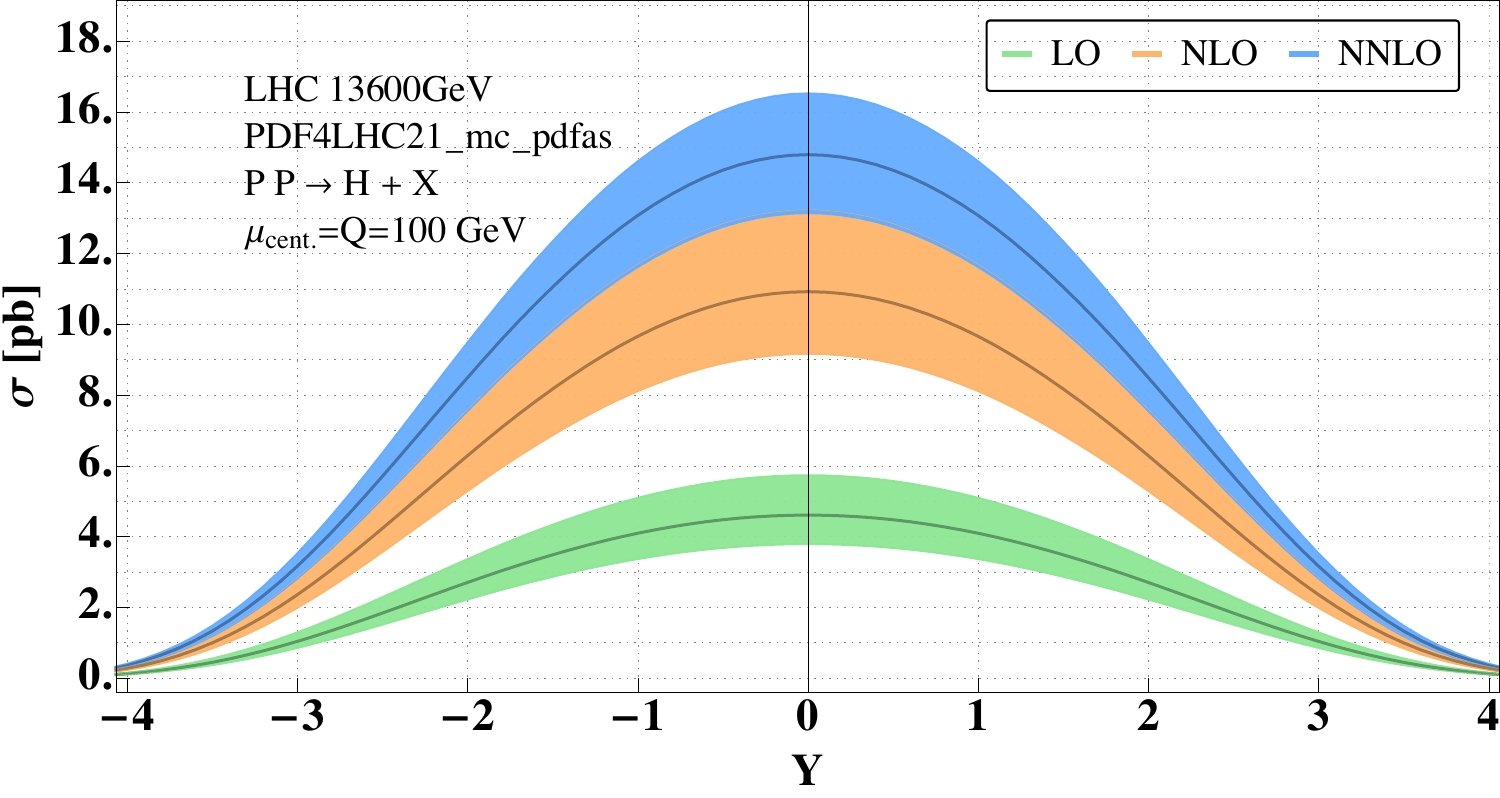}
\includegraphics[width=0.49\textwidth]{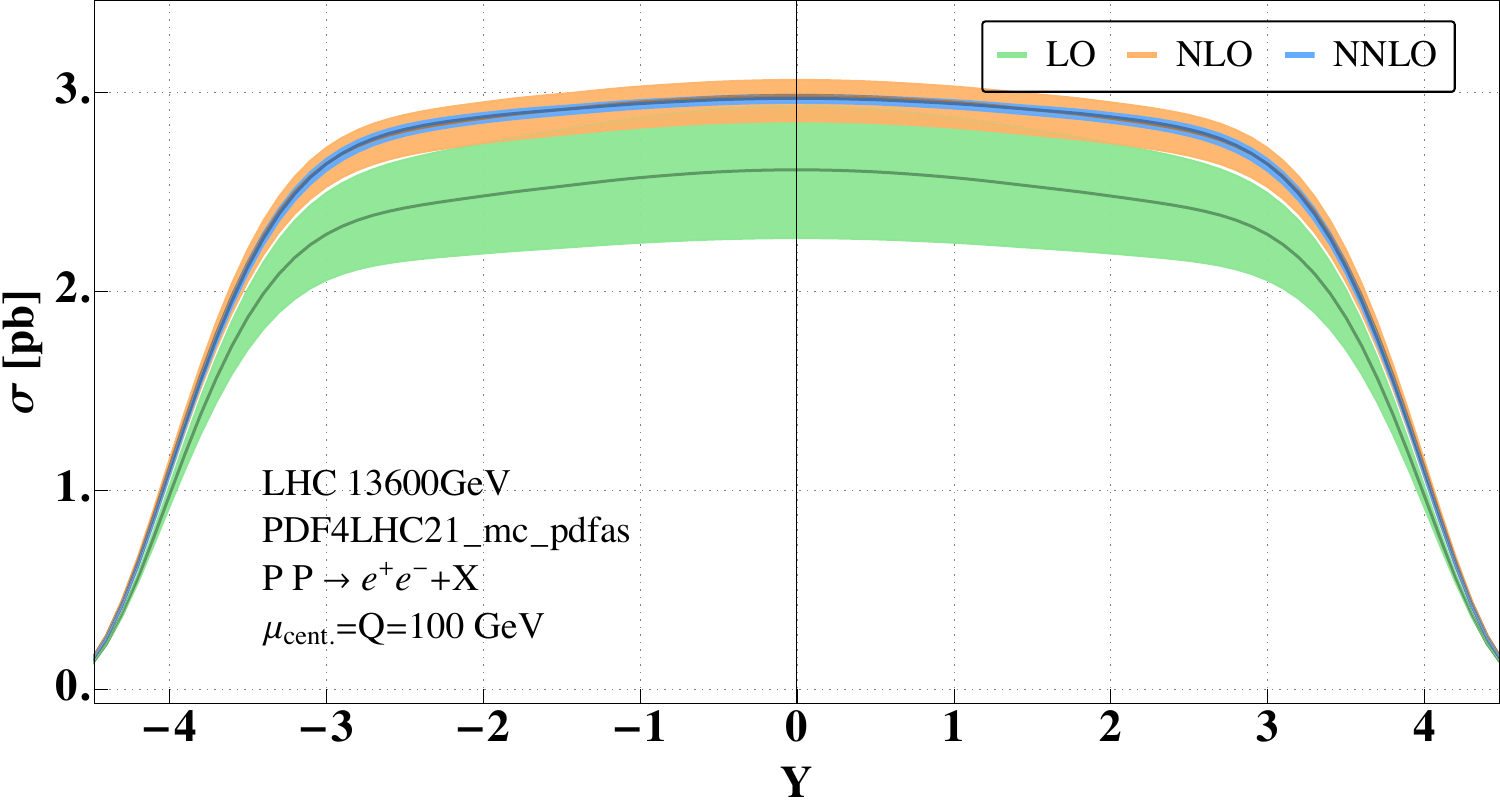}
\caption{\label{fig:NNLORap}
Rapidity distributions up to NNLO in QCD perturbation theory at the LHC. Predictions in green, orange and blue are computed at LO, NLO and NNLO respectively. The left and right panel show the Higgs boson and DY rapidity distributions respectively.
}
\end{figure*}
Figure~\ref{fig:NNLORap} shows the rapidity distributions for gluon fusion Higgs boson production and the photon mediated Drell-Yan process and introduce the short-hand
\beq
\sigma=Q^2 \frac{\df \sigma}{\df Q^2 \df Y}.
\eeq
We compute here both the rapidity distributions for a virtuality of $Q=100$ GeV through NNLO in QCD perturbation theory.
The bands indicate an uncertainty estimate and are obtained by varying the perturbative renormalization scale $\mu_R$ and factorization scale $\mu_F$ by a factor of two around their central value $\mu_F^0=\mu_R^0=Q$ such that
\beq
\label{eq:sevenpt}
0.5\leq \frac{\mu_i}{\mu_i^0}\leq 2,\hspace{1cm} 
0.5\leq \frac{\mu_F}{\mu_R}\leq 2
\eeq
The second condition in the above equation excludes extreme ratios of scales and corresponds to a common choice (7-point scale variations).

\begin{figure*}[!h]
\centering
\includegraphics[width=0.49\textwidth]{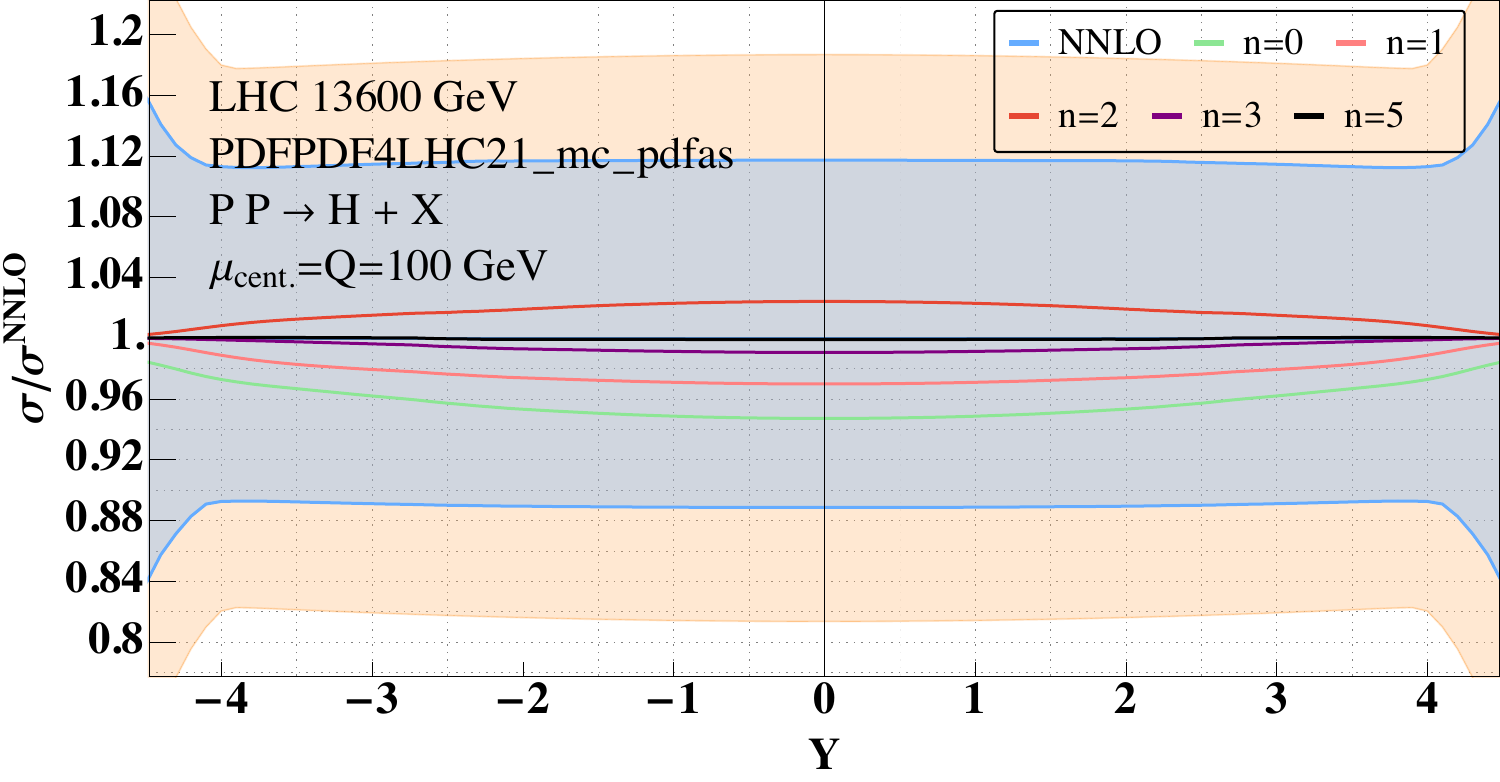}
\includegraphics[width=0.49\textwidth]{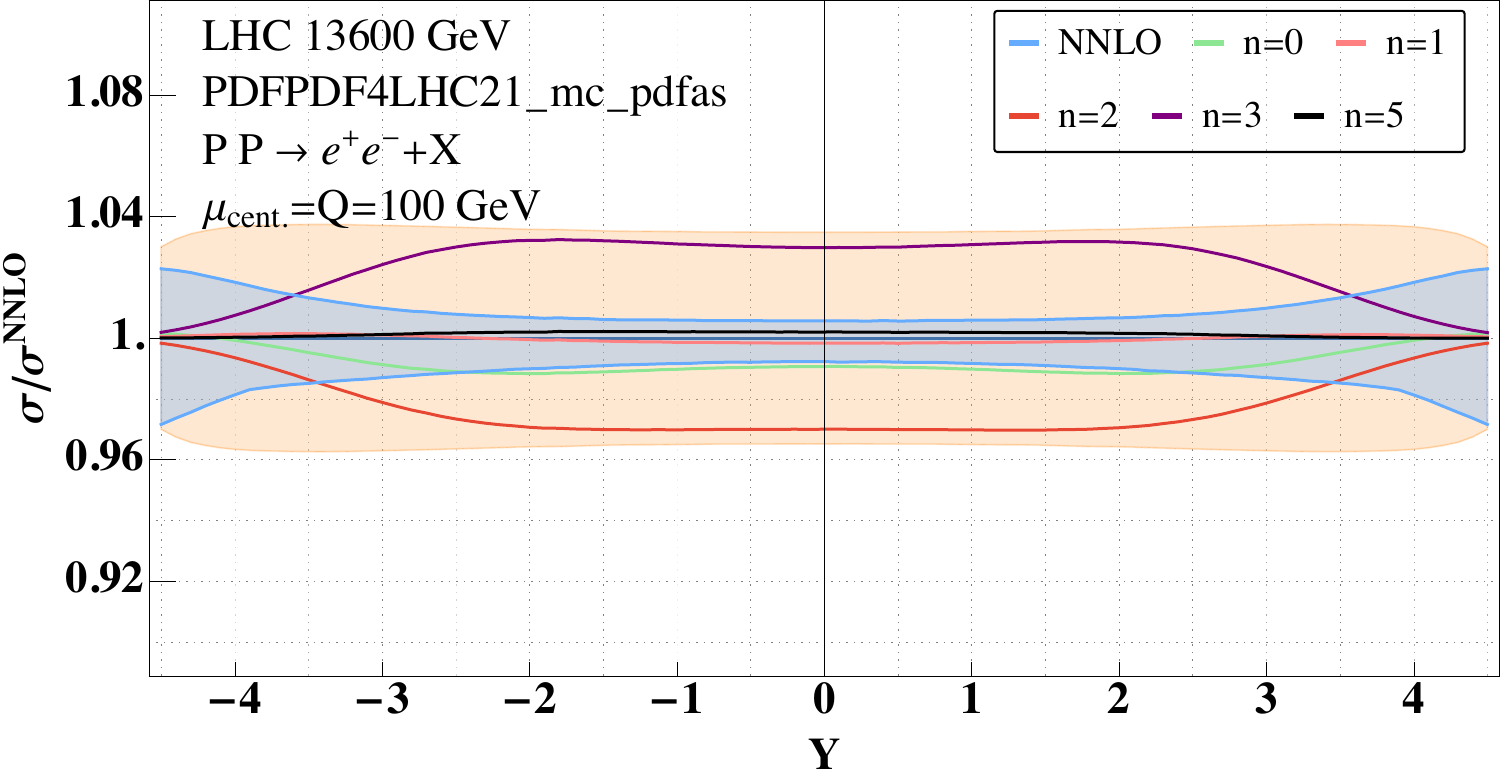}
\caption{\label{fig:NNLORapExp100}
Comparison of the expanded and exact rapidity distribution for Higgs boson (left) and Drell-Yan (right) production.
Expansions are only performed for the NNLO coefficient.
Solid lines in green, pink, red, purple and black corrspond to truncation of the collinear expansion at $n=0$, $n=1$, $n=2$, $n=3$ and $n=5$ respectively. The plots are normalized to the exact rapidity distribution at NNLO and the bands blue (orange) are derived by NNLO (NLO) scale variation.
}
\end{figure*}
Figure~\ref{fig:NNLORapExp100} compares the expansion around the collinear limit defined in eq.~\eqref{eq:expansiondef} to the exact fixed order prediction. 
We label the order at which we truncate the expansion by $n$. 
In the figure, we included LO and NLO cross sections exactly approximated only the partonic NNLO coefficient function.
We then normalized the resulting predictions to the exact NNLO contribution. 
Differently colored solid lines correspond to different orders in the truncation at power $n$.
The blue bands correspond to exact fixed order scale variations and the pink bands correspond the NLO scale variation bands centered around the NNLO prediction.

Interpreting fig.~\ref{fig:NNLORapExp100}, we find that the first few orders in our expansion alternate around the full prediction at NNLO\footnote{Note, that a similar behavior was already found for threshold expansions~\cite{Anastasiou:2015ema,Dulat:2017prg,Herzog:2014wja}.}. 
Interestingly, the relative deviation in the bulk of the rapidity distribution ($Y\sim 0$) is the same in the quark- (DY) and in the gluon (Higgs) initiated process. 
In the forward (large Y) and backward (small Y) limit of the distribution the quality of the approximation improves and even just the first term in the expansion provides excellent results. 
The stark difference in the DY and Higgs process is the residual perturbative uncertainty estimated by scale variations. 
For the DY process very small residual scale uncertainty is estimated and the collinear approximation cannot reliably approximate the true NNLO correction using only the first term. 
Predictions including the second and third term even deviate further from the true NNLO result in comparison to the first term only. 
However, this does not lend additional credence to the first term but is rather an indication of the size of power corrections to our approximation. 
On the other hand, all approximations for the Higgs boson production cross section are contained in the NNLO scale variation bands.
Knowing even the first term would have yielded an improved description of the process in the absence of NNLO corrections. 
Overall, we see that the the leading term in our expansion approximates the true NNLO rapidity distribution for both processes at the level of $\sim 4\%$.

\begin{figure*}[!h]
\centering
\includegraphics[width=0.49\textwidth]{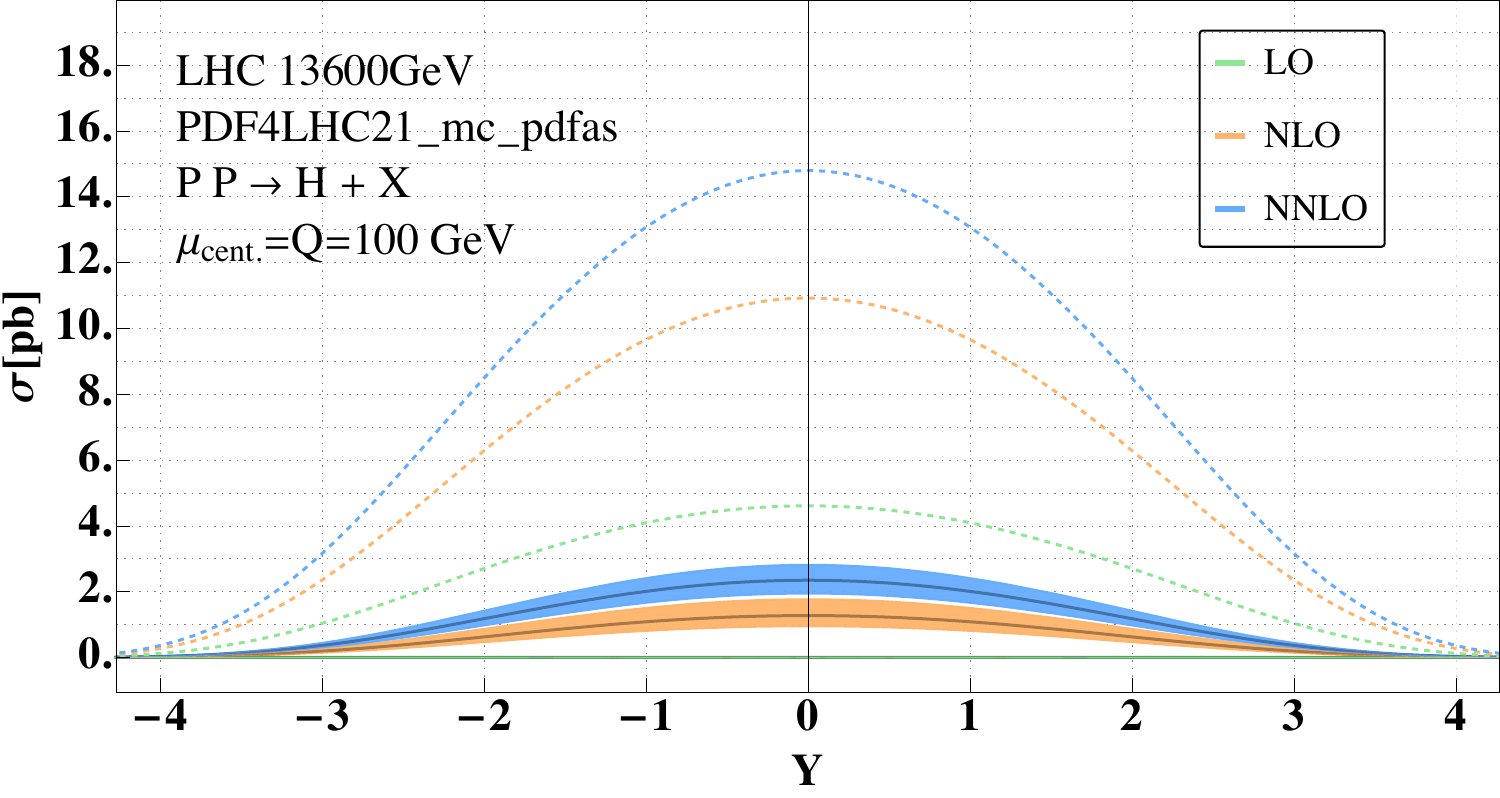}
\includegraphics[width=0.49\textwidth]{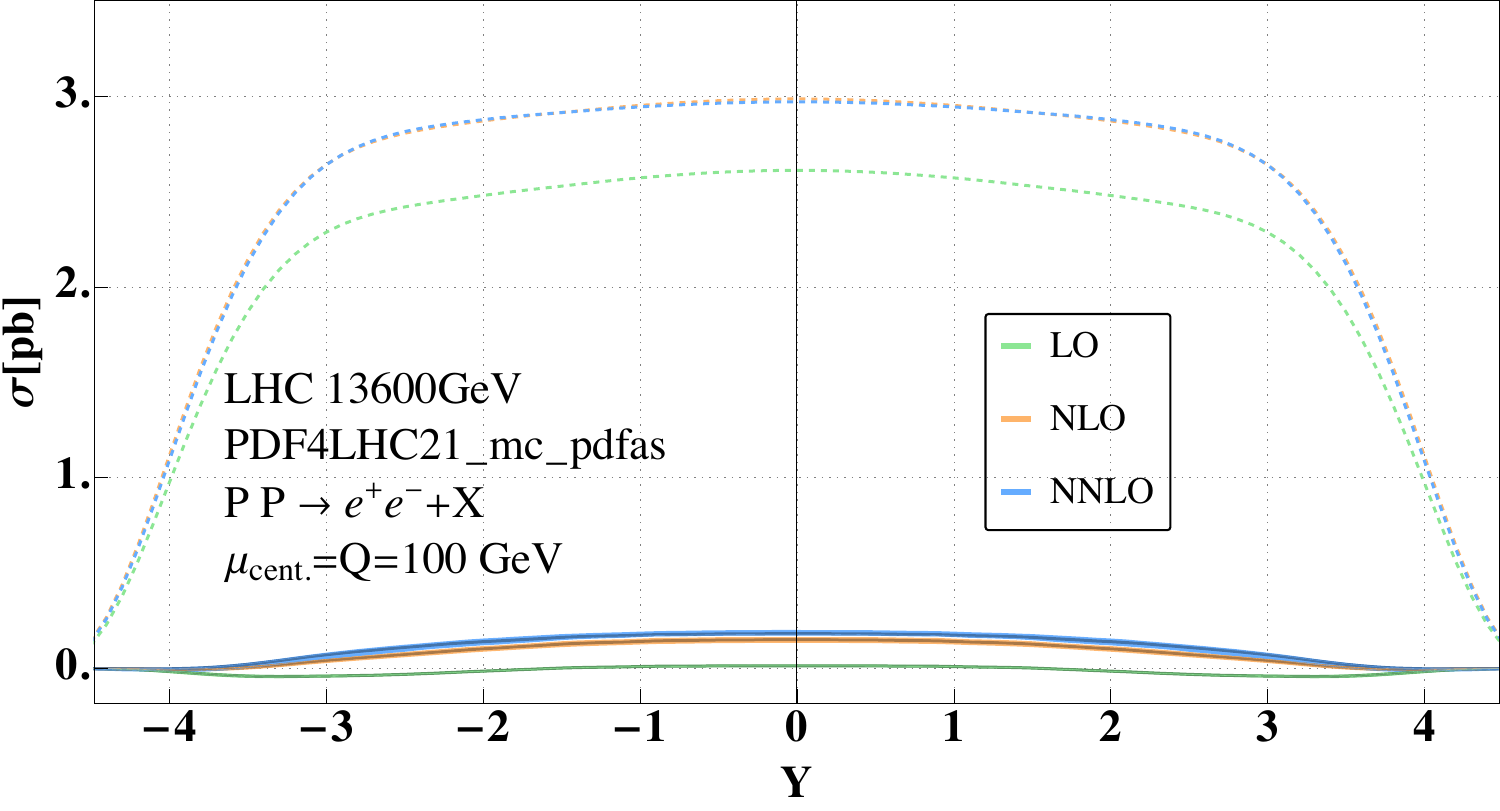}
\caption{\label{fig:RapResidual}
Rapidity distribution for the production of a Higgs boson (left) or a electron-positron pair via the Drell-Yan cross section (right). 
Dashed lines represent fixed order predictions with a scale choice of $\mu_F=\mu_R=Q$.
Solid predictions are based on subtracting our approximation from the exact NNLO result.
N$^n$LO predictions for $n=0$, $n=1$ and $n=2$ are shown in green, orange and blue respectively. 
Bands are derived based on factorization and renormalization scale variations (see text).
}
\end{figure*}
Figure~\ref{fig:RapResidual} shows the residual contribution of fixed order cross sections beyond the accuracy of our approximation.
We observe that these are overall small corrections depending on the envisioned target accuracy.
Combined with our study of higher power corrections to our approximation above we find that NNLO corrections can be approximated within $\sim\mathcal{O}(5\%)$ for a color neutral final state with $Q=100$ GeV.

\begin{figure*}[!h]
\centering
\includegraphics[width=0.49\textwidth]{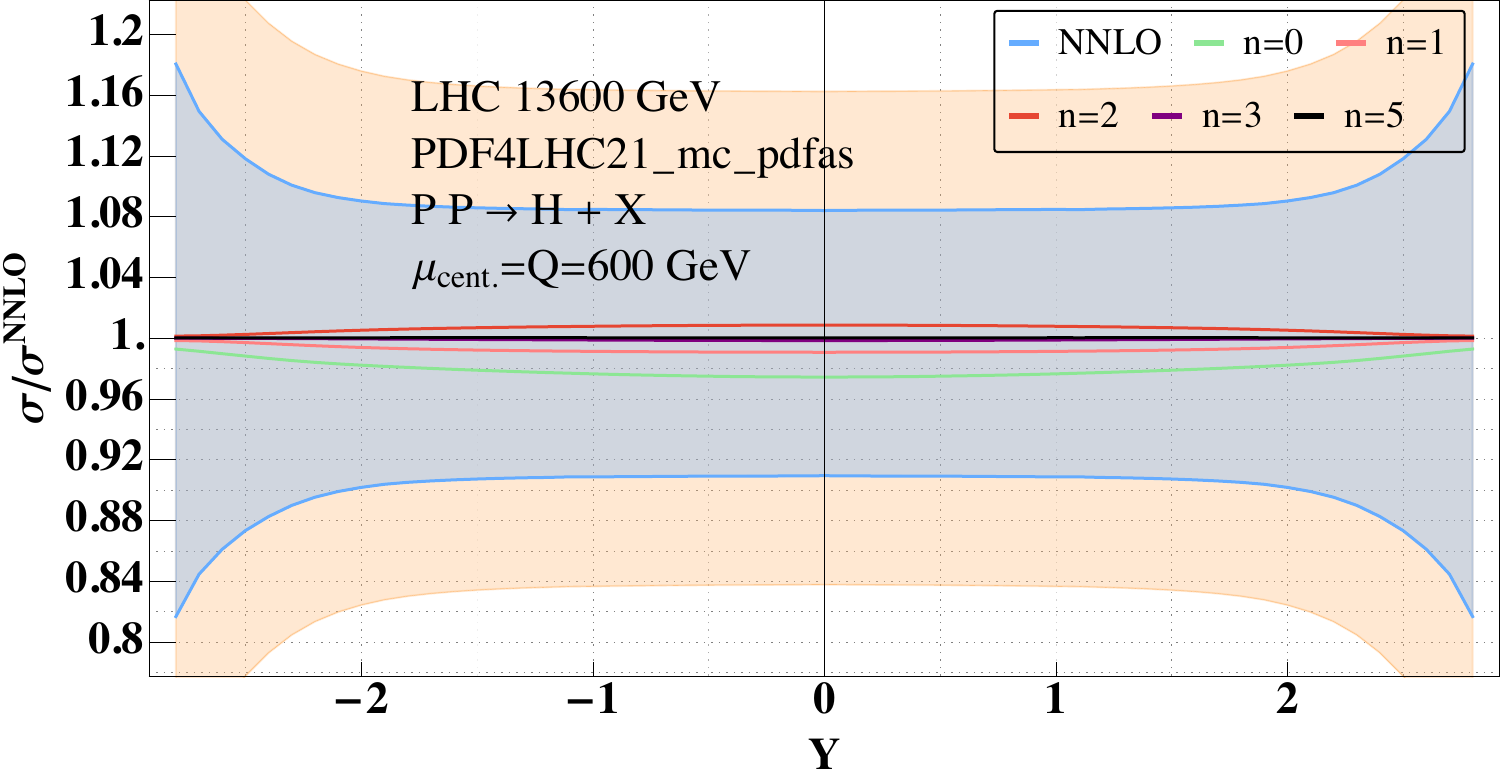}
\includegraphics[width=0.49\textwidth]{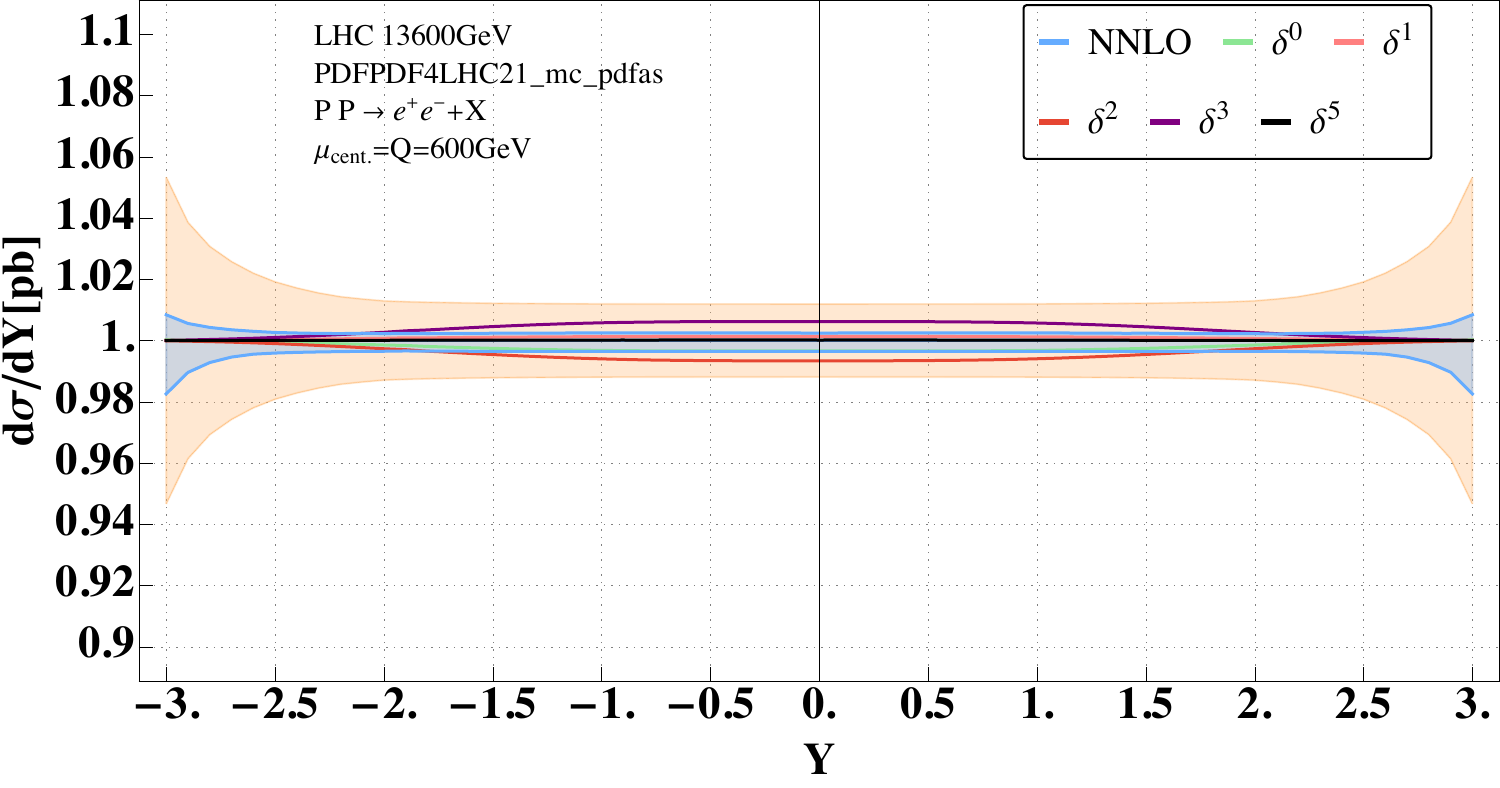}
\caption{\label{fig:NNLORapExp600}
Same as fig.~\ref{fig:NNLORapExp100} but for $Q=600$ GeV.
}
\end{figure*}
\begin{figure*}[!h]
\centering
\includegraphics[width=0.49\textwidth]{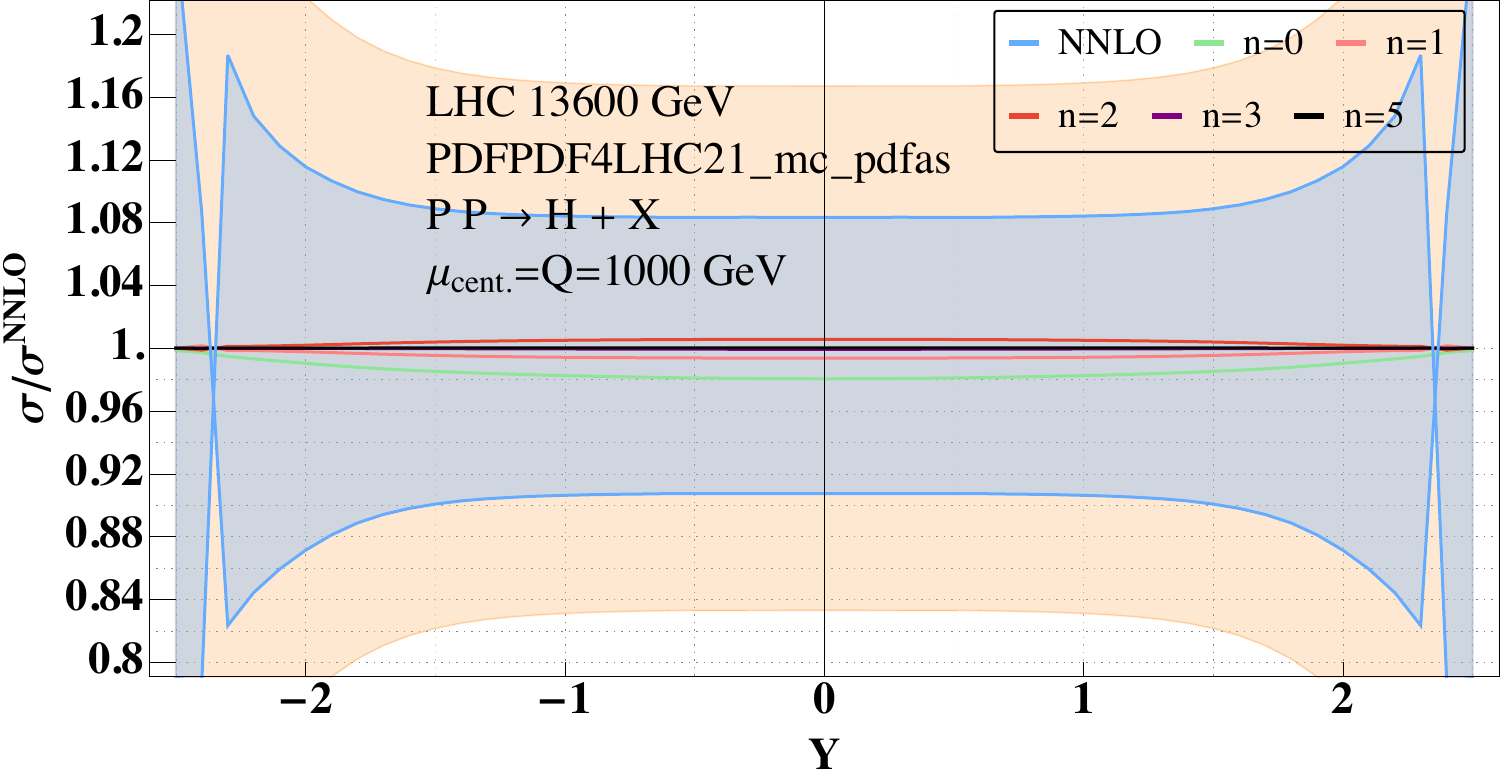}
\includegraphics[width=0.49\textwidth]{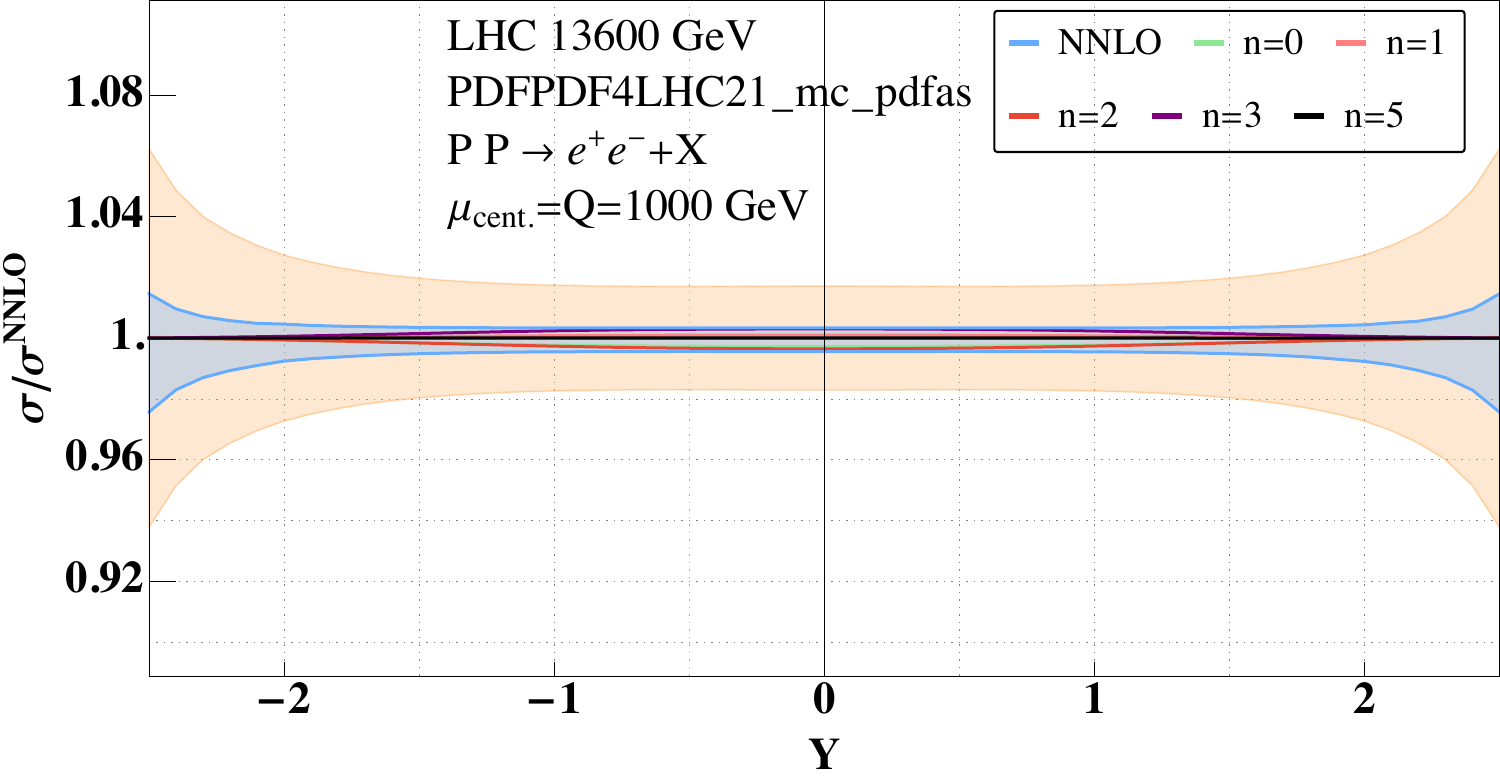}
\caption{\label{fig:NNLORapExp1000}
Same as fig.~\ref{fig:NNLORapExp100} but for $Q=1000$ GeV.
}
\end{figure*}
Figures~\ref{fig:NNLORapExp600} and ~\ref{fig:NNLORapExp1000} show the same analysis as fig.~\ref{fig:NNLORapExp100} but for $Q=600$ GeV and $Q=1000$ GeV respectively. 
We see that as we increase $Q$ the collinear approximation becomes better. 
In particular, NNLO corrections to the DY cross section are well described by our approximation at $Q=1000$ GeV. 
The range of predictions based on the different orders where the expansion is truncated may serve as an estimator of the quality of the approximation.
Alternatively, variation of power suppressed terms similar to the ones considered in ref~\cite{Herzog:2014wja} could be considered in order to derive uncertainty estimates.

\begin{figure*}[!h]
\centering
\includegraphics[width=0.49\textwidth]{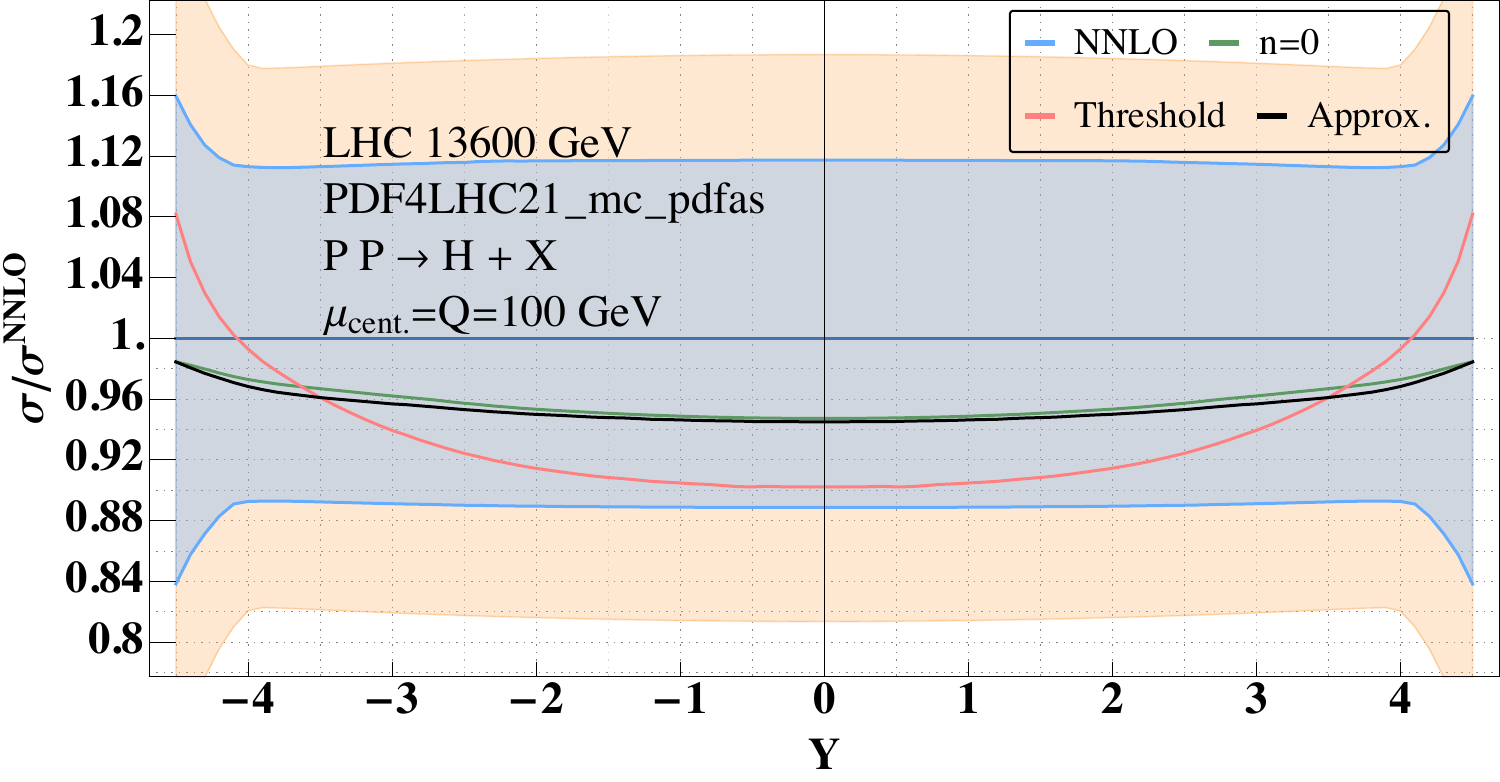}
\includegraphics[width=0.49\textwidth]{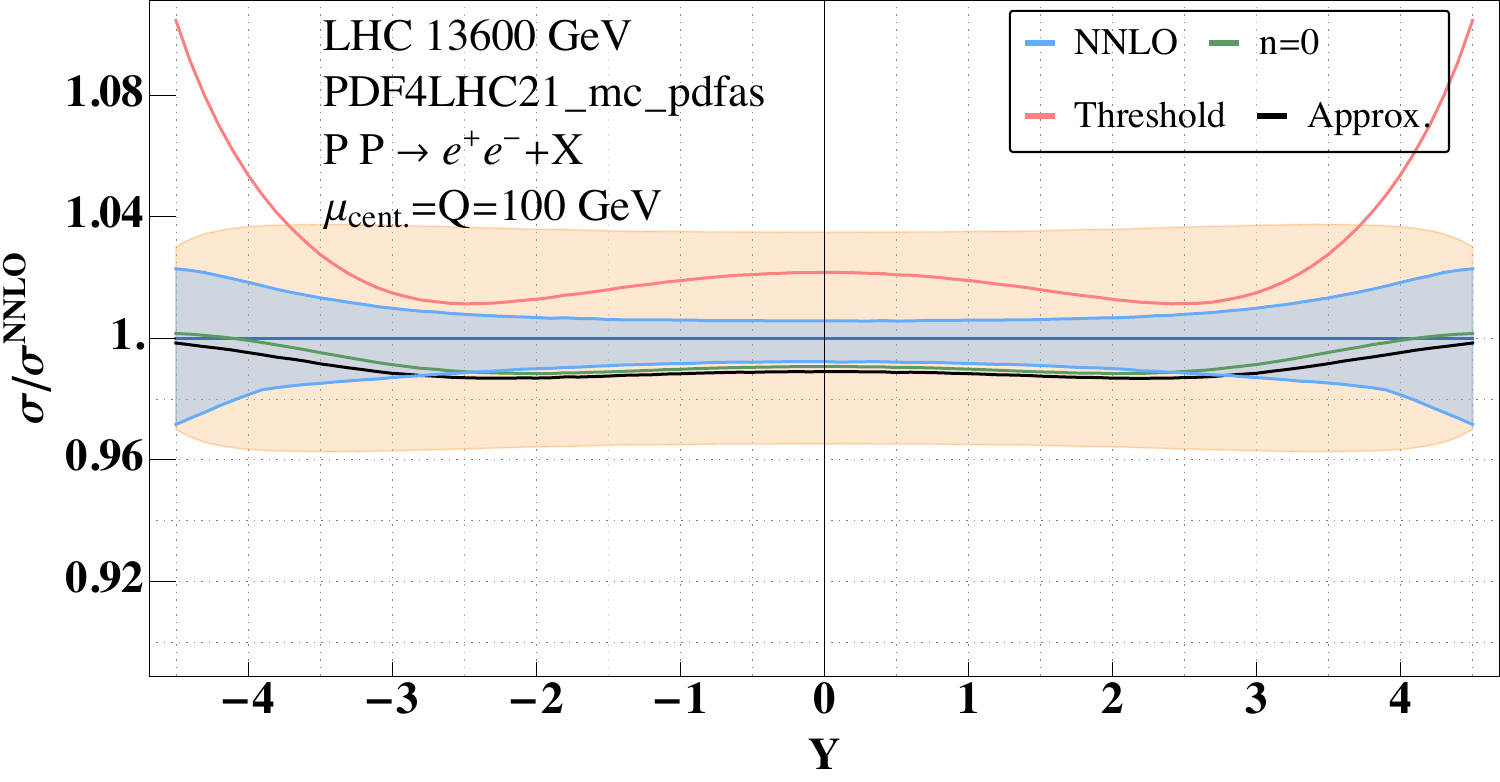}
\caption{\label{fig:NNLORapApprox100}
Comparison of the expanded and exact rapidity distribution for Higgs boson (left) and Drell-Yan (right) production.
Expansions are only performed at NNLO.
The green line correspond to truncation of the collinear expansion at $n=0$.
The pink line correspond to approximating the partonic coefficient function in the threshold limit.
The black line corresponds to the approximation derived in eq.~\eqref{eq:raplimitdef} truncated strictly at NNLO.
 The plots are normalized to the exact rapidity distribution at NNLO and the bands blue (orange) are derived by NNLO (NLO) scale variation.
}
\end{figure*}
Figure~\ref{fig:NNLORapApprox100} compares different approximations of the rapidity distributions to the exact fixed order prediction. 
In the figure, we included LO and NLO cross sections exactly approximated only the partonic NNLO coefficient function.
We then normalized the resulting predictions to the exact NNLO contribution. 
The blue bands correspond to exact fixed order scale variations and the pink bands correspond the NLO scale variation bands centered around the NNLO prediction.
Differently colored solid lines correspond to different approximations.
The pink line includes only the threshold approximation of the partonic coefficient function, i.e. the limit where both $\xi_1\to 1$ and $\xi_2\to 1$.
The green line represents the $n=0$ approximation of eq.~\eqref{eq:expansiondef}. 
The black line is the result of expanding eq.~\eqref{eq:raplimitdef} to fixed order. 
The green and black prediction differ only by power suppressed terms beyond the collinear approximation and are remarkably similar.
The threshold approximation is visibly a worse approximation of the hadronic cross section and indicates larger power suppressed terms. 

Naturally, our approximation can also be used to approximate the inclusive cross section using eq.~\eqref{eq:increl}.
The pattern displayed is of course very similar to the one observed in the bulk of the rapidity distribution. 
The first few orders display a varying pattern around the exact result and after about five terms the approximation is close to perfect.

\subsection{Channel by Channel comparison}
In building approximations for higher order corrections, it is crucial to be able to correctly capture as many channels as possible.
As a matter of fact, when considering corrections at NNLO and beyond all channels start contributing and the numerical impact of each one needs to be carefully addressed.
This is particularly true if one wants to build a theoretical framework applicable to several processes. 
The prototypical test for this interplay between different channels are the NNLO corrections to the Drell-Yan rapidity spectrum at the LHC for invariant masses around the electroweak scale.
In this case, the off-diagonal $qg$ channel is strongly anticorrelated with the diagonal $q\bar{q}$ channel~\cite{Duhr:2020sdp,Cruz-Martinez:2025ffa}.
The NNLO corrections in each of these channels are sizable, but because of this anti-correlation their sum turns out to be very small.
\begin{figure*}[!h]
\centering
\includegraphics[width=0.7\textwidth]{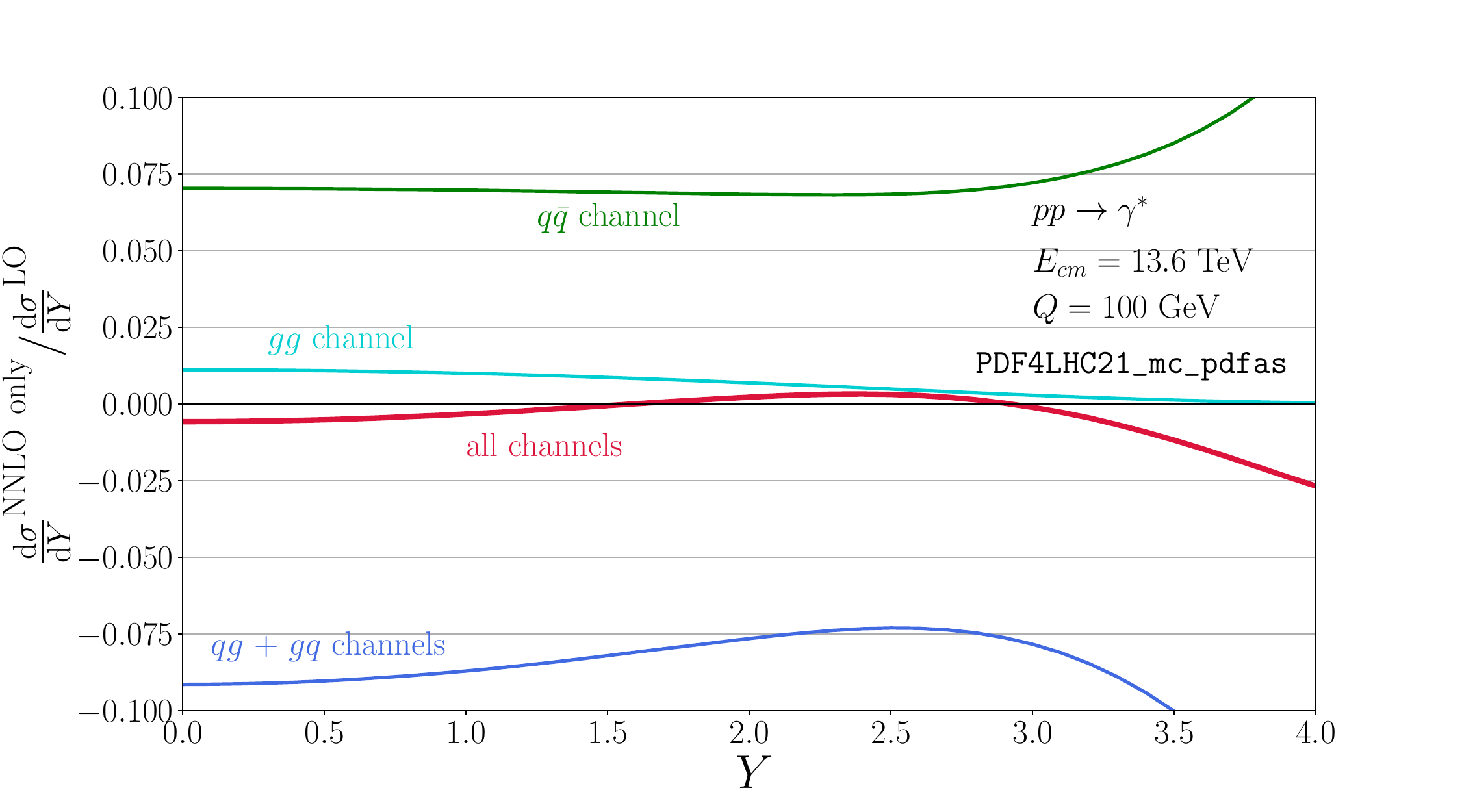}
\caption{\label{fig:NNLOChannelsDY}
Channel decomposition of the NNLO corrections to the Drell-Yan rapidity distribution}
\end{figure*}

This mechanism is the main origin of the tiny corrections (and scale uncertainty estimate) appearing at NNLO for this process, but this delicate cancellation is reduced at N$^3$LO~\cite{Duhr:2021vwj,Duhr:2020sdp,Duhr:2020seh}.
\begin{figure*}[!h]
\centering
\includegraphics[width=0.52\textwidth]{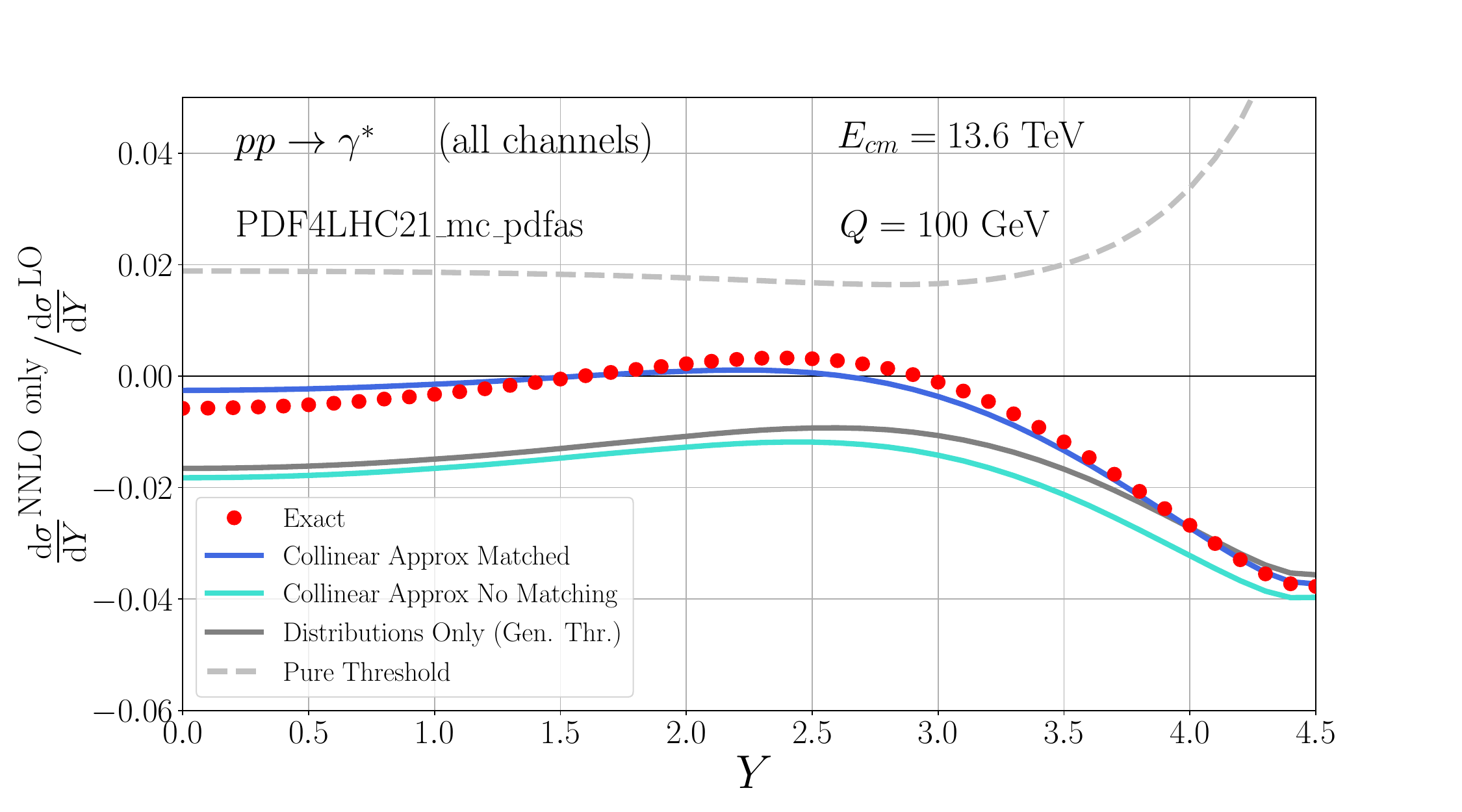}
\!\!\!\!\!\!\!\!\!\!\!\!\!\!\!\!
\includegraphics[width=0.52\textwidth]{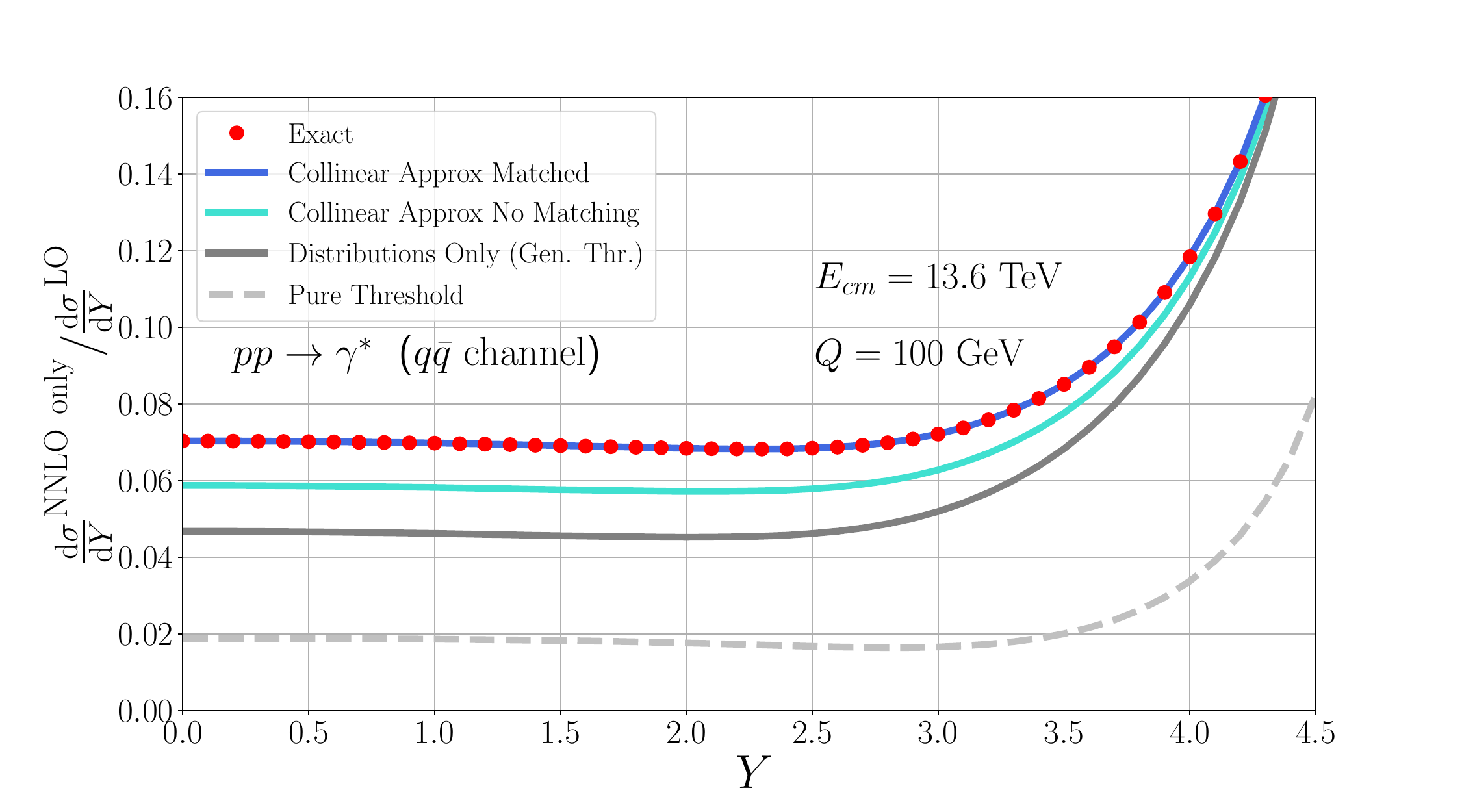}
\\
\includegraphics[width=0.52\textwidth]{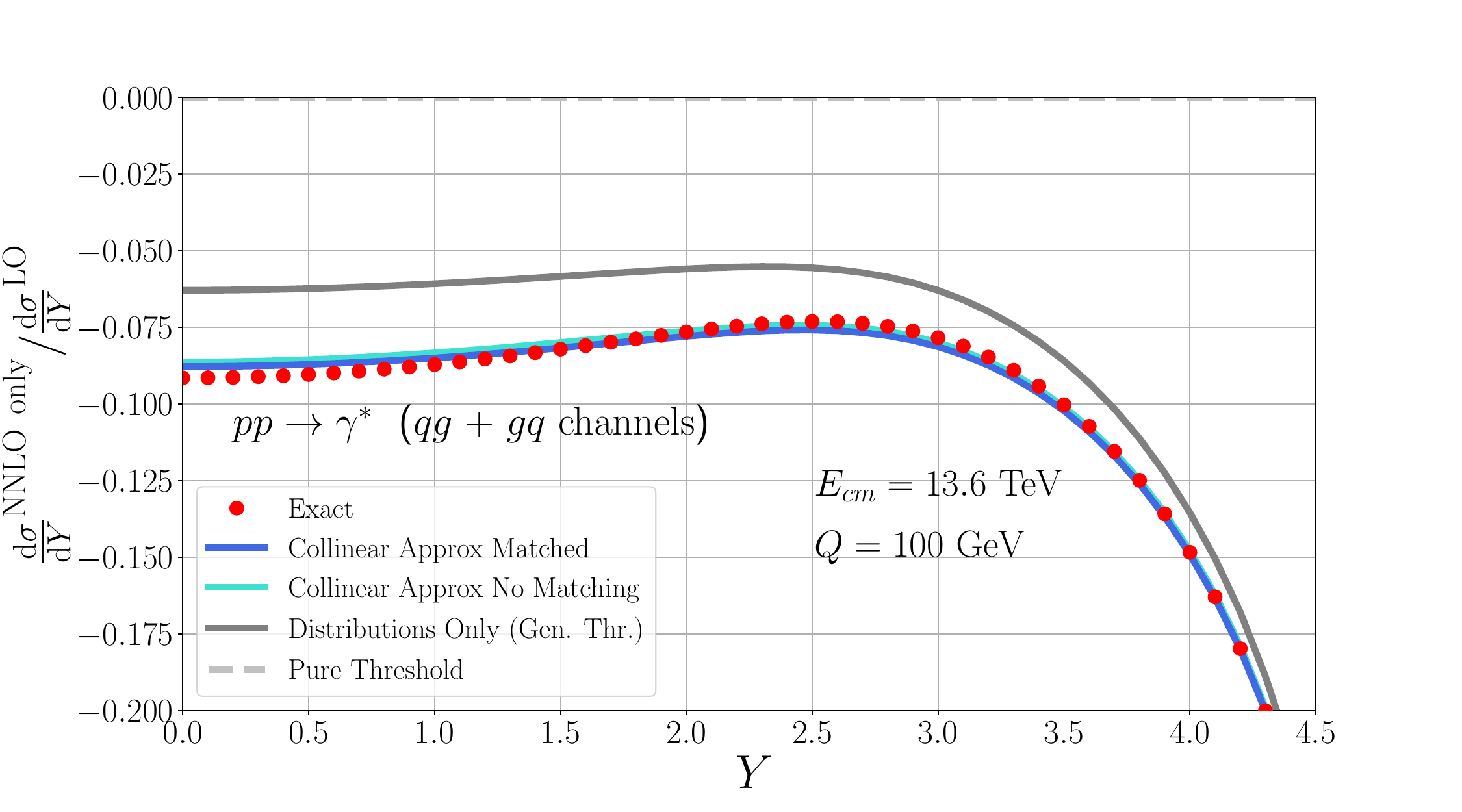}
\!\!\!\!\!\!\!\!\!\!\!\!\!\!\!\!
\includegraphics[width=0.52\textwidth]{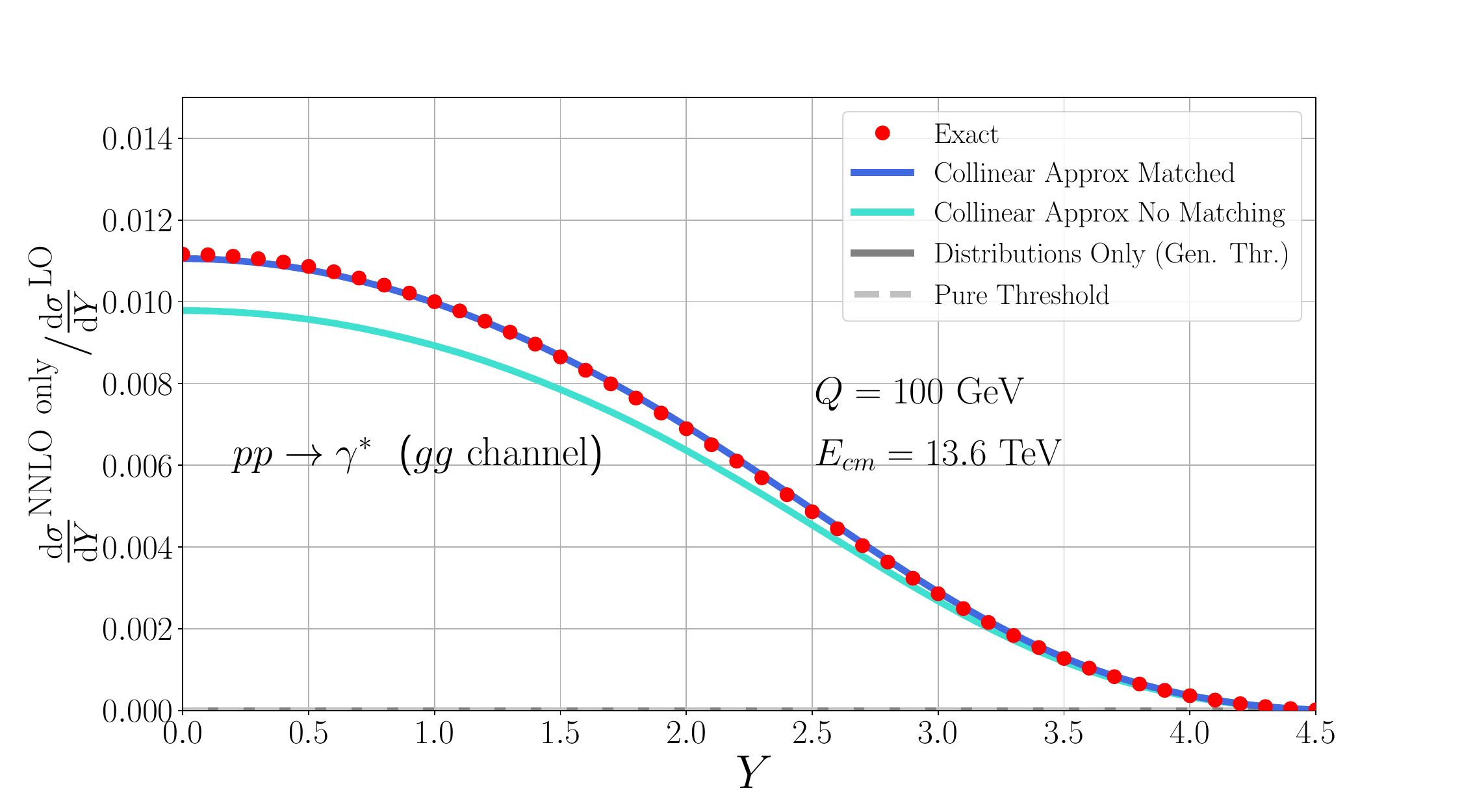}
\caption{\label{fig:NNLOChannelCompDY}
Channel by channel comparison Drell-Yan}
\end{figure*}
In \fig{NNLOChannelCompDY} we show the break down by channel of our collinear approximation compared to the exact result from \texttt{Vrap} \cite{Anastasiou:2003ds} and other types of approximation of rapidity distributions.
In ref.~\cite{Dulat:2018bfe} an additional approximation was introduced in which power suppressed terms (beyond the validity of the soft approximation considered in that work) are added such that the partonic coefficient function of the rapidity distribution integrates to the partonic coefficient function of the inclusive cross section.
Here, following the same approach and the relation to the inclusive partonic cross section of \ref{sec:relationtoInc}, we extend this procedure to the case of the collinear approximation. We present the results for this approximation in our figures, using the label \emph{matched}. 
Looking at \fig{NNLOChannelCompDY}, we see that the collinear approximation provides a great approximation for \emph{every} partonic channel. 
Compared to the threshold approximation, our collinear approximation not only correctly accounts for all higher order partonic channels, but also captures more terms overall in each channel. 
We also notice that our collinear approximation accounts very well for the contribution of the double off-diagonal channel $gg$, which is completely missed by both the threshold and generalized threshold approximations, even though it contributes to a non negligible part of the full NNLO coefficient. 
Overall, these results support that that our approach will provide a good approximation for a wide range of processes. Moreover, the matching to the exact inclusive partonic cross section further improves the agreement with the exact differential result, yielding spectacular agreement for each channel and throughout the entire range of rapidity.

%% file: Chapters/Resummation.tex
\section{Resummation}
\label{sec:resummation}
In this section we first derive a formula for the resummation of logarithms of $1-\xi_i$.
Next, we study the phenomenological implications of this resummation formula at the example of the hadronic Drell-Yan and gluon fusion Higgs boson production cross section.

\subsection{Evolution and Resummation of individual Functions}
We begin by deriving the scale evolution and resummation of the hard, soft and beam functions separately.
Our approach is based on well-established techniques used throughout resummation and Soft and Collinear Effective Theory \cite{Bauer:2000ew, Bauer:2000yr, Bauer:2001ct, Bauer:2001yt}.
The general idea is pretty straightforward: the renormalization factors of \sec{HFsetup}, \sec{SFsetup} and \sec{BFsetup}, encode divergences that in SCET are of UV nature. We can therefore derive RG equations by demanding the invariance on the renormalization scale of the bare Hard, Beam and Soft functions, in a similar fashion as one would do for the running coupling in massless QCD.
In this way each function obey an RGE in terms of an anomalous dimension directly related to its renormalization factor. 
The resummed cross section is then given by \eq{raplimitdef} where now each function is obtained as the solution of its own RG equation.
Note that the dependence on $\mu$ cancels in \eq{raplimitdef} and, provided we run each function to the same common scale $\mu$, one can pick $\mu$ to be \emph{any} arbitrary scale.
It is convenient and customary, although not necessary, to choose $\mu$ to be one of the physical scales in the problem. 
This has the benefit, for example, that at least one between the Hard, beam and soft functions have no RGE running as its default scale equals to $\mu$.
In this article, we will work with the choice $\mu=Q$.

\subsubsection{Resummation of the  Hard Function}

The hard function obeys the renormalization group equation (RGE)
\beq
\label{eq:HRGE}
	\mu^2 \frac{\df }{\df \mu^2} H_{ij}\left(Q^2,\mu^2\right) = \left[\GammaC^{r_{ij}}(\alpha_s(\mu))\ln \frac{Q^2}{\mu^2} + \frac{1}{2}\gamma_H^{r_{ij}}(\alpha_s(\mu^2)) \right]H_{ij}(Q^2,\mu^2),
\eeq
Above $\Gamma_\text{cusp}^r(\alpha_S(\mu^2))$ is the cusp anomalous dimension~\cite{Korchemsky:1993uz} which is known to fourth loop order~\cite{vonManteuffel:2020vjv,Henn:2019swt}.
The cusp anomalous dimension does not depend on the particular flavors $i$ or $j$ but on the color representation of the partons, which we indicate with the superscript $r$.
Furthermore, $\gamma_H^r[\alpha_s(\mu^2)]$ is the non-cusp part of the anomalous dimension.

The evolution equation is readily solved by the following solution expressing the hard function at a scale $\mu$ in terms of a boundary evaluated at the hard scale $\mu_H$ and an evolution factor
\beq\label{eq:HardRGE_sol}
H_{ij}(Q^2,\mu^2) =H_{ij}(Q^2,\mu_H^2) \times \exp\left\{\int_{\mu_H^2}^{\mu^2} \frac{\df \nu^2}{\nu^2}
\GammaC^{r_{ij}}[\alpha_s(\nu)]\ln \frac{Q^2}{\nu^2} + \frac{1}{2}\gamma_H^{r_{ij}}[\alpha_s(\nu^2)] 
\right\}.
\eeq
A solution is easily obtained numerically or by perturbative expansion. 
Note that the boundary function $H_{ij}(Q^2,\mu_H^2)$ contains fixed order logarithms of the form $\ln^k Q/\mu_H$. 
The choice of $\mu_H$ is in principle arbitrary, but given the presence of these logarithms in the boundary function we want to choose $\mu_H \sim Q$ such that large logarithms get removed.
With $\mu_H \sim Q$ and our choice $\mu = Q$, we are effectively turning off the resummation of logarithms in the hard function, but it is clear from \eq{HardRGE_sol} how to resum them in case one makes a different set of choices.

\subsubsection{Resumming the Rapidity Soft Function}\label{sec:resSoft}
The soft function is most easily studied after performing a Laplace transformation, eq.~\eqref{eq:LTrafo}.
We find that 
\beq
\tilde S^r\left(\log \frac{\kappa_1 \kappa_2 \mu^2}{Q^2},\mu^2\right) =\mathcal{L}\left[\mathcal{L}\left[S(\mu^2;\bar \xi_1\bar \xi_2),\bar \xi_2,\kappa_2\right],\bar \xi_1,\kappa_1\right].
\eeq
In Laplace space, the soft function only depends on the product of $\kappa_1$, $\kappa_2$ and $\mu^2$ as indicated above. This is consequence in SCET of reparametrization invariance (RPI)~\cite{Manohar:2002fd,Marcantonini:2008qn}.
This means we can determine the functional dependence of the soft function on the variables $\kappa_i$ by solving its evolution equation with respect to the perturbative scale $\mu$.
The renormalization group equation of the soft function is given by
\beq
\frac{\df \log \tilde  S^r\left(\log \frac{\kappa_1 \kappa_2 \mu^2}{Q2},\mu^2\right)}{\df \log (\mu^2)} =\GammaCS[\alpha_S(\mu^2)] \log\left(\frac{\kappa_1\kappa_2\mu^2}{Q^2}\right)+\frac{1}{2} \gammaS[\alpha_S(\mu^2)].
\eeq
The evolution of the soft function is consequently determined by the two anomalous dimensions $\Gamma^r_{\text{S}}$ and $\gammaS$ and we summarize their relation to known quantities below in sec.~\ref{sec:anomdimrels}
The solution to the RGE consequently takes the form
\bea
\tilde S^r\left(\ln \frac{\kappa_1 \kappa_2\mu^2}{Q^2},\mu^2\right)=\tilde  S^r\left(\ln \frac{\kappa_1 \kappa_2 \mu_S^2}{Q^2},\mu_S^2\right)e^{\int_{\mu_S^2}^{\mu^2}\frac{\df \nu^{2}}{\nu^2} \left[\GammaCS[\alpha_S(\nu^2)] \ln\left(\frac{\kappa_1\kappa_2\nu^2}{Q^2}\right)+\frac{1}{2} \gammaS[\alpha_S(\nu^2)]\right]}
\eea
We introduce the short hand notation
\bea
A(\gamma,\mu^2,\mu_0^2))&=&\int_{\mu_0^2}^{\mu^2} \frac{\df \nu^2}{\nu^2} \gamma[\alpha_S(\nu^2)] \log\left(\frac{\mu_0^2}{\nu^2}\right),\nonumber\\
B(\gamma,\mu^2,\mu_0^2))&=&\int_{\mu_0^2}^{\mu^2} \frac{\df \nu^2}{\nu^2} \gamma[\alpha_S(\nu^2)].
\eea
With this we may re-arrange the exponential in our RGE to find
\bea
\tilde  S^r\left(\ln \frac{\kappa_1 \kappa_2\mu^2}{Q^2},\mu^2\right)&=& \tilde S^r\left( \ln \frac{\kappa_1 \kappa_2\mu_S^2}{Q^2},\mu_S^2\right)
 \left(\frac{\kappa_1\kappa_2 \mu_S^2}{Q^2}\right)^{\eta}\Bigg|_{\eta=B\left(\GammaCS,\mu^2,\mu_S^2\right)}\\
&\times &\exp\left\{A\left(\GammaCS,\mu^2,\mu_S^2\right)+\frac{1}{2}B\left(\gammaS,\mu^2,\mu_S^2\right)\right\}\nn\,.
 \eea
Because the soft function only depends on the $\kappa_i$ via logarithms we may promote it to an operator depending on the derivative with respect to $\eta$.
\bea
 \tilde S^r\left(\ln \frac{\kappa_1 \kappa_2 \mu_S^2}{Q^2},\mu^2\right)&=&\tilde  S^r\left(\frac{\df }{\df \eta},\mu_S^2\right)
 \left(\frac{\kappa_1\kappa_2 \mu_S^2}{Q^2}\right)^{\eta}\Bigg|_{\eta=B\left(\GammaCS,\mu^2,\mu_S^2\right)}\\
&\times &\exp\left\{A\left(\GammaCS,\mu^2,\mu_S^2\right)+\frac{1}{2}B\left(\gammaS,\mu^2,\mu_S^2\right)\right\}\nn\,.
 \eea
Now we would like to perform the inverse of the Laplace transform.
To this end we make use of the formula
\beq
\mathcal{L}\left[x^{-1-\eta},x,\kappa\right]= e^{-\gamma_E \eta} \kappa^\eta\Gamma(-\eta).
\eeq
We create a dictionary allowing us to transform back to momentum space using the following identities.
\bea
\mathcal{L}^{-1}\left[ \kappa^\eta,\kappa,x\right]&=&\frac{e^{\gamma_E \eta}}{\Gamma(-\eta)} x^{-1-\eta}.\\
\mathcal{L}^{-1}\left[ \log^n(\kappa) \kappa^\eta,\kappa,x\right]&=&\mathcal{L}^{-1}\left[ \frac{\df^n}{\df \eta^n} \kappa^\eta,\kappa,x\right]=
\frac{\df^n}{\df \eta^n}\left(\frac{e^{\gamma_E \eta}}{\Gamma(-\eta)} x^{-1-\eta}\right).
\eea
With this we find the resummed soft function
\bea
\label{eq:softresummed}
 S^r\left(\mu^2,\mu_S^2;\bar\xi_1\bar\xi_2\right)&=& \tilde{S}^r\left(\frac{\df }{\df \eta},\mu_S^2\right)
\frac{e^{2\eta\gamma_E}}{ \bar \xi_1\bar \xi_2 \Gamma(-\eta)^2} \left(\frac{Q^2 \bar \xi_1 \bar \xi_2}{\mu_S^2} \right)^{-\eta}\Bigg|_{\eta=B\left(\GammaCS,\mu^2,\mu_S^2\right)}\nonumber\\
&\times &\exp\left\{A\left(\GammaCS,\mu^2,\mu_S^2\right)+\frac{1}{2}B\left(\gammaS,\mu^2,\mu_S^2\right)\right\}
 \eea

\subsubsection{Resummation of the Rapidity Beam Function}\label{sec:resBeam}
To discuss the resummation of the rapidity beam function we first define the Laplace transform with respect to $\bar \xi_1$ of the beam function matching kernel.
\beq
\tilde{I}^Y_{ij}\left(\log\frac{\kappa_1\mu^2}{Q^2},\xi_2,\mu^2\right)=\mathcal{L}\left[ I^Y_{ij}(\xi_2,\mu^2;\bar \xi_1) ,\bar \xi_1,\kappa_1\right].
\eeq
After Laplace transformation, the logarithms in $\mu^2$ and $\kappa_1$ combine and the matching kernel has no other explicit dependence on $\kappa_1$.
The renormalization group equation for matching kernel is easily derived by consistency from eq.~\eqref{eq:raplimdef}.
\bea\label{eq:Ikernel_RGE}
&&\frac{\df }{\df \log \mu^2} \tilde{I}^Y_{ij}\left(\log\frac{\kappa_1\mu^2}{Q^2},\xi_2,\mu^2\right)=
-\sum\limits_{k}\tilde{I}^Y_{ik}\left(\log\frac{\kappa_1\mu^2}{Q^2},\xi_2,\mu^2\right)\otimes_{\xi_2} P_{kj}(\xi_2,\mu^2)
\nonumber\\
&&+\left[\GammaCB[\alpha_S(\mu^2)] \log\left(\frac{\kappa_1\mu^2}{Q^2}\right)+\frac{1}{2} \gammaB[\alpha_S(\mu^2)]\right]\tilde{I}^Y_{ij}\left(\log\frac{\kappa_1\mu^2}{Q^2},\xi_2,\mu^2\right).
\eea
Above, $P_{kj}$ is the Altarelli-Parisi splitting function and $\gammaB$ is the non-cusp anomalous dimensions governing the evolution of the beam function. This anomalous dimension is the same for several SCET I collinear object. As a matter of fact is the same anomalous dimension as the thrust jet function and the $N$-Jettiness beam function \cite{Stewart:2009yx,Becher:2010tm} and it has been extracted to 4 loops \cite{Duhr:2022cob} from the 4 loop collinear and virtual anomalous dimensions \cite{Das:2019btv,Agarwal:2021zft} using consistency relations.
The first term on the right hand side of the above equation compensates the scale evolution of the parton distribution function in the rapidity beam function.
We consequently find that the Laplace transform of the rapidity beam function,
\beq
\tilde{B}^Y_{ij}\left(\log\frac{\kappa_1\mu^2}{Q^2},\xi_2,\mu^2\right)=\mathcal{L}\left[ B^Y_{ij}(\xi_2,\mu^2;\bar \xi_1) ,\bar \xi_1,\kappa_1\right],
\eeq
 satisfies the following evolution equation
\beq
\frac{\df }{\df \log \mu^2} \log \tilde{B}^Y_{i}\left(\log\frac{\kappa_1\mu^2}{Q^2},\xi_2,\mu^2\right)=
\left[\GammaCB[\alpha_S(\mu^2)] \log\left(\frac{\kappa_1\mu^2}{Q^2}\right)+\frac{1}{2} \gammaB(\alpha_S(\mu^2))\right]\,.
\eeq
Note that the RHS is $\xi_2$ independent. The fact that the full beam function anomalous dimension depends only on its observable and not on the momentum fraction variable is a well known and general feature of the RGE of beam functions \cite{Stewart:2009yx} and it appears also for $0$-Jettiness and the $q_T$ beam functions \cite{Stewart:2009yx,Becher:2011xn,Chiu:2012ir}. For the rapidity beam function this was already observed in \cite{Lustermans:2019cau}.
Following the same derivation as in the previous section for the rapidity soft function we arrive at the resummed result for the rapidity beam function.
\bea
 B_{i}\left(\xi_2,\mu^2,\mu_{C_1}^2;\bar \xi_1 \right)&=& \tilde{B}_i\left(\frac{\df }{\df \eta},\xi_2,\mu_{C_1}^2\right)
\frac{e^{\eta\gamma_E}}{ \bar \xi_1 \Gamma(-\eta)} \left(\frac{Q^2 \bar \xi_1 }{\mu_{C_1}^2} \right)^{-\eta}\Bigg|_{\eta=B\left(\GammaCB,\mu^2,\mu_{C_1}^2\right)}\nonumber\\
&\times &\exp\left\{A\left(\GammaCB,\mu^2,\mu_{C_1}^2\right)+\frac{1}{2}B\left(\gammaB,\mu^2,\mu_{C_1}^2\right)\right\}
 \eea
 The Laplace space boundary conditions of the beam function can be expanded perturbatively and take the following form.
 \beq
 \tilde{B}^{Y}_i\left(\frac{\df }{\df \eta},\xi_2,\mu_{C_1}^2\right)=\sum_{o=0}^\infty \sum_{l=0}^{2o} \left(\frac{\alpha_S\left(\mu_{C_1}^2\right)}{\pi}\right)^o \tilde{B}_i^{Y,(o,l)}(\xi_2,\mu_{C_1}^2) \frac{\df^l}{\df\eta^l}.
 \eeq

\subsubsection{Summary of Anomalous Dimensions} 
\label{sec:anomdimrels}

In summary, in Laplace space the hard, soft, and beam functions contributing to our inclusive cross section satisfy an evolution equation of the form
 \beq
\frac{\df}{\df \log \mu^2}\tilde F\left(\log\frac{\kappa \mu^2 }{Q^2},\mu^2\right)=\left[\Gamma(\alpha_S(\mu^2)) \log\left(\frac{\kappa \mu^2}{Q^2}\right)+\frac{1}{2}\gamma(\alpha_S(\mu^2))\right]\tilde F\left(\log\frac{\kappa \mu^2 }{Q^2},\mu^2\right)
 \eeq
Here, $\tilde F\in \{H_{ij},\tilde B_i^Y,\tilde S_{r}\}$.
 In the case of the hard function no Laplace transform is necessary and $\kappa=1$ in the above equation.
 In the following, we summarize the specific anomalous dimensions contributing to each function.

\begin{table}[!h]
\begin{center}
\begin{tabular}{ | c || c | c | c |  }
 \hline
 & $H_{ij}$ & $\tilde B^Y_i$ & $\tilde S_r$ \\
 \hline
 \hline
 $\Gamma_F^r$ & $- \Gamma_{\text{cusp}}^r$ & $ \Gamma_{\text{cusp}}^r$  & $- \Gamma_{\text{cusp}}^r$ \\
 \hline
 $\gamma^{\text{F,r}}$ & $\gamma_{\text{H}}^r$ &  $\gamma_{J}^r=\frac{1}{2}\left(\gamma_{\text{thr.}}^r-\gamma_{\text{coll.}}^r\right) $ &  $- \gamma_{\text{thr.}}^r$ \\
  \hline
\end{tabular}
\caption{ \label{tab:anomdim} Summary of the anomalous dimensions used for the hard, soft and beam function.}
\end{center}
\end{table}
Here, the superscript $r$ refers to the color representation of the initial state partons. 
 Above $ \Gamma_{\text{cusp}}^r$ is the cusp anomalous dimension~\cite{Korchemsky:1993uz} which is known to fourth loop order~\cite{vonManteuffel:2020vjv,Henn:2019swt} and approximated using a Padé expansion to fifth loop order in ref.~\cite{Herzog:2018kwj}.
 Furthermore, $\gamma_{\text{thr.}}^r$ is the threshold anomalous dimension which was computed recently to N$^4$LO in QCD perturbation theory in refs.~\cite{Duhr:2022cob,Das:2020adl}.
The hard function anomalous dimension $\gamma_H^{r}$ is typically equal to the collinear anomalous dimension.
 The collinear anomalous dimension was computed in refs.~\cite{vonManteuffel:2020vjv,Agarwal:2021zft} as well through fourth loop order.
 In case of Higgs boson gluon fusion production in the limit of infinite top quark mass, the hard function also contains a Wilson coefficient and its anomalous dimension needs to be added to the collinear anomalous dimension.
 $\gamma^{i}_J$ is the thrust anomalous dimension (see for example ref.~\cite{Duhr:2022cob})

\subsection{Resummation Formula}
In this subsection we collect the resumed rapidity soft function and rapidity PDF to present a combined resummed formula for the hadronic cross section.
The obtained result is analogous to the approximate hadronic cross section of eq.~\eqref{eq:raplimitdef} and is given by
\bea
\label{eq:rapresummed}
&&Q^2 \frac{\df \sigma_{P\,P\rightarrow h+X}}{\df Y\df Q^2}(\xi_1,\xi_2,\mu^2, \mu_S^2,\mu_{C_1}^2,\mu_{C_2}^2)\nonumber\\
&&=\tau \sum_{i,j}  H_{ij}(\mu^2,\mu_H^2) \left[
B^{Y}_{i}(\xi_1,\mu^2,\mu_{C_1}^2;\bar \xi_2) \otimes_{\xi_1;\xi_2} S^{r_{ij}}(\mu^2,\mu_{S}^2;\bar \xi_1\bar \xi_2) \otimes_{\xi_2;\xi_1} B^{Y}_{i}(\xi_2,\mu^2,\mu_{C_2}^2;\bar \xi_1)
 \right]\nonumber\\
&&+\mathcal{O}\left(\bar \xi_1^0,\bar\xi_2^0\right).
\eea
We refrain from resumming the logarithms appearing in the hard function by choosing $\mu_H=\mu$. 
The individual functions in the above resummation formula are independent of the scales $\mu_{C_1}^2$, $\mu_{C_2}^2$ and $\mu_{S}^2$ up to a given logarithmic order.
Any residual dependence on these resummation scales of the resummed cross section indicates terms beyond the logarithmic accuracy at which the resummation is performed. 
The hadronic cross section itself is independent of the common scale $\mu^2$.
In order to resum logarithms that become large as $\xi_i\to 1$ we make the following canonical scale choices.
\beq
\label{eq:scalechoice}
\hat \mu_{S}^2=Q^2(1-\xi_1)(1-\xi_2),\hspace{1cm}
\hat \mu_{C_1}^2=Q^2(1-\xi_2),\hspace{1cm}
\hat \mu_{C_2}^2=Q^2(1-\xi_1).
\eeq

To specify the quality of our resummation we count the number of exponentiated logarithms. 
Including the first $n$ logarithms gives then the (Next-To)$^n$-Leading-Logarithm (N$^n$LL).
To achieve this accuracy we must include anomalous dimensions and boundary terms of our factorization theorem (hard, soft and beam functions) up to a given order in the strong coupling constant. 
All functions and anomalous dimensions X have a perturbative expansion
\beq
X=\sum_{o=0}^n \left(\frac{\alpha_S}{\pi}\right)^o X^{(o)}.
\eeq
Table~\ref{tab:resprec} summarizes which is the highest order for each quantity that is included for N$^n$LL resummation.
We always perform the evolution of the strong coupling constant including the QCD through five loops~\cite{Baikov:2016tgj,Czakon:2004bu,Herzog:2017ohr,Larin:1993tp,vanRitbergen:1997va,Tarasov:1980au}.

\begin{table}[!h]
\begin{center}
 \begin{tabular}{l|c|c|c|c|c} \hline\hline
  Accuracy & $H$, $B^Y$, $S^Y$ &   $\gamma^{\text{non-cusp}}_{H,B,S}(\as)$ & PDF Evolution ($\Gamma_{ij}$)&  $\beta(\as)$ & $\GammaC(\as)$ \\\hline
  LL           & Tree level & --       & --       & $1$-loop & $1$-loop \\\hline
  NLL          & Tree level & $1$-loop & $1$-loop & $2$-loop & $2$-loop \\\hline
  NLL$^\prime$ & $1$-loop   & $1$-loop & $1$-loop & $2$-loop & $2$-loop \\\hline
  NNLL         & $1$-loop   & $2$-loop & $2$-loop & $3$-loop & $3$-loop \\\hline
  NNLL$^\prime$& $2$-loop   & $2$-loop & $2$-loop & $3$-loop & $3$-loop \\\hline
  N$^3$LL         & $2$-loop & $3$-loop & $3$-loop & $4$-loop & $4$-loop \\\hline
  N$^3$LL$^\prime$& $3$-loop & $3$-loop & $3$-loop & $4$-loop & $4$-loop \\\hline
  N$^4$LL         & $3$-loop & $4$-loop & $4$-loop & $5$-loop & $5$-loop \\\hline
  N$^4$LL$^\prime$& $4$-loop & $4$-loop & $4$-loop & $5$-loop & $5$-loop \\\hline
 \hline
 \end{tabular}
\caption{\label{tab:resprec} 
Summary of the perturbative order to which ingredients of the resummation formula are required to obtain certain degree of precision.
}
\end{center}
\end{table}

\subsection{Implementation}
To facilitate the computation of the resummed rapidity distribution of eq.~\eqref{eq:rapresummed} we want to perform part of the integrals analytically. 
To this end, we perform the Laplace convolutions by multiplying the soft and beam functions in Laplace space and then performing a double inverse Laplace transformation.
This approach leads to the following factor $\mathcal{F}$.
\bea
&&\mathcal{F}(\mu^2,\mu_{C_1}^2,\mu_{C_2}^2,\mu_{S}^2,\bar\xi_1,\bar \xi_2)\nonumber\\
&&=\mathcal{L}^{-1}\left[\mathcal{L}^{-1}\left[ \left(\frac{\kappa_2 \mu_{C_1}^2}{Q^2}\right)^{\eta_{C_1}}  \left(\frac{\kappa_1\kappa_2 \mu_S^2}{Q^2}\right)^{\eta_S} \left(\frac{\kappa_1 \mu_{C_2}^2}{Q^2}\right)^{\eta_{C_2}} ,\kappa_1,\bar \xi_1\right] ,\kappa_2,\bar \xi_2\right]\nonumber\\
&&=\frac{e^{\gamma_E(2\eta_S+\eta_{C_1}+\eta_{C_2})}}{\bar \xi_1\bar \xi_2 \Gamma(-\eta_S-\eta_{C_1})\Gamma(-\eta_S-\eta_{C_2})}\left(\frac{ \mu_{C_1}^2}{\bar \xi_2 Q^2}\right)^{\eta_{C_1}}  \left(\frac{ \mu_S^2}{\bar \xi_1\bar  \xi_2Q^2}\right)^{\eta_S} \left(\frac{ \mu_{C_2}^2}{\bar \xi_1 Q^2}\right)^{\eta_{C_2}}
\eea
The beam and soft functions act on this factors with derivatives with respect to the $\eta_i$ and get Mellin convoluted with the result of the derivatives. 
\bea
\label{eq:rapresummed2}
&&\frac{\df \sigma_{P\,P\rightarrow h+X}}{\df Y \df Q^2}(\xi_1,\xi_2,\mu^2, \mu_S^2,\mu_{C_1}^2,\mu_{C_2}^2)=\tau \sum_{i,j}  H_{ij}(\mu^2,\mu_H^2) 
\int_{\xi_1}^1\frac{\df x_1}{x_1} \int_{\xi_2}^1\frac{\df x_2}{x_2}
\nonumber\\
&&\times \left[  
 \tilde{B}^{Y}_i\left(\frac{\df }{\df \eta_{C_1}},\frac{\xi_1}{x_1},\mu_{C_1}^2\right) 
  \tilde{S}^{r_{ij}}\left(\frac{\df }{\df \eta_{S}},\mu_S^2\right)
\tilde{B}^{Y}_i\left(\frac{\df }{\df \eta_{C_2}},\frac{\xi_2}{x_2},\mu_{C_2}^2\right) 
\mathcal{F}(\mu^2,\mu_{C_1}^2,\mu_{C_2}^2,\mu_{S}^2,\bar x_1,\bar x_2) 
 \right]\nonumber\\
&&\times U(\mu^2,\mu_{C_1}^2,\mu_{C_2}^2,\mu_{S}^2,\mu_H^2).
\eea
The factor $U$ includes the individual exponentials relating the scales of the individual functions to the common perturbative scale.
\bea
U(\mu^2,\mu_{C_1}^2,\mu_{C_2}^2,\mu_{S}^2,\mu_H^2)&=&
e ^{A\left(\GammaCB,\mu^2,\mu_{C_1}^2\right)+\frac{1}{2}B\left(\gammaB(\alpha_S(\mu^2)),\mu^2,\mu_{C_1}^2\right)}\nonumber\\
&&\times e ^{A\left(\GammaCB,\mu^2,\mu_{C_2}^2\right)+\frac{1}{2}B\left(\gammaB,\mu^2,\mu_{C_2}^2\right)}\nonumber\\ 
&&\times e ^{A\left(\GammaCS,\mu^2,\mu_{S}^2\right)+\frac{1}{2}B\left(\gamma^{S,r},\mu^2,\mu_{S}^2\right)}\nonumber \\
&&\times e ^{A\left(-\GammaC,\mu^2,\mu_{H}^2\right)+\frac{1}{2}B\left(\gamma^{H,r},\mu^2,\mu_{H}^2\right)}.
\eea
The Mellin convolutions with $\mathcal {F}$ can be sharply peaked as the integration variables $x_i$ go to one.
To facilitate the numerical integration we perform the following replacement in our implementation.
\beq
\int_\xi^1\frac{\df x}{x} (1-x)^{-1-\eta} f\left(\frac{\xi}{x}\right)=\int_\xi^1\frac{\df x}{x} \left(\frac{\delta(1-x)}{-\eta}+\left[(1-x)^{-1-\eta}\right]_+\right) f\left(\frac{\xi}{x}\right).
\eeq
In the above equation the distribution act in the usual way. 
Furthermore, we truncate the product of soft and beam functions in eq.~\eqref{eq:rapresummed2} to maximally include 2n derivatives for N$^{n+1}$LL resummation.
We finally perform the remaining Mellin convolutions of PDFs and matching kernels and of the beam functions with the derivatives of $\mathcal{F}$ numerically using the CUBA library~\cite{Hahn:2004fe}.

\subsection{Numerical Results for Rapidity Beam Functions at NNLO}
To show numerical results for the rapidity beam functions we first define the cumulant of the beam function $C$ at fixed order and including resummation.
\bea
C_i(\xi_2,\mu^2;\bar \xi_1)&=&\int_{\xi_1}^1 dx_1 B_{i}(\xi_2,\mu^2;\bar x_1).\nonumber\\
C_i(\xi_2,\mu^2,\mu_{C_2};\bar \xi_1)&=&\int_{\xi_1}^1 dx_1 B_{i}(\xi_2,\mu^2,\mu_{C_2};\bar x_1).
\eea
We implement the $C_i$ in a numerical c++ code and evaluate them using PDF4LHC21 PDFs.

\begin{figure*}[!h]
\centering
\includegraphics[width=0.49\textwidth]{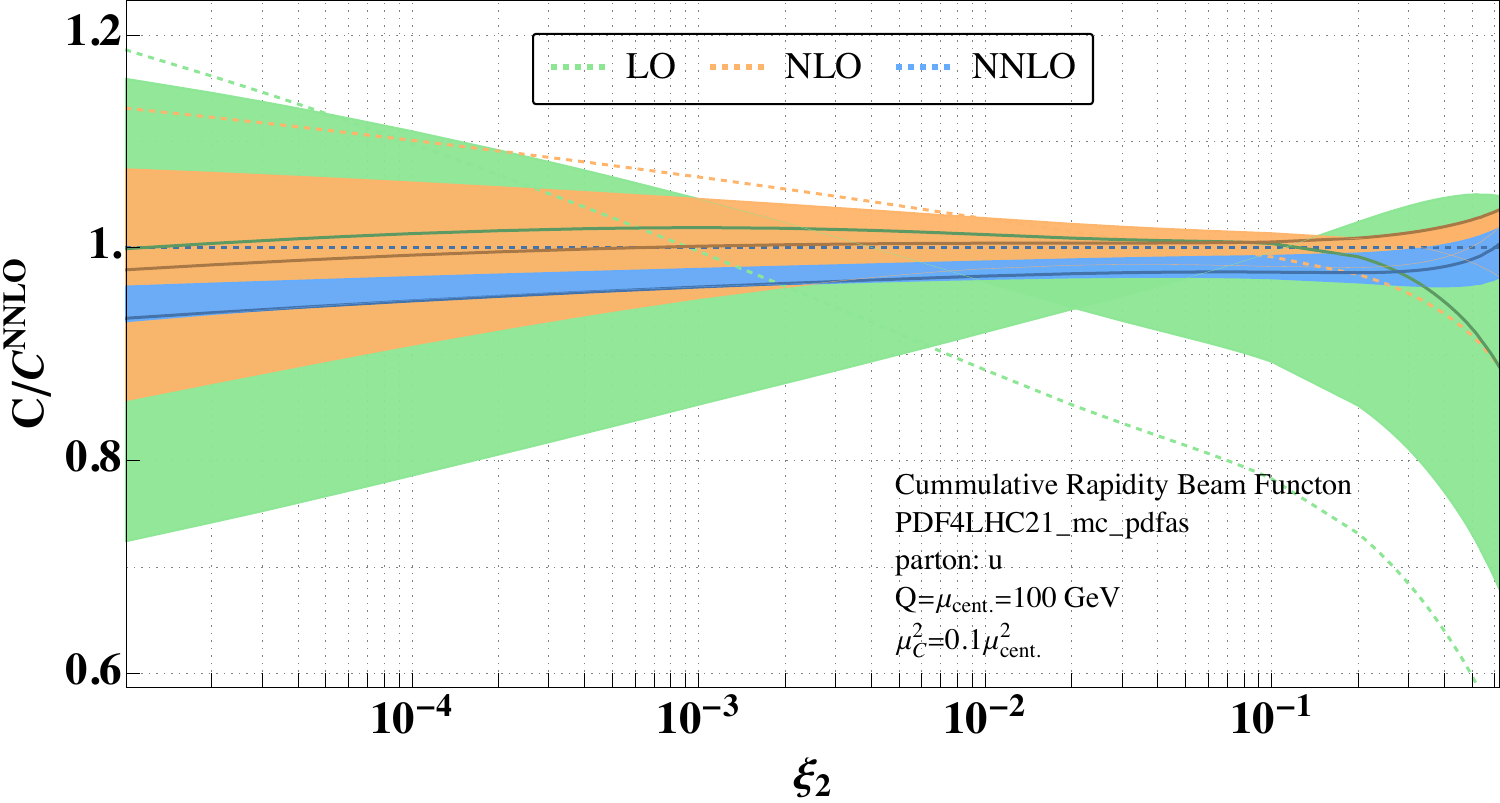}
\includegraphics[width=0.49\textwidth]{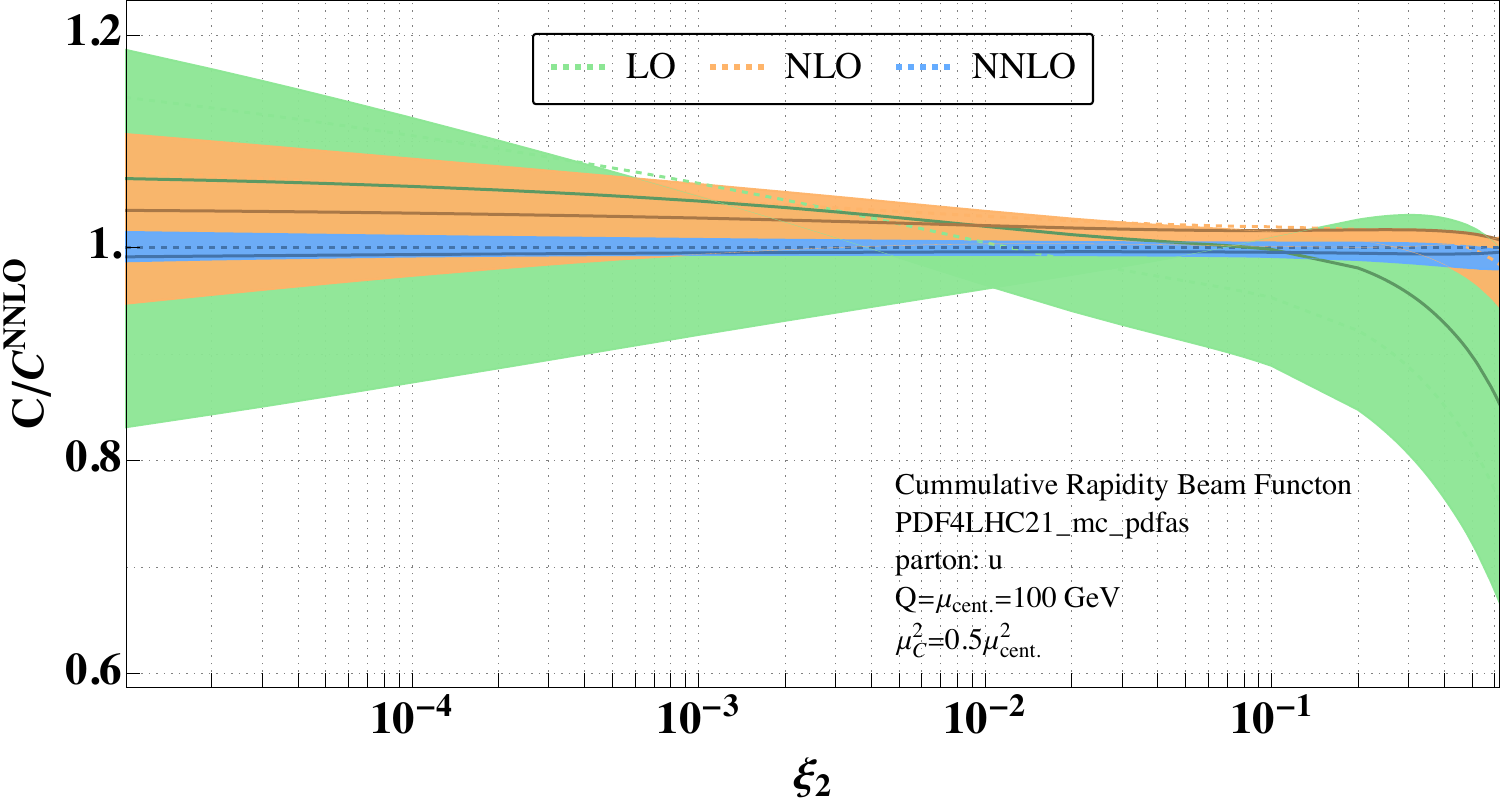}
\caption{\label{fig:uBF}
The figure shoes the cummulant of the beam function for the up quark at $Q=100$ GeV as a function of the the variable $\xi_2$ normalized to its fixed order prediction at NNLO.
The dashed lines represent fixed order results and the solid lines represent resummed results. 
N$^n$LO and N$^{n+1}$LL results for $n=0$, $n=1$ and $n=2$ are shown in green, orange and blue respectively. 
The left plot displays results for $\bar \xi_1=0.1$ and the right plot for $\bar \xi_1=0.5$.
}
\end{figure*}
\begin{figure*}[!h]
\centering
\includegraphics[width=0.49\textwidth]{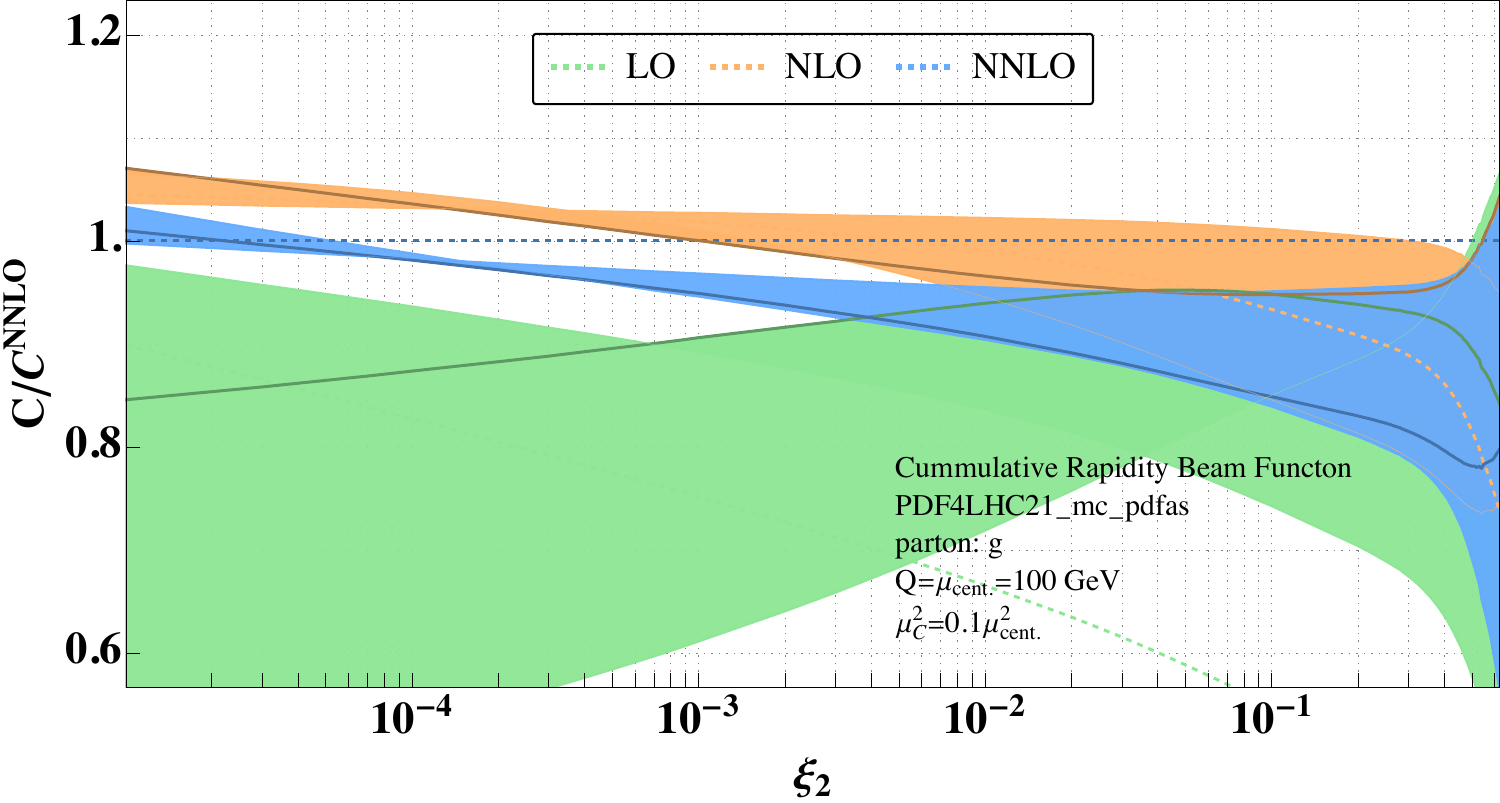}
\includegraphics[width=0.49\textwidth]{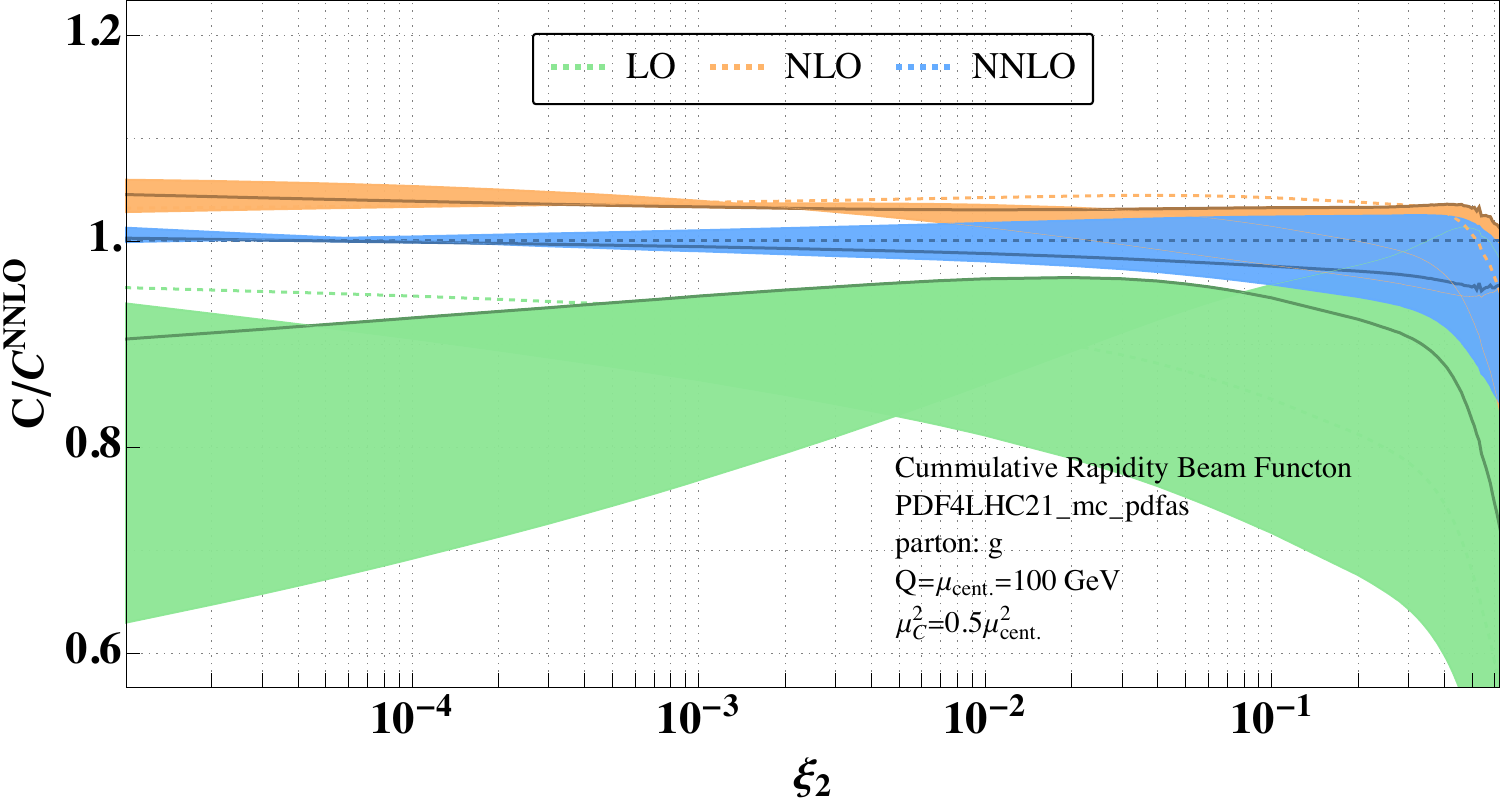}
\caption{\label{fig:gBF}
Same as fig.~\ref{fig:uBF} but for the gluon rapidity beam function.
}
\end{figure*}
Figures~\ref{fig:uBF} and ~\ref{fig:gBF}  show for example the rapidity beam function for the up quark and gluon respectively. 
We show results truncating the beam function matching kernel at N$^n$LO or N$^{n+1}$LL for n=0, n=1 and n=2 in green, orange and blue respectively. 
The dashed lines represent fixed order results and the solid lines resmmued results. 
The bands indicate the variation of the resummed results with the collinear scale $\mu_{C_2}^2$ by factors of two around our central choices of eq.~\eqref{eq:scalechoice}.
The left and right plots represent the two choices $\bar \xi_1=0.1$ and $\bar \xi_1=0.5$ respectively. 
We observe that for larger values of $\bar \xi_1$ the resummation has less of an effect and agrees very well with fixed order predictions. 
For small values of $\bar \xi_1$ the resummation leads to significant deviations from fixed order as expected. 
The deviation is frequently not covered by the collinear scale variation bands shown here but would mainly be covered by factorization scale variations at fixed order.
For the gluon beam function the variation of the collinear scale $\mu_{C_2}^2$ does not provide reliable uncertainty estimates for the truncation of the beam function matching kernel as bands do not overlap.

\subsection{Numerical Results for the Resummed Rapidity Distribution at N$^3$LL}
We implemented eq.~\eqref{eq:rapresummed2} in a numerical c++ code to perform the remaining Mellin convolutions.
In this section we present our findings for resummed predictions for the Higgs boson and Drell-Yan rapidity distribution.

\begin{figure*}[!h]
\centering
\includegraphics[width=0.49\textwidth]{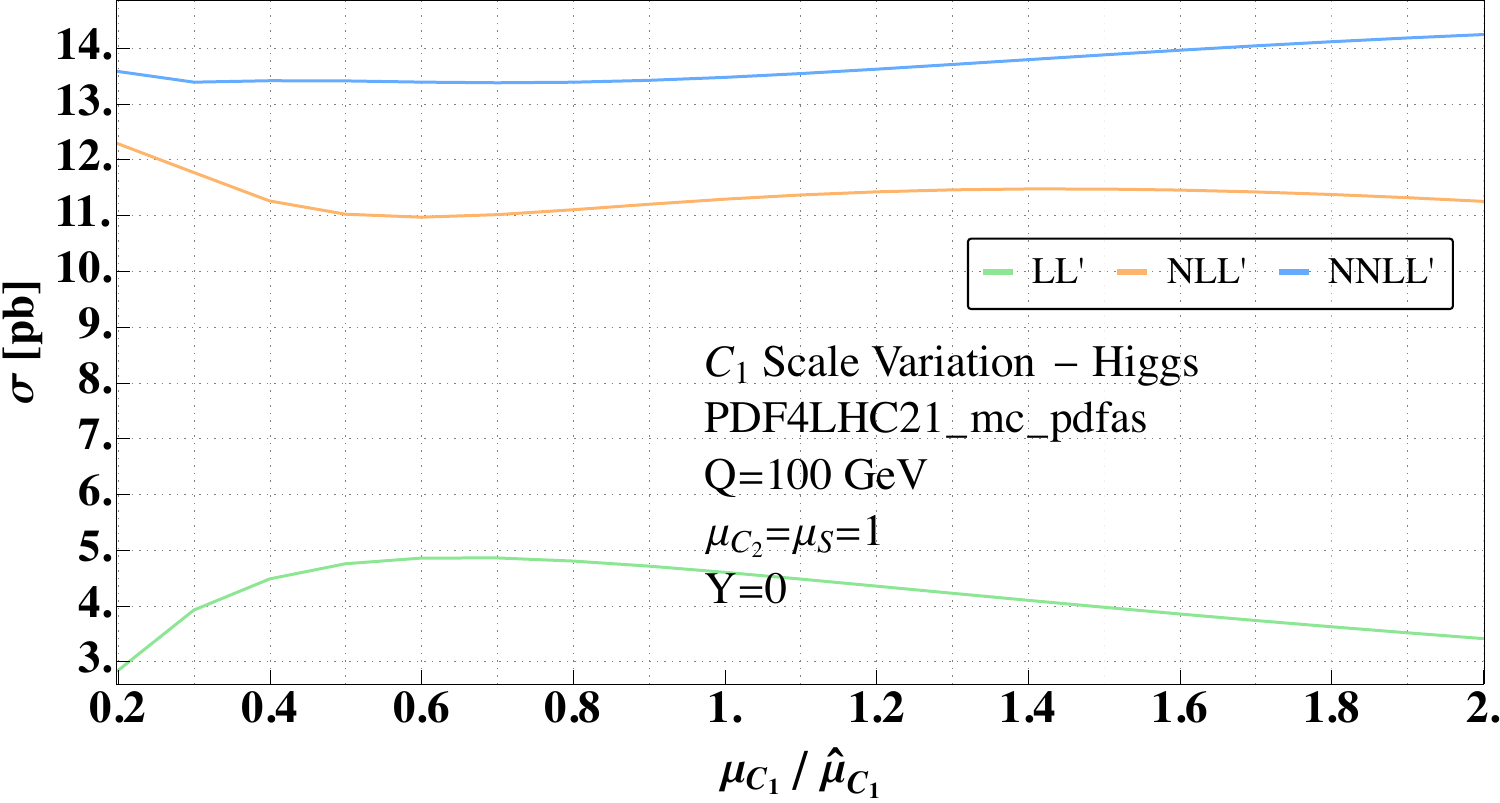}
\includegraphics[width=0.49\textwidth]{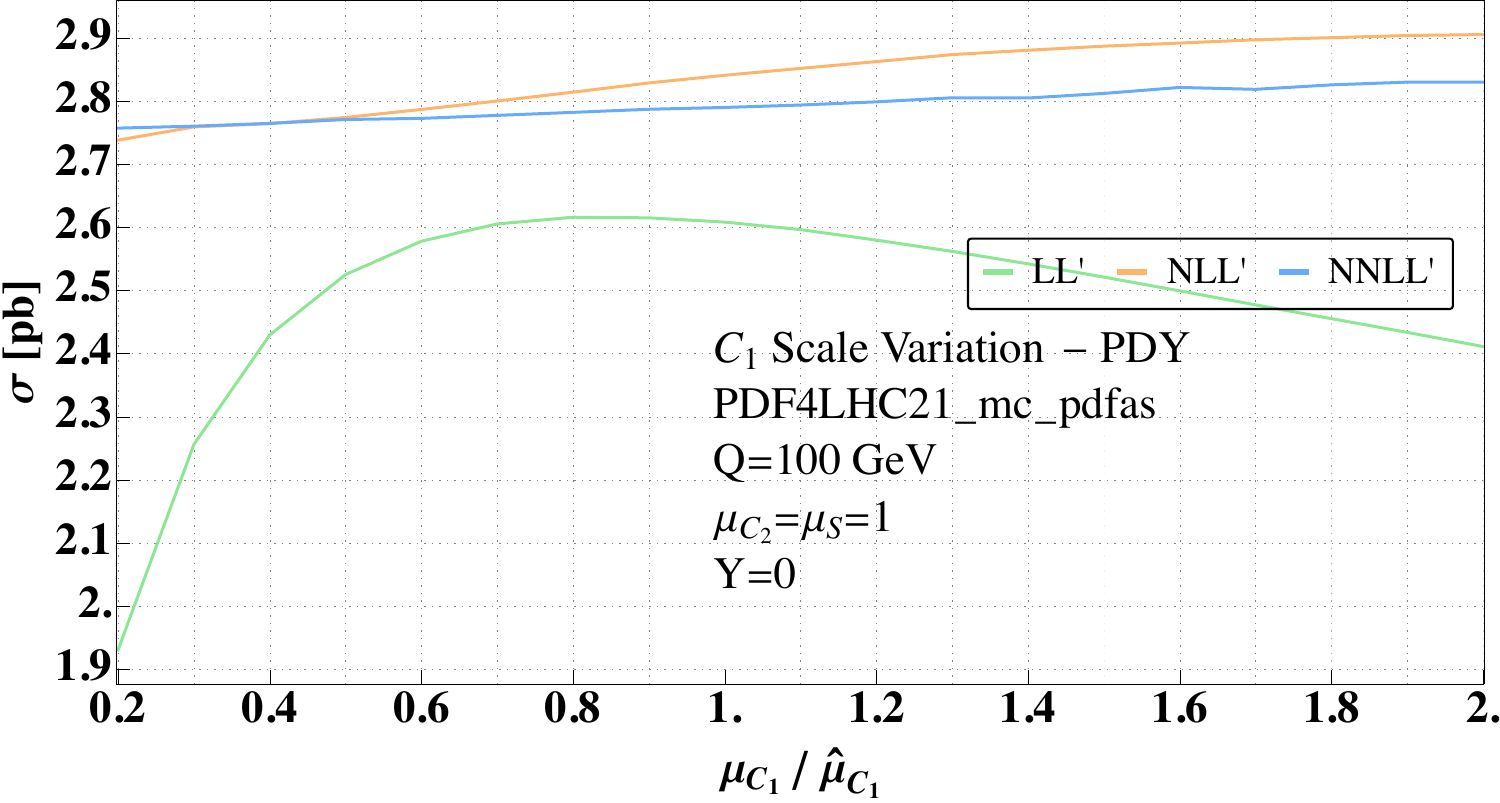}
\caption{\label{fig:C1Variation}
Predictions for the production cross section of a Higgs boson (left) and an $e^-e^+$ pair via the Drell-Yan cross section (right) at $Y=0$ and $Q^2=100$ GeV.
The plots show the variation of the cross section as a function of the variation of the scale $\mu_{C_1}$ at NLL (green), NNLL (orange) and N3LL (blue).
}
\end{figure*}
\begin{figure*}[!h]
\centering
\includegraphics[width=0.49\textwidth]{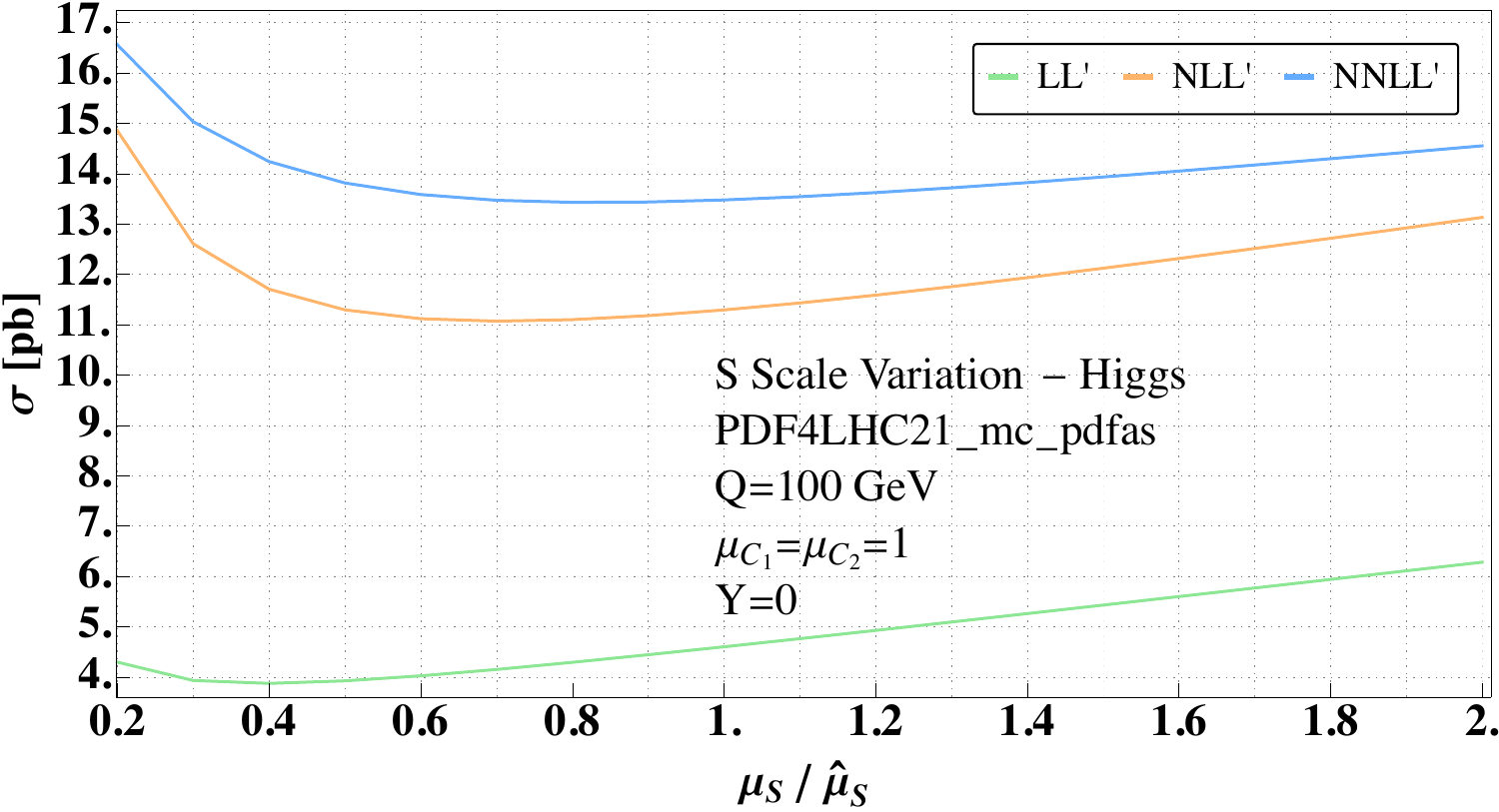}
\includegraphics[width=0.49\textwidth]{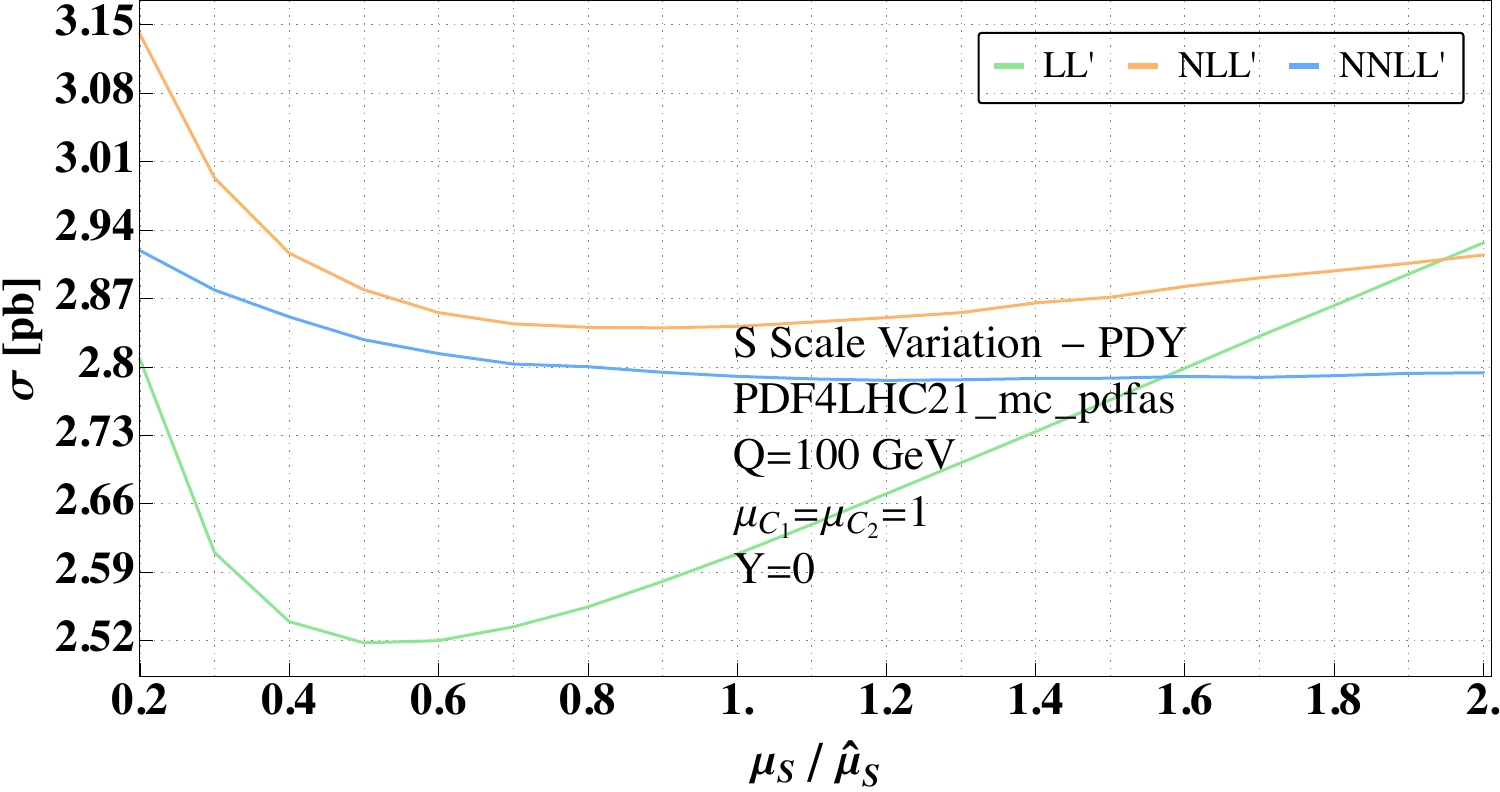}
\caption{\label{fig:SVariation}
Same as figure~\ref{fig:C1Variation} but as a function of the scale $\mu_S$.
}
\end{figure*}
Figures~\ref{fig:SVariation} and~\ref{fig:C1Variation} show the variation of the prediction for the Higgs boson and Drell-Yan cross section as a function of the scale $\mu_{C_1}$ and $\mu_S$ respectively.
In the figures all but this one scale are held fixed to their canonical value.
Different colors correspond to different logarithmic precision in the resummation formalism. 
As expected, achieving higher logarithmic accuracy in resummation reduces the dependence of on the individual starting scales. 
The variation of the scale $\mu_{C_2}$ is identical to the variation of $\mu_{C_1}$ due to the symmetry of the approach.

\begin{figure*}[!h]
\centering
\includegraphics[width=0.49\textwidth]{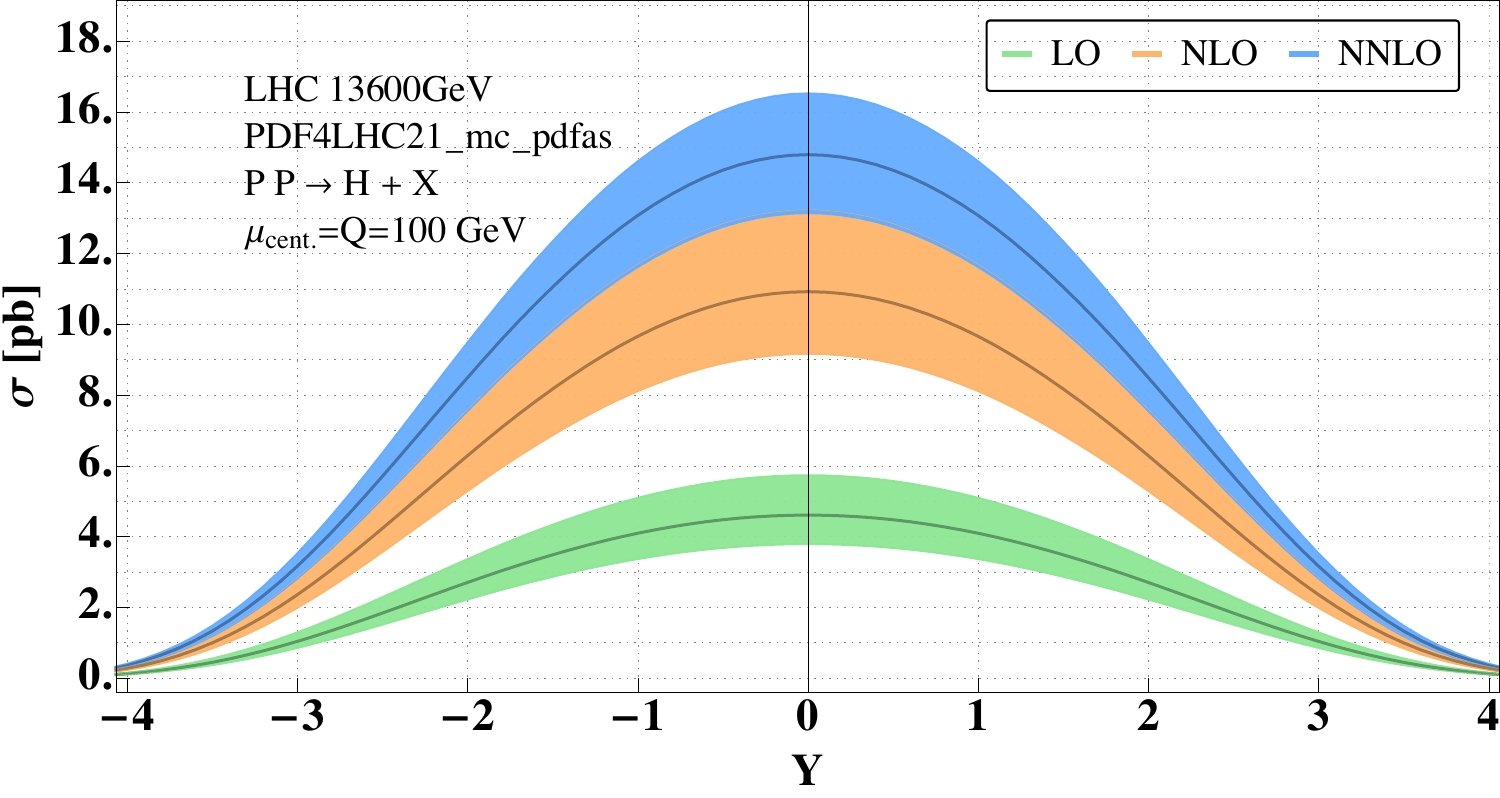}
\includegraphics[width=0.49\textwidth]{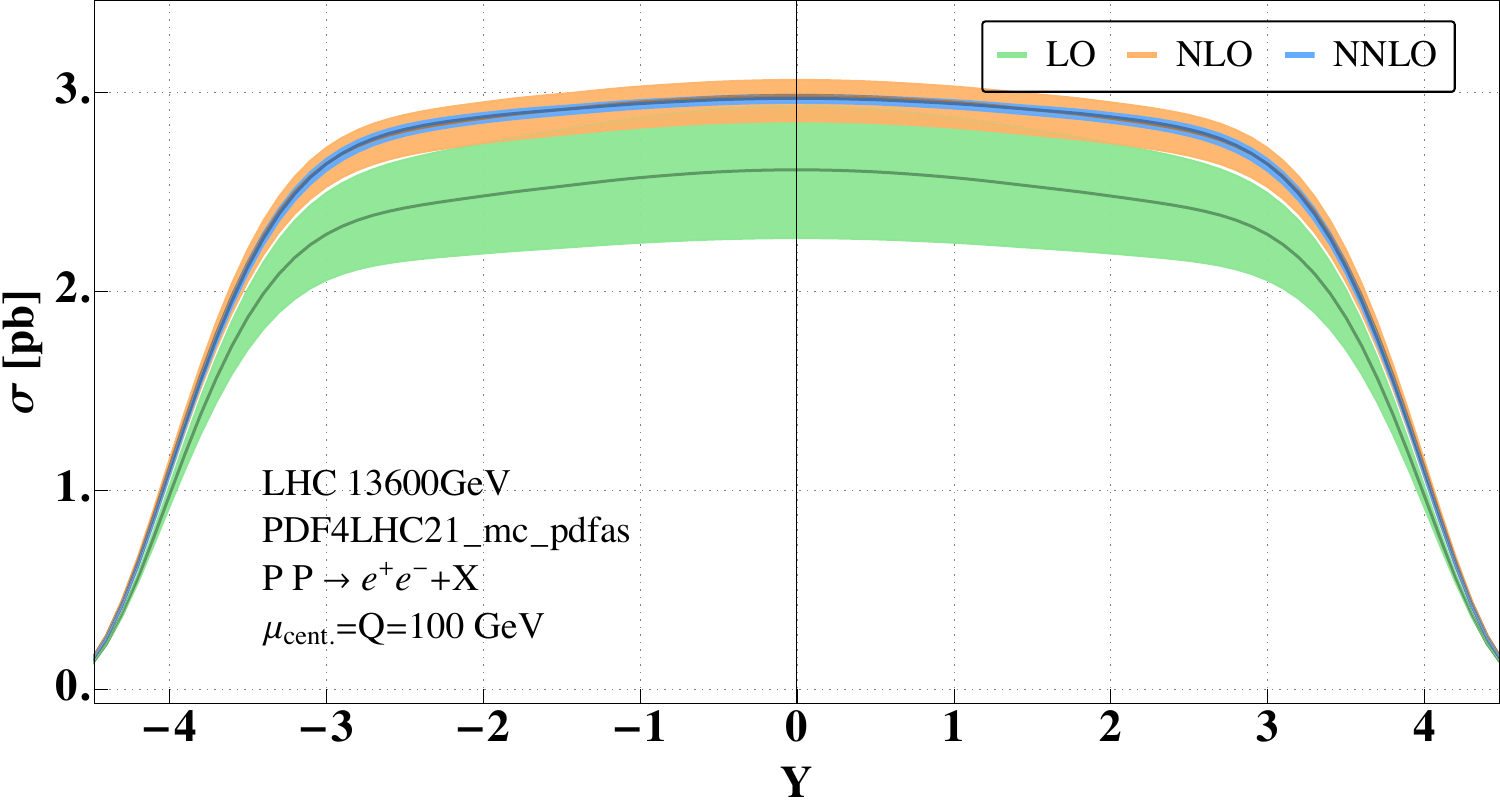}
\caption{\label{fig:RapRes}
Rapidity distribution for the production of a Higgs boson (left) or a electron-positron pair via the Drell-Yan cross section (right). 
Dashed lines represent fixed order predictions with a scale choice of $\mu_F=\mu_R=Q$ and solid lines represent resummed predictions with canonical choices for the resummation scales.
N$^n$LO and N$^n$LL$^\prime$ predictions for $n=0$, $n=1$ and $n=2$ are shown in green, orange and blue respectively. 
Bands are derived based on scale variations (see text).
}
\end{figure*}
Figure~\ref{fig:RapRes} shows the rapidity distribution for our two example processes based on our collinear approximation at fixed order (dashed lines) and with resummation (solid lines). 
We see that resummed predictions and fixed order predictions are almost identical for the Drell-Yan case. 
For Higgs boson production the resummed predictions tend to be larger than the fixed order predictions. 
The bands are derived based on a generalization of the scale variations used at fixed order we discussed in eq.~\eqref{eq:sevenpt}.
We vary the soft and collinear scales by factors of two around the canonical choice of eq.~\eqref{eq:scalechoice} and exclude points where any ratio of two scales is larger than two.
The scale bands at N$^3$LL are very asymmetric in the case of the Higgs boson production cross section.
Furthermore, the fixed order prediction at N$^2$LO is outside these bands. 
However, if we add perturbative uncertainties to the fixed order prediction the two prediction we see that resummation and fixed order are compatible within uncertainties.
Given the fact that the N$^3$LO predictions for Higgs boson production are known and are positive, one could interpret this as anticipating the N$^3$LO correction~\cite{Cieri:2018oms,Dulat:2018bfe}.

Finally, we want to combine resummed predictions with fixed order predictions.
To achieve this we define the following matched cross section.
\beq
\label{eq:matched}
Q^2 \frac{\df \sigma^{\text{N$^{n+1}$LL+N$^m$LO}}_{P\,P\rightarrow h+X}}{\df Y \df Q^2}=Q^2 \frac{\df \sigma^{\text{N$^{n+1}$LL}}_{P\,P\rightarrow h+X}}{\df Y \df Q^2}-Q^2 \frac{\df \sigma^{\text{N$^m$LO approx.}}_{P\,P\rightarrow h+X}}{\df Y \df Q^2}+Q^2 \frac{\df \sigma^{\text{N$^m$LO}}_{P\,P\rightarrow h+X}}{\df Y \df Q^2}.
\eeq
The above formula ensures that we do not double count any terms that are known without approximation and are accounted for in the exact fixed order cross section.
The first term on the right hand side includes the resummed contributions computed using eq.~\eqref{eq:rapresummed2}. 
The second term corresponds to the rapidity distribution computed with the approximate partonic coefficient function of eq.~\eqref{eq:etaaprox}.
The third term is simply the N$^m$LO fixed order prediction for the Higgs boson rapidity distribution.
We assume that $n\geq m$.

\begin{figure*}[!h]
\centering
\includegraphics[width=0.49\textwidth]{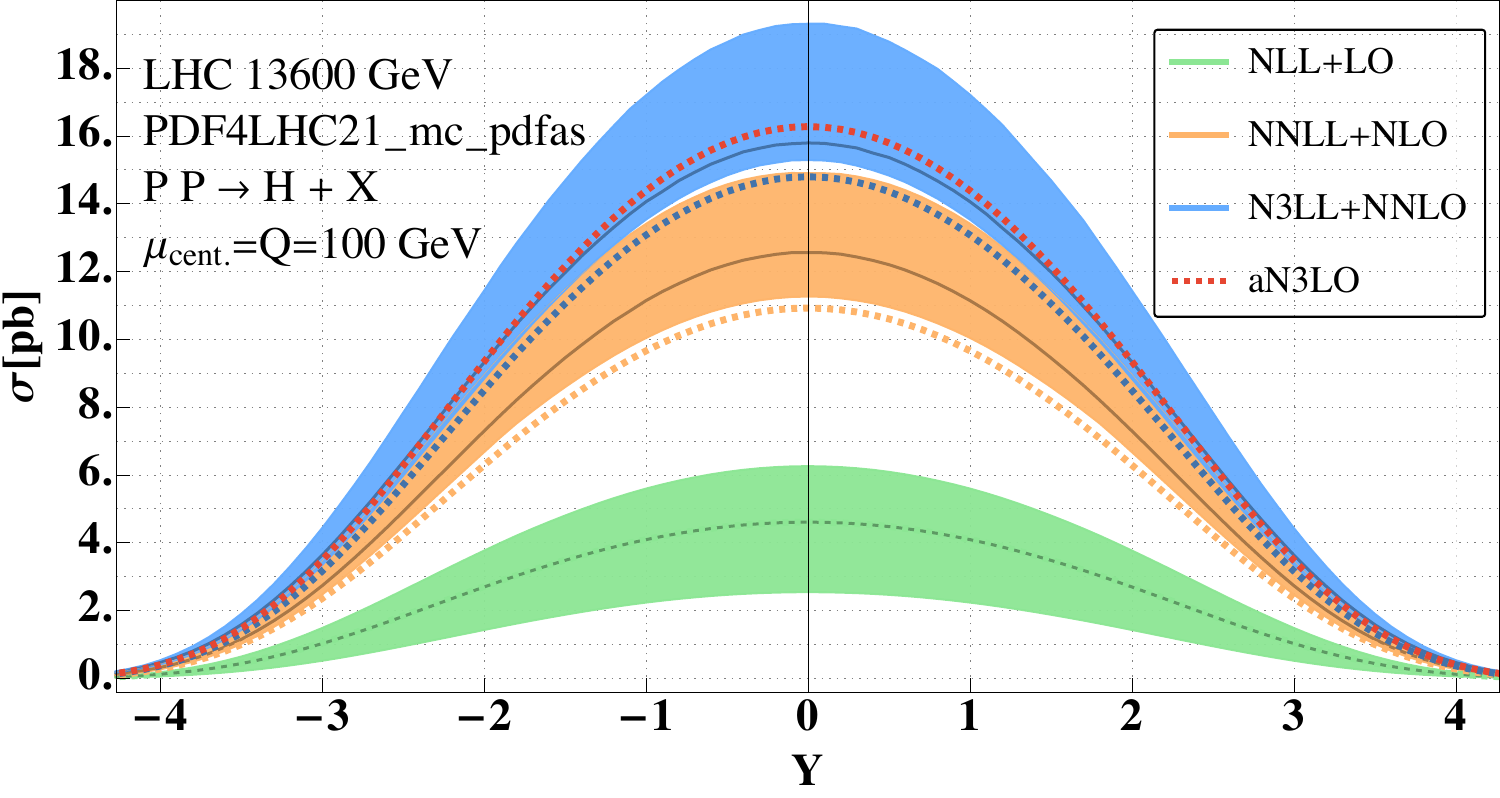}
\includegraphics[width=0.49\textwidth]{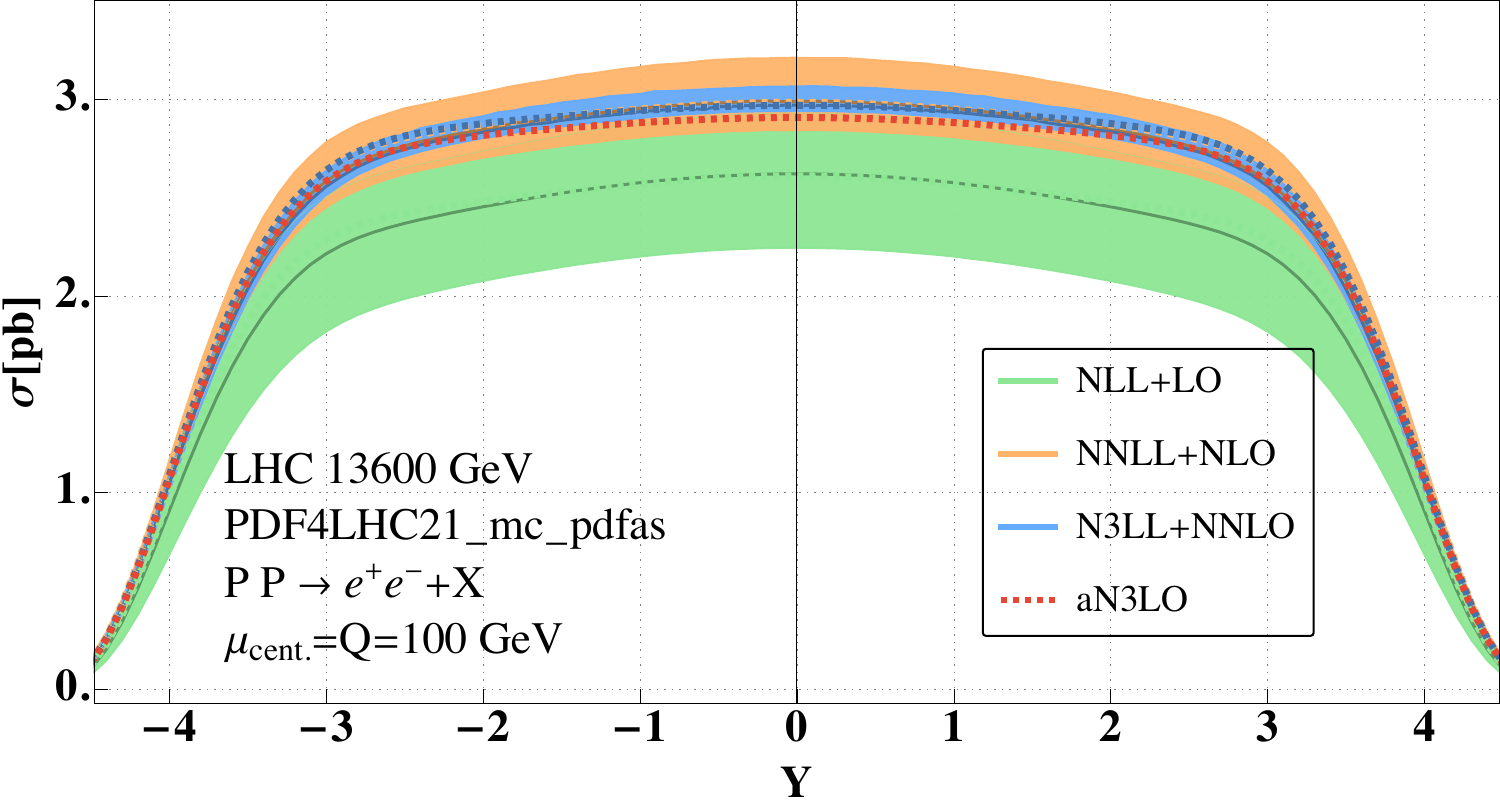}
\caption{\label{fig:RapMatched}
Rapidity distribution (eq.~\eqref{eq:matched})  for the production of a Higgs boson (left) or a electron-positron pair via the Drell-Yan cross section (right). 
Dashed lines represent fixed order predictions with a scale choice of $\mu_F=\mu_R=Q$ and solid lines represent resummed predictions matched to fixed order with canonical choices for the resummation scales.
N$^n$LL$^\prime$ + N$^n$LO predictions for $n=0$, $n=1$ and $n=2$ are shown in green, orange and blue respectively. 
Bands are derived based on scale variations (see text).
}
\end{figure*}
Figure~\ref{fig:RapMatched} shows the rapidity distribution matched to fixed order predictions following eq.~\eqref{eq:matched} for our two example processes. 
Dashed lines correspond to fixed order only predictions and solid lines correspond to the matched predictions. 
The bands are derived as above using scale variations. 
The resummation scale variations are added linearly to the fixed order scale variations to derive the combined bands.
We can see that resummation has very little effect on the Drell-Yan cross section while resummation corrections to Higgs boson production are sizable and positive. 
However, as for fig.~\ref{fig:RapRes}, the matched and fixed order predictions are comparable within uncertainties. 

In addition, we show in fig.~\ref{fig:RapMatched} an approximation to the N$^3$LO rapidity distribution which is simply given by multiplying the NNLO rapidity distribution with the ratio of the N$^3$LO to the NNLO fixed order inclusive cross section~\cite{Anastasiou:2015ema,Mistlberger:2018etf,Duhr:2020seh} (i.e. the integral over the rapidity distribution).
This approximation was empirically shown to work extremely well~\cite{Cieri:2018oms,Dulat:2018bfe,Chen:2021vtu}.
We observe that for Higgs boson production the resummed N$^3$LL +N$^2$LO prediction is very close to the approximate N$^3$LO prediction. 
On the other hand for Drell-Yan production the impact of resummation is negligible.

Our resummation framework overall resums logarithms $\log(1-\xi_i)$. 
As these logarithms don't represent a dominant contribution to the cross section in the bulk of the rapidity distribution, it is understandable that no significant improvement of including resummation is present.
Event at very forward rapidity of $Y=4.5$ for the production of a Higgs boson with $m_H=125$ GeV we find $\log(1-\xi)=\sim -1.7$, which is simply not very large and doesn't lead to substantial enhancements.
In fact, fixed order and matched predictions are compatible within their uncertainties. 
The difference of fixed order predictions and resummed predictions in fig.~\ref{fig:RapRes} is by and large due to the fact that the hard function in eq.~\eqref{eq:rapresummed2} was factored out and enters as a multiplicative prefactor. 
In contrast, in fixed order predictions the hard functions is multiplied with the remaining partonic coefficient function and the product is truncated at a specific fixed order.
The resummation itself contributes understandably very little.
Rather than an improvement on fixed order predictions, resummed predictions can here be used to derive an alternative representation of the perturbative expansion and therefor check if perturbative uncertainty estimates are reliable.
This check is clearly not exhaustive as the correction from NNLO to N$^3$LO predictions for DY is not covered by either fixed order or resummation scale uncertainties.
This is was already studied at hands of the inclusive cross section, see refs.~\cite{Duhr:2020sdp,Duhr:2020seh,Duhr:2021vwj}.
As in the case of threshold resummation, different choices of setting the resummation scales can be considered (see for example ref.~\cite{Beneke:2019mua}) and we leave this for future research.

\begin{table}[!h]
\begin{center}
\begin{tabular}{|c || c | c || c | c |}
\hline
 & $\sigma_{P\, P\to H+X}^{N^{n+1}LL+N^nLO}$ [pb]& $\sigma_{P\, P\to H+X}^{N^nLO}$   [pb] &  $\sigma_{P\, P\to e^+e^-+X}^{N^{n+1}LL+N^nLO}$   [pb] & $\sigma_{P\, P\to e^+e^-+X}^{N^nLO}$   [pb] \\
\hline
\hline
 n=0 &  $13.7^{+38.7 \% }_{-45.2\%}$ & $13.8 ^{+23.5\%}_{-17.3\%} $ & $18.9^{+12.7\%}_{-16.1\%}$ & $ 19.1 ^{+10.2\%}_{-11.2\%} $ \\
\hline
 n=1 & $36.8^{+15.3\% }_{-6.7\%} $ &  $31.7 ^{+19.8\%}_{-15.3\%} $ &  $22.5^{6.2\%}_{-4.2\%}$ & $ 22.6 ^{+2.8\%}_{-4.3\%} $ \\
\hline
 n=2 &  $48.2^{+11.8\% }_{-8.64\%}$ & $42.5 ^{+10.2\%}_{-10.2\%} $ &  $21.8^{+3.8\%}_{-1.8\%}$ & $ 21.97 ^{+0.8\%}_{-1.1\%} $ \\
\hline
 n=3 & -  & $46.6 ^{+3.4\%}_{-5.1\%} $  & -  & $ 21.51 ^{+0.7\%}_{-0.8\%} $\\
\hline
\end{tabular}
\caption{\label{tab:incpred} 
Predictions for the inclusive gluon fusion Higgs boson production cross section and Drell-Yan production cross section. 
Both, fixed order results through N$^n$LO and resummed results through N$^{n+1}LL$ match to N$^n$LO, are presented.
}
\end{center}
\end{table}
Table~\ref{tab:incpred} shows the inclusive cross section obtained with rapidity resummation at N$^{n+1}LL$ matched to N$^n$LO for the inclusive gluon-fusion Higgs boson and Drell-Yan $e^+e^-$ production via photon exchange. 
The inclusive cross section is obtained simply by integrating over the corresponding rapidity distribution. 
Here, we evaluate the Higgs boson cross section at the physical point $Q=m_H=125.09$ GeV.
We evaluate the fixed order predictions for both processes with a central scale choice of $\hat \mu_F=\hat \mu_R= Q$.
Tab.~\ref{tab:incpred} furthermore compares the resummed predictions with pure fixed order predictions.
Similar to the differential case, we find that resummed and fixed order predictions are compatible within uncertainties.

%% file: Chapters/Conclusions.tex
\section{Conclusions}
\label{sec:conclusions}
In this article, we studied a factorization formula for production cross sections of colorless final states in LHC collisions. 
The cross sections are differential in the rapidity and invariant mass of the final state. 
The factorization formula describes the limit of the hadronic cross section where all radiation produced on top of the colorless final state is collinear to either of the scattering protons or very low energetic. 
We refer to the factorization approach, which is made concrete by eq.~\eqref{eq:raplimitdef}, as factorization of the collinear approximation. 
Our work here builds on and extends the "generalized threshold" studied in ref.~\cite{Lustermans:2019cau}, which already realizes factorization in this limit and the possibility to approximate cross sections.
We state explicitly the definition of all perturbative ingredients necessary for the factorization formula.
We recompute explicitly all quantities through NNLO in QCD perturbation theory, find agreement with the literature and attach them in electronically readable form together with arXiv submission of this article. 

We argue that inclusive cross sections and rapidity distributions can be approximated using the factorization of the collinear approximation.
We claim that our formula should provide good approximation for LHC processes due to kinematic enhancement of the partonic coefficient function and rapidly falling PDFs away from the considered limit, in particular for gluon-initiated processes. 
To test these statements we study the gluon-fusion Higgs boson production and Drell-Yan $e^+e^-$ production cross section at the LHC through NNLO in QCD perturbation theory.
First, we find that analytic partonic coefficient functions obtained by the factorization formula contain a remarkable amount of information of their exact counterpart. 
For example, the leading terms in a threshold expansion of {\it every} partonic initial state is reproduced correctly by our factorization formula through NNLO.

Numerically, we show that NNLO cross sections obtained with this factorization formula are within five percent of the exact NNLO prediction for our two example processes at typical virtuality for LHC processes. 
To arrive at this conclusion we specifically investigated the size of power corrections to our example processes.
Furthermore, we demonstrate that our approximation performs outstandingly for very large absolute values of rapidity and high values of invariant mass. 
Consequently, we conclude that depending on how well a processes is known, the factorization formula of the collinear limit may provide an improved description over the current state of the art. 
Our factorization formula contains all information of the widely used threshold factorization formula for inclusive processes and includes in addition all information about collinear limits of hadronic cross sections.
Consequently, this factorization formula should be used for any and all applications of threshold factorization.

Finally, our factorization formula provides the opportunity to resum logarithms that become large when all radiation becomes collinear to one of the colliding protons (or very low energetic). 
We explicitly derive resummed predictions in this approximation through N$^3$LL accuracy. 
We find that the impact of resummation on fixed order cross sections is relatively mild and compatible with fixed order predictions. 
This is to be expected as no hadronic kinematic enhancement is present for typical, central LHC cross sections.

The factorization formula for the collinear approximation represents an exciting new avenue to explore the universal structure of LHC cross sections and can be used to improve our description of certain LHC processes beyond the state of the art. 
The analytic structures we observe in partonic cross sections obtained with our approximation promise to shed new light on factorization beyond the leading power.

%% file: Chapters/Transforms.tex
\section{Convolutions and Integral Transformations}
\label{app:trafos}

In this article, we use three types of convolutions.
We indicate a Laplace convolution over the variable x using the symbol $\circ_x$ and its definition is then given by the following equation.
\beq
f(x)\circ_x g(x)=\int_{x}^1 \df z f(z) g(1+x-z).
\eeq
We indicate a Mellin convolution over the variable x by the symbol $\otimes_x$ and use the following definition.
\beq
f(x)\otimes_x g(x)=\int_{x}^1 \frac{\df z}{z} f(z) g\left(\frac{x}{z}\right).
\eeq
The Laplace convolution yields the limit of $x\to 1$ of the Mellin convolution.
\beq
\lim_{x\to 1}f(x)\otimes_x g(x)=f(x)\circ_x g(x).
\eeq
Furthermore, we introduce a double convolution over two variables in eq.~\eqref{eq:dconvdef}, which is a combination of a Mellin and a Laplace convolution and indicate it by the symbol $\otimes_{x_1;x_2}$.
\beq
f(x_1,x_2)\otimes_{x_1;x_2} g(x_1,x_2)=\int_{x_1}^1 \frac{\df y_1}{y_1} \int_{x_2}^1 \df y_2 f(y_1,y_2) g\left(\frac{x_1}{y_1},1+x_2-y_2\right).
\eeq
All our convolutions are commutative. 

Integral transformations play an important role to simplify convolutions by transforming the convolution in $x$-space to a product of associated functions in a conjugate space $\kappa$-space. 
The Laplace transform is defined as
\beq
\label{eq:LTrafo}
\tilde f(\kappa e^{-\gamma_E}) =\mathcal{L}\left[f(x),x,\kappa\right]=\int_0^\infty \df x e^{-\kappa x e^{-\gamma_E}} f(x).
\eeq
In particular we will make use of the transformation
\beq
\mathcal{L}\left[x^{-1+a \epsilon},x,\kappa\right]= e^{\gamma_E a \epsilon} \kappa^{-a \epsilon} \Gamma(a \epsilon).
\eeq
The inverse is therefore given by
\beq
\mathcal{L}^{-1}\left[\kappa^{-a\epsilon}\right]=\frac{e^{-\gamma_E a\epsilon}}{\Gamma(a \epsilon)} x^{-1+a \epsilon}
\eeq
The Laplace transform of a Laplace convolution is a product of the convoluted functions.
\beq
\mathcal{L}\left[f(z)\circ_z g(z),z,\kappa\right]=f(\kappa e^{-\gamma_E} ) g(\kappa e^{-\gamma_E}).
\eeq

%% file: Chapters/OpDefandMellin.tex
\section{Results in Mellin Space}
A significant portion of the literature on threshold resummation is formulated in Mellin space, see for example \cite{Sterman:1986aj,Catani:1989ne,Catani:1990rp,Catani:1996yz,Catani:1996vz,Catani:1998tm,Catani:2003zt,Magnea:1990zb,Kidonakis:1997gm,Bonvini:2012an,Bonvini:2014joa,Bonvini:2014qga}. 
Since this work shows a generalization of threshold resummation that includes collinear dynamics beyond DGLAP, we think it is interesting to have a brief presentation of our results in Mellin space. We do this for illustration purposes as substituting the Laplace convolutions with Mellin ones introduce subleading terms that may change the phenomenological accuracy of our approximation discussed in \sec{fo}, as well as introducing complications related to performing the inverse Mellin transform over functions such as the rapidity beam functions or matching kernels that beyond one loop order involve generalized polylogarithms and worse. 

We take the factorization formula of \eq{raplimitdef} and we promote the Laplace convolutions to Mellin convolutions. This introduces subleading power terms that go beyond the accuracy at which we are working, see \app{trafos}. Given the structure of \eq{raplimdef} it is clear that a single Mellin variable will not suffice and we will need a double Mellin transform to factorize the convolutions, in a similar way in which we obtained both $\kappa_1$ and $\kappa_2$ when dealing with the Laplace transforms, see \sec{resummation}. 
Similarly, we promote the RHS of the RGEs for the Rapidity Beam and Soft functions defined in \secs{resSoft}{resBeam} to Mellin convolutions, such that they factorize in double Mellin space
\begin{align}\label{eq:MellinRGEs}
	\mu \frac{\df}{\df \mu} B^Y\Bigl(\ln \frac{N_a \mu^2}{Q^2},N_b,\mu\Bigr) &= \gamma_B\Bigl(\ln \frac{N_a \mu^2}{Q^2}, \mu \Bigr) B^Y\Bigl(\ln \frac{N_a \mu^2}{Q^2},N_b,\mu\Bigr)\,, \nn \\
	\mu \frac{\df}{\df \mu} S\Bigl(\ln \frac{N_a N_b \mu^2}{Q^2},\mu\Bigr) &= \gamma_S\Bigl(\ln \frac{N_a N_b \mu^2}{Q^2}, \mu \Bigr) S\Bigl(\ln \frac{N_a N_b \mu^2}{Q^2},\mu\Bigr) \,.
\end{align} 
Given the multiplicative structure of the RGEs in \eq{MellinRGEs}, the solution is trivial
\begin{align}\label{eq:RGE_sol_Mellin}
	B^Y\Bigl(\ln \frac{N_1 \mu^2}{Q^2},N_2,\mu\Bigr) &= B^Y_i\Bigl(\ln \frac{N_1 \mu_{B_n}^2 }{Q^2},N_2,\mu_{B_n}\Bigr) 
	\\ &\quad \times
	\exp\left\{\int_{\mu_{B_n}}^{\mu} \frac{\df \mu^\prime}{\mu^\prime} \left[2 \GammaC[\as(\muprime)] \ln \left( \frac{N_1 \mu^{\prime \, 2}}{Q^2} \right) + \gamma_B[\alpha_s(\muprime)] \right] \right\} \nn\,, \\ 
	S\Bigl(\ln \frac{N_1 N_2 \mu^2}{Q^2},\mu\Bigr) &= S\Bigl(\ln \frac{ N_1N_2 \mu_{S}^2}{Q^2},\mu_{S}\Bigr)
	\nn\\ &\quad \times
	\exp\left\{\int_{\mu_{S}}^{\mu} \frac{\df \mu^\prime}{\mu^\prime} \left[ -2 \GammaC[\as(\muprime)] \ln \left( \frac{N_1 N_2 \mu^{\prime \, 2}}{Q^2} \right) + \gamma_S[\alpha_s(\muprime)] \right] \right\} \nn\,,
\end{align}
where, with a slight abuse of notation, we distinguished the full anomalous dimension $\gamma_{B,S}\bigl( \ln \dot \,, \mu \bigr)$ from its non-cusp piece $\gamma_{B,S}[\alpha(\mu)]$ by the number of arguments as often done in SCET literature. 
From \eq{RGE_sol_Mellin} is clear the separation between the boundary functions $B^Y,S$ and the evolution factors that encode the RG evolution to the common scale $\mu$. 
The running of the Hard function is the same as in  \eq{HardRGE_sol}, therefore we can use \eq{HardRGE_sol} for its resummation. 
We are now ready to obtain the expression for the resummed cross section in double Mellin space, which reads
\begin{align}\label{eq:resMellin_general}
	\frac{\df \sigma^\text{res}}{\df N_1 \df N_2} &= \sum_{i,j} H^\text{res}_{ij}(Q^2,\mu) B^\text{res}_i(N_1,N_2,\mu) S^\text{res}(N_1 N_2, \mu) B^\text{res}_j(N_2,N_1,\mu) + \cO\left(\frac{1}{N_1},\frac{1}{N_2}\right) \nn \\
	&=\sum_{i,j}H_{i,j}\Bigl(\ln \frac{\mu_H^2}{Q^2},\mu_H\Bigr) \exp\left\{\int_{\mu_H}^{\mu} \frac{\df \mu^\prime}{\mu^\prime}\left[ -4 \GammaC[\as(\muprime)] \ln \left( \frac{\muprime}{Q} \right) + \gamma_H[\alpha_s(\muprime)] \right]\right\}  \nn\\
	&\quad\times B^Y_i\Bigl(\ln \frac{N_1 \mu_{B_n}^2 }{Q^2},N_2,\mu_{B_n}\Bigr) \exp\left\{\int_{\mu_{B_n}}^{\mu} \frac{\df \mu^\prime}{\mu^\prime} \left[2 \GammaC[\as(\muprime)] \ln \left( \frac{N_1 \mu^{\prime \, 2}}{Q^2} \right) + \gamma_B[\alpha_s(\muprime)] \right] \right\} \nn \\
	&\quad\times S\Bigl(\ln \frac{ N_1N_2 \mu_{S}^2}{Q^2},\mu_{S}\Bigr) \exp\left\{\int_{\mu_{S}}^{\mu} \frac{\df \mu^\prime}{\mu^\prime} \left[ -2 \GammaC[\as(\muprime)] \ln \left( \frac{N_1 N_2 \mu^{\prime \, 2}}{Q^2} \right) + \gamma_S[\alpha_s(\muprime)] \right] \right\} \nn \\
	&\quad\times B_j^Y\Bigl(\ln \frac{N_2\mu_{B_\bn}^2}{ Q^2},N_1,\mu_{B_\bn}\Bigr) \exp\left\{\int_{\mu_{B_\bn}}^{\mu} \frac{\df \mu^\prime}{\mu^\prime}\left[ 2 \GammaC[\as(\muprime)] \ln \left( \frac{ N_2 \mu^{\prime \, 2}}{Q^2} \right) + \gamma_B[\alpha_s(\muprime)]  \right]\right\} 
\end{align}
This is the general form for the resummed cross section and it is valid at all logarithmic orders (assuming the absence of factorization breaking effects due to Glauber interactions).
Note also that in each beam function there is one Mellin variables whose functional dependence is simple, but the dependence on the other one is in general much more complicated. This is not a flaw of this framework, but it is a general feature of the rapidity distribution. Clearly, the level of complexity depends on the perturbative order. At $\cO(\alpha_s)$, which is the order needed for $B^Y$ for NNLL resummation, the expression are simple Mellin transform of dilogarithms, but starting from $\cO(\as^2)$ (which is needed for NNLL$^\prime$ and N$^3$LL resummation), $B^Y$ would include complicated functions coming, for example, from the Mellin transforms of weight 3 GPLs.
In \eq{resMellin_general} the dependence on $\mu$ cancels exactly, however there is a freedom in the choice of the boundary scales $\{\mu_H,\mu_{B_n},\mu_{B_\bn},\mu_S\}$ which is where the boundary conditions of the RGEs are chosen to be evaluated and one may evaluate resummation uncertainties by performing variation of these scales around a central value%
\footnote{
Note that in general, an explicit choice of scale in conjugate space before inverting to the physical space may lead to issues related to integration over the Landau pole during the inversion. 
While several prescriptions exist to address this issue, one can also treat the scales as $N$-independent quantities when inverting and set them only later in the physical space. For a thorough study of the impact of these choices within both direct and EFT methods of QCD resummation see \refcite{Almeida:2014uva}.
}%
. 
For each of these scales there is clearly a particular choice, which is the one that minimizes the logarithmic terms contained in the boundary functions. We refer to this choice as \emph{canonical scale setting} and in this case it corresponds to
\beq
 \mu^{*\,2}_H = Q^2\,,\qquad\mu^{*\,2}_{B_n} = \frac{Q^2}{N_1} \,,\qquad \mu^{*\,2}_{B_\bn} =  \frac{Q^2}{N_2}\,,\qquad\mu^{*\,2}_S = \frac{Q^2}{N_1 N_2}\,.
\eeq 
Starting from \eq{resMellin_general} we can proceed in making several simplifications to cast the equation in different forms.
For example, taking $\mu=Q$ and the boundary scales exactly to their canonical values we obtain
\begin{align}\label{eq:resMellin_canonical}
	\frac{\df \sigma^\text{res}}{\df N_1 \df N_2} &=\sum_{i,j}H_{i,j}(\as(Q)) B^Y_i(N_2,\mu^*_{B_n}) B^Y_j(N_1,\mu^*_{B_\bn}) S(\as(\mu^*_S))\\ 
	&\quad\times  \exp\left\{\int_{\frac{Q}{\sqrt{N_1}}}^{Q} \frac{\df \mu^\prime}{\mu^\prime} \left[2 \GammaC[\as(\muprime)] \ln \left( \frac{N_1 \mu^{\prime \, 2}}{Q^2} \right) + \gamma_B[\alpha_s(\muprime)] \right] \right\} \nn \\
	&\quad\times  \exp\left\{\int_{\frac{Q}{\sqrt{N_1 N_2}}}^{Q} \frac{\df \mu^\prime}{\mu^\prime} \left[-2 \GammaC[\as(\muprime)] \ln \left( \frac{N_1 N_2 \mu^{\prime \, 2}}{Q^2} \right) + \gamma_S[\alpha_s(\muprime)] \right] \right\} \nn \\
	&\quad\times  \exp\left\{\int_{\frac{Q}{\sqrt{N_2}}}^{Q} \frac{\df \mu^\prime}{\mu^\prime} \left[2 \GammaC[\as(\muprime)] \ln \left( \frac{N_2 \mu^{\prime \, 2}}{Q^2} \right) + \gamma_B[\alpha_s(\muprime)] \right] \right\} \nn 
\end{align}
where we used the short hand notation
\beq
	B^Y_i(N_2,\mu^*_{B_n}) \equiv B^Y_i\Bigl(\ln \frac{N_1 \mu_{B_n}^{*\,2} }{Q^2},N_2,\mu^*_{B_n}\Bigr) = B^Y_i\Bigl(0,N_2,\mu^*_{B_n}\Bigr)\,,
\eeq
indicating the rapidity beam function boundary evaluated at its canonical scale.

Note that \eqs{resMellin_general}{resMellin_canonical} are for the hadronic cross section, since the rapidity beam functions contain the PDFs. 
One can factor out the PDFs to obtain an expression for the partonic cross section using \eq{beam_master}. 
In Mellin space \eq{beam_master} takes the form
\beq\label{eq:beam_master_Mellin}
	 B_i\left(\ln \frac{N_a \mu^2}{Q^2},N_b,\mu\right)= \sum_k \cI_{i,k}\left(\ln \frac{N_a \mu^2}{Q^2},N_b,\mu\right)f_k(N_b,\mu)\,,
\eeq
so that 
\begin{align}
	B^Y_i(N_2,\mu^*_{B_n}) &= \sum_k \cI_{i,k}(0,N_2,\mu^*_{B_n})f_k(N_2,\mu^*_{B_n})\,,\nn\\ 
	B^Y_j(N_1,\mu^*_{B_\bn}) &= \sum_\ell \cI_{j,\ell}(0,N_1,\mu^*_{B_\bn})f_\ell(N_1,\mu^*_{B_\bn})\,.
\end{align}
It is then an easy exercise to rewrite \eq{resMellin_general} or \eq{resMellin_canonical} as
\beq\label{eq:resMellin_PDFout}
	\frac{\df \sigma^\text{res}}{\df N_1 \df N_2} = \sum_{k,\ell} f_k(N_1,Q) \hat{\sigma}^\text{res}_{k,\ell}(N_1,N_2,Q) f_\ell(N_2,Q)\,.
\eeq
It is important to notice that in writing \eq{resMellin_PDFout} one must use DGLAP to evolve the PDFs from the scales $\mu^*_{B_n}$ and $\mu^*_{B_\bn}$ in \eq{beam_master_Mellin} to the hard scale $Q$.
This is evident also from the RG equation of the matching kernels, see \eq{Ikernel_RGE}, that in double Mellin space takes the form
\beq
	\mu \frac{\df}{\df \mu} \cI_{ij}\left(\ln \frac{N_a \mu^2}{Q^2},N_b,\mu\right) = \left[\gamma_B\left(\ln \frac{N_a \mu^2}{Q^2},\mu\right)\delta_{jk} - \Gamma_{kj}(N_b,\mu)\right] \cI_{ik}\left(\ln \frac{N_a \mu^2}{Q^2},N_b,\mu\right)\,,
\eeq
with $\Gamma_{kj}(N_b,\mu)$ being the Mellin transform of the Altarelli-Parisi splitting functions.